\documentclass[authoryear,12pt]{elsarticle}
\makeatletter
   \def\ps@pprintTitle{%
      \let\@oddhead\@empty
      \let\@evenhead\@empty
	  \def\@oddfoot{\centerline{\thepage}}%
      \let\@evenfoot\@oddfoot
   }
\makeatother
\usepackage[T1]{fontenc}
\usepackage[normalem]{ulem}
\usepackage[utf8]{inputenc}
\usepackage{amssymb}
\usepackage{subfiles} 
\usepackage{subcaption}
\usepackage{multicol}
\usepackage{bm}
\usepackage{graphicx}
\usepackage[export]{adjustbox}
\usepackage[font=footnotesize]{caption}
\usepackage{float} 
\usepackage{placeins}
\usepackage[dvipsnames]{xcolor}
\usepackage{longtable}

\usepackage{amsmath,stmaryrd,footmisc}
\usepackage{esint}
\usepackage{placeins}
\usepackage{algorithm}
\usepackage{algpseudocode}
\usepackage{esvect}
\MakeRobust{\vv}

\usepackage{xcolor}
\usepackage{epstopdf}
\usepackage{natbib}
\usepackage{nohyperref}
\usepackage{graphicx}
 \usepackage{setspace}
\usepackage{color}
\usepackage{verbatim}
 \usepackage[usenames,dvipsnames]{pstricks}
 \usepackage{epsfig}
 \usepackage{pst-grad} 
 \usepackage{pst-plot} 
 \usepackage{fancyref}

\usepackage{etoolbox}
\usepackage{mathtools}
\preto\subequations{\ifhmode\unskip\fi}

\usepackage{physics}
\usepackage{amsmath}

\usepackage{placeins}
\usepackage{siunitx}

\newlength{\tleft}
 \newlength{\tright}
\def\nten#1{\mathbf{#1}}

\renewcommand{\vec}[1]{\mathbf{#1}}

\def\dD0{\mathcal{\partial D}_0}
\def\D0{\mathcal{D}_0}

\def\dV0{\dd{V_0}}

\def\dA0{\dd{A_0}}
\def\lam{\lambda}

\journal{Journal of the Mechanics and Physics of Solids}
\usepackage{cleveref}

\begin{document}

\begin{frontmatter}

\title{A large deformation theory for coupled swelling and growth with application to growing tumors and bacterial biofilms}

\author[label1]{S. Chockalingam}
\author[label2,label3]{T. Cohen\corref{cor1}}
\cortext[cor1]{Corresponding author: talco@mit.edu}

\address[label1]{Massachusetts Institute of Technology, Department of Aeronautics and Astronautics, Cambridge, MA, 02139, USA}
\address[label2]{Massachusetts Institute of Technology, Department of Civil and Environmental Engineering, Cambridge, MA, 02139, USA}
\address[label3]{Massachusetts Institute of Technology, Department of Mechanical Engineering, Cambridge, MA, 02139, USA}

\begin{abstract}
There is significant interest in modelling the mechanics and physics of growth of soft biological systems such as tumors and bacterial biofilms. Solid tumors account for more than 85$\%$ of cancer mortality and bacterial biofilms account for a significant part of all human microbial infections. 
These growing biological systems are a mixture of fluid and solid components and increase their mass by intake of diffusing species such as fluids and nutrients (swelling) and subsequent conversion of some of the diffusing species into solid material (growth). Experiments indicate that these systems swell by large amounts and that the swelling and growth are intrinsically coupled, with the swelling being an important driver of growth. However, \textcolor{black}{many} existing theories for swelling coupled growth employ linear poroelasticity, which is limited to small swelling deformations, and employ phenomenological prescriptions for the dependence of growth rate on concentration of diffusing species and the stress-state in the system. In particular, the termination of growth is enforced through the prescription of a critical concentration of diffusing species and a homeostatic stress.  In contrast, by developing a fully coupled swelling-growth theory that accounts for large swelling through nonlinear poroelasticity, we show that the emergent driving stress for growth automatically captures all the above phenomena. Further, we show that for the soft growing systems considered here, the effects of the homeostatic stress and critical concentration can be encapsulated under a single notion of a critical swelling ratio. The applicability of the theory is shown by its ability to capture experimental observations of growing tumors and biofilms under various mechanical and diffusion-consumption constraints. Additionally, compared to generalized mixture theories, our theory is amenable to relatively easy numerical implementation with a minimal physically motivated parameter space.
\end{abstract}


\end{frontmatter}

\section{Introduction}
\label{sec:Intro}
Understanding the mechanics and physics of growth of soft biological systems ranging from cellular systems such as tumors and bacterial biofilms to tissues and organs such as arteries, lungs, skin, and the brain \citep{araujo2004history,kuhl2014growing,jain2014role,mattei2018continuum}, can aid in development of clinical therapies and diagnosis that have important societal implications. 
 For example, solid tumors account for more than 85\% of cancer mortality \citep{jain2005normalization}. Similarly, bacterial biofilms, aggregates of bacterial cells held together by an extracellular matrix, account for a significant part of all human microbial infections \citep{bryers2008medical}. These growing biological systems are a mixture of  solid components such as cells and extracellular matrix, and diffusing fluid components with dissolved solutes such as nutrients, oxygen, and growth factors. They increase their mass by intake of diffusing species from their surroundings, the process of swelling, and subsequent conversion of some of the diffusing species into additional solid material, a process henceforth referred to here as growth. \\

Experiments on bacterial biofilms indicate that swelling and growth are intrinsically coupled and that swelling is an important driver of growth \citep{seminara2012osmotic,yan2017extracellular}. The diffusing fluids supply the mass for growth and dissolved species such as nutrients, oxygen and growth factors significantly affect the growth rate. In particular, the growth rate typically increases with the concentration of diffusing species, saturating at high concentrations \citep{monod1949growth,casciari1992variations,narayanan2010silico} while the growth rate at small concentrations is well captured using a critical concentration below which growth is assumed to stop \citep{greenspan1972models,hlatky1988joint,ward1997mathematical,bertuzzi2010necrotic,araujo2004history,xue2016biochemomechanical}. There are other considerations which also make it important to account for swelling during the growth process.  For example, it is known that cell packing density of tumor cells (which is inversely related with the amount of swelling) increases when they grow against a stiff medium \citep{helmlinger1997solid} or through application of external pressure \citep{koike2002solid,alessandri2013cellular,chalut2014clamping} and such compaction can set the basis for a multicellular-dependent mechanism of increased radiation resistance and both intrinsic
and acquired drug resistance \citep{olive1994drug,kobayashi1993acquired,kerbel1994multicellular,croix1996reversal}. Similarly, compactness of bacterial biofilms, which also increases with confining stiffness (SI of \cite{zhang2021morphogenesis}), is a key determinant of underlying resistance to invader cells \citep{nadell2015extracellular,yan2017extracellular}. A fundamental understanding of the physics of coupled swelling-growth in these systems can aid potential development of clinical therapies, for example, \cite{croix1996reversal} showed that a decrease in cell compaction, as induced by hyaluronidase treatment of tumor spheroids, alleviated multicellular-dependent drug resistance. Similarly, \cite{yan2017extracellular} demonstrate the possibility of using external osmolytes to control the compactness of pathogenic biofilms, which is critical to the penetration of antibiotic molecules.  \\

Further, the extracellular matrix in these cellular systems are akin to hydrogels and can swell by significant amounts \citep{yan2017extracellular} where the solid volume fraction can be as low as 20\% (SI of \cite{zhang2021morphogenesis} and \Cref{subsec:biofilm_model} of this manuscript). Thus a large deformation coupled swelling-growth theory is required to accurately characterize and elucidate the physics in these growing systems. However, current growth theories that account for diffusing species typically either model the diffusion as an auxiliary process that neglects swelling, mechanics coupling, and mass balance \citep{ambrosi2002mechanics,ambrosi2004role, ambrosi2007growth,kim2011role,ciarletta2013mechano}, or employ linear poroelasticity \citep{roose2003solid,sarntinoranont2003interstitial,xue2016biochemomechanical,sacco2017poroelastic, xue2018biochemomechanical,carpio2019biofilms} based on the seminal works of Biot \citep{biot1941general,biot1972theory}, which is limited to relatively small deformations \citep{bouklas2012swelling}. Mixture theories of growth on the other hand account for the mechanics and motion of every phase in the growing body {\citep{humphrey2002constrained,ambrosi2002closure,byrne2003two,garikipati2004continuum,ateshian2007theory,azeloglu2008heterogeneous,narayanan2009micromechanics,
ateshian2010multigenerational,chatelain2011emergence,ciarletta2011radial,amar2011contour,ateshian2012mechanics,ateshian2014computational,myers2014interstitial,faghihi2020coupled} }and can account for large deformations and the coupling between swelling and growth. However they also suffer from several drawbacks - (i) The requirement for specification of a large number of constitutive relations and associated material parameters \citep{ambrosi2011perspectives,oden2016toward} that can be both impractical to experimentally determine and to model, (ii) Difficulties associated with specifying boundary conditions and in dealing with partial stresses and mass exchanges between the single phases \citep{ambrosi2010insight}, and (iii) Complexity of numerical implementation with more equations being modelled than might be necessary to describe the required physics.\\

In addition to the shortcomings discussed above, a major limitation  of the vast majority of growth theories in the literature is the use of phenomenological prescriptions to capture important experimental observations. For example, most growth theories capture the earlier discussed dependence of the growth rate on concentration of diffusing species using phenomenological prescriptions including a critical concentration that halts growth. Similarly, when it comes to the effects of mechanical constraints, it is generally known that applied compressive stresses deter volumetric growth. A widely established concept in the literature is that of `homeostatic stress', which prescribes a preferred stress state that growing systems tend towards and upon reaching which the growth process halts \citep{rodriguez1994stress,taber1998model,fung2013biomechanics}. In terms of theory development, the homeostatic stress is mostly introduced in an ad hoc phenomenological fashion \citep{rodriguez1994stress,taber1998model,dicarlo2002growth,lubarda2002mechanics,ambrosi2007stress,menzel2012frontiers,mpekris2015stress,xue2016biochemomechanical,curatolo2017swelling}. Under compressive stress states, which is generally the case for many of these growing systems, this ad hoc homeostatic term often becomes the primary positive driving term in evolution laws for the growth process. This  kinematic modelling of the physics  does not explain the kinetic origins of these phenomena starting from the underlying mechanisms. It further limits the confidence in general applicability of the theories to scenarios beyond the specific experimental data they are calibrated to capture.
\\

In light of all the above considerations, the goal of this manuscript is \textit{to formulate a homogenized single phase continuum theory for fully coupled swelling and growth that accounts for large swelling deformations and mass balance, while capturing important coupled swelling-growth phenomena using underlying kinetics instead of ad hoc kinematic prescriptions}. To do so, we will augment growth modelling frameworks \citep{rodriguez1994stress,dicarlo2002growth,ambrosi2002mechanics,ambrosi2007growth,ambrosi2007stress,menzel2012frontiers,kuhl2014growing} with \textcolor{black}{nonlinear poroelasticity formulations developed for soft swelling elastomers and polymeric gels \citep{hong2008theory,doi2009gel,duda2010theory,chester2010coupled,chester2011thermo,lucantonio2013transient,liu2015advances} and prescribe the swelling and growth evolution using the kinetic driving terms that arise naturally. {We note in passing that a nonlinear poroelastic swelling growth model has also been developed in \cite{fraldi2018cells}, assuming moderate variations in fluid content.} 
A large deformation swelling-growth theory has also been developed by \cite{curatolo2017swelling}, in the context of bulking of wood, wherein growth is simply a change in relaxed state instead of an increase in mass of the solid as considered here, while employing phenomenological prescriptions including for the homeostatic stress.} \textcolor{black}{See also \cite{dervaux2011buckling} where instability arising from large deformation growth is compared with that arising from large swelling.}\\  

The paper is organized as follows. In  \Cref{sec:Overview_tumor}, we begin by discussing a set of tumor growth experiments that encapsulate all the phenomena discussed above
and use it to motivate the theory development and analysis to follow in subsequent sections. The governing equations of the theory are then developed in \Cref{sec:theory}. The theory is specialized in \Cref{sec:spec_constit} by choosing specific forms of the constitutive functions and non-dimensionalized to allow for ease of analysis and to make limiting approximations. Subsequently, in \Cref{sec:Results} we analyze the theory and use it to solve boundary value problems and model experiments of growing tumor and bacterial biofilms. We provide some concluding remarks in \Cref{sec:Conclusions}.

\section{Motivating example of tumor growth experiments}
\label{sec:Overview_tumor}
To motivate the theory development in the following sections, we begin by discussing a set of tumor growth experiments that encapsulates the complete spectrum of phenomena that the theory should capture.

  \begin{figure}[H]
    \centering
    \includegraphics[width=\textwidth]{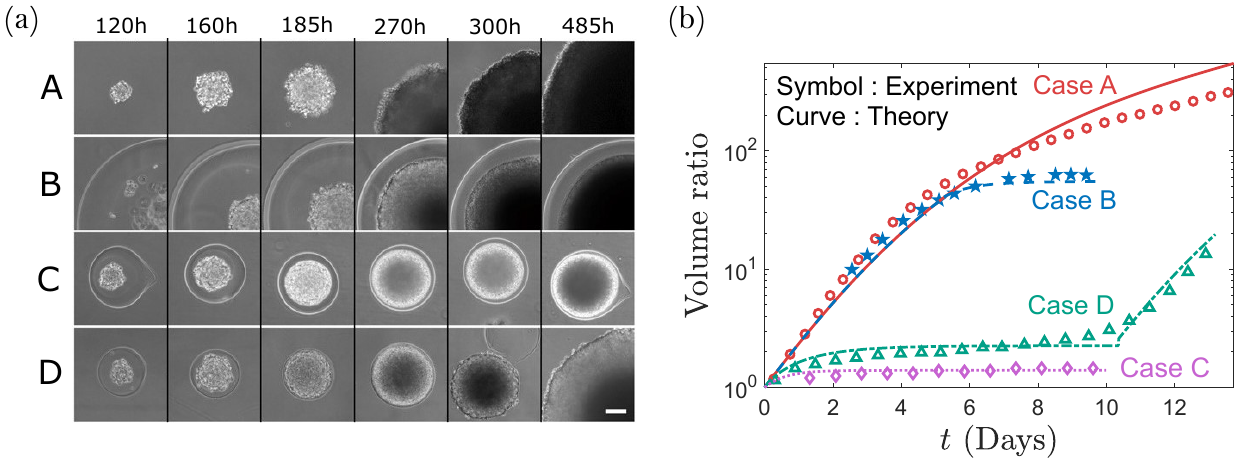}
    \caption{(a) Experiments of \textit{in vitro} spheroid tumor growth in four different environments: (A) free growth, (B) large and thick microcapsule confinement, (C) small and thick microcapsule confinement, and (D) small and thin microcapsule confinement that breaks. Scale bar: 50 $\mu$m. Reprinted from \cite{alessandri2013cellular}. (b) Comparison of experimental data of time evolution of tumor volume ratio (ratio of tumor volume at time $t$ to its value at $t=0$) with predictions of our swelling growth theory (see \Cref{subsec:tumorresults}). Volume of tumor at $t=0$ is 0.003 mm$^3$.} \label{fig:tumor}
\end{figure}

\cite{alessandri2013cellular} performed microfluidic experiments to study tumor growth in response to mechanical confinement using  spheroids of CT26 mouse colon carcinoma cells. The experiments studied multiple scenarios: free growth with no confinement (Case A) and growth under the confinement of alginate microcapsules of different radii and thicknesses (Cases B - D). Images from the experiments are shown in \Cref{fig:tumor}(a) while the experimental data for the evolution of the tumor volume ratio (ratio of volume at time $t$ to its value at $t=0$) is plotted in \Cref{fig:tumor}(b). The experimental data was time-shifted so that in all cases, the volume at $t=0$ is 0.003 mm$^3$ which is the inner volume of the smallest confining microcapsule. The
microcapsules were permeable and thus the surface of the tumors were supplied with required fluids, nutrients and oxygen for growth. Below are the observations for these four typically studied cases.

\begin{enumerate}
  \item In Case A, the free growth is initially exponential followed by a power-law volume increase. Further growth leads to a necrotic/dying core while growth is confined to thin rim of proliferating cells. 
  \item In Case B, the tumor undergoes free growth until it comes in contact with a large and thick microcapsule following which the growth rate is rapidly suppressed. 
  \item In Case C, the tumor is confined by a small and thick microcapsule, and the growth is inhibited by the confinement as it deforms.
\item In Case D, the tumor is confined by a small and thin microcapsule. The growth rate falls as the tumor expands against the confinement but the large deformation causes the microcapsule to burst following which the tumor resumes exponential growth similar to free-growth case.
\end{enumerate}
Conventional growth theories would capture these experiments using phenomenological prescriptions for the established phenomena discussed in \Cref{sec:Intro}, namely the growth rate dependence on concentration of diffusing species and applied stresses, and in particular the existence of a critical concentration and homeostatic stress \citep{xue2016biochemomechanical}. The diffusion of fluids and dissolved nutrients from the outside of the tumor to its core, combined with consumption for growth leads to a radially decreasing concentration profile from the outside to inside. While the outer edge of the growing body remains replenished with diffusing species, the spatial decrease of concentration towards the core becomes more pronounced as the body grows exponentially to larger sizes. Initially the concentration everywhere is high and the growth rate is at a saturated value leading to the exponential growth. But at larger sizes the growth rates are spatially decreasing towards the core leading to the decrease in overall growth rate. The growth in the core halts when the concentration reaches the critical value. Eventually most of the tumor stops growing and the growth localizes to a very thin proliferating ring near the outside. Further, increasing compressive stresses when growing against mechanical confinement causes the growth rate to fall off, till the homeostatic value is reached when growth halts and a steady size is reached. In the case of thin confinement, the confinement breaks before homeostatic stress is reached, leading to resumption of free-growth.\\

Such modelling of the above experiments is fundamentally handicapped by the reliance on  phenomenological prescriptions and raises several questions. Why does the growth process accelerate with increasing concentration of diffusing species while saturating at high values and what causes it to halt below a critical concentration? Why is there a mechanical stress state that halts growth? Can these seemingly unlinked diffusion-consumption and mechanical phenomena be described by a unified kinetic theory? We will answer these questions in the following sections by developing a fully coupled large deformation swelling-growth theory where we use the kinetic driving stress that naturally arises to model the growth process.

\section{Theory}
\label{sec:theory}

The key ingredients and assumptions of the swelling-growth theory are summarized below following which we develop the theory.

\begin{enumerate}
    \item Solid components such as cells and the \textcolor{black}{extracellular matrix} are abstracted and their behavior is assumed to be described by models for hydrogels. 
    \item While biological growth generally depends on several diffusing components such as {water/}fluids, nutrients, oxygen, glucose, and growth factors, we consider a single representative fluid phase whose concentration reflects the concentration of dissolved solutes, which we henceforth refer to as the diffusing species. {Though the physical or mechanical swelling is primarily a result of the diffusing water, the bio-chemical energy of the dissolved components such as nutrients and growth factors is an important driver of growth}. While the theory here can be generalized to multiple diffusing components, we will show that a single representative species is sufficient to capture the essential physics.
    \item The consumption of the diffusing species supplies the mass for solid growth. The remaining diffusing species 
    swells the grown solid. While growth can involve various complex cellular processes including cell division and production of extracellular matrix, considering continuum scale phenomena here, 
    it is abstracted as conversion of the diffusing species into additional solid material. The chemical energy for the conversion process is thus a homogenized value from all the underlying cellular processes.
    \item The growing body is described by a fixed set of material points and growth and swelling are modelled as an increase in volume of these material points in appropriate relaxed spaces mapped from a reference space through the widely employed decomposition of deformation gradient \citep{rodriguez1994stress,ambrosi2002mechanics,chester2010coupled}. The volume increases correspond to an increase in mass of the homogenized continuum through closure of mass balance. This is in contrast to several growth theories in the literature where an intrinsic mass supply is assumed and open system thermodynamics is employed \citep{ambrosi2004role,kuhl2014growing,xue2016biochemomechanical}.
    \item We neglect any viscoelastic and inertial effects at the timescale of growth. 
\end{enumerate}

\subsection{Kinematics}
\label{subsec:Fdecomp}

\begin{figure}
    \centering
    \includegraphics[width=0.7\textwidth]{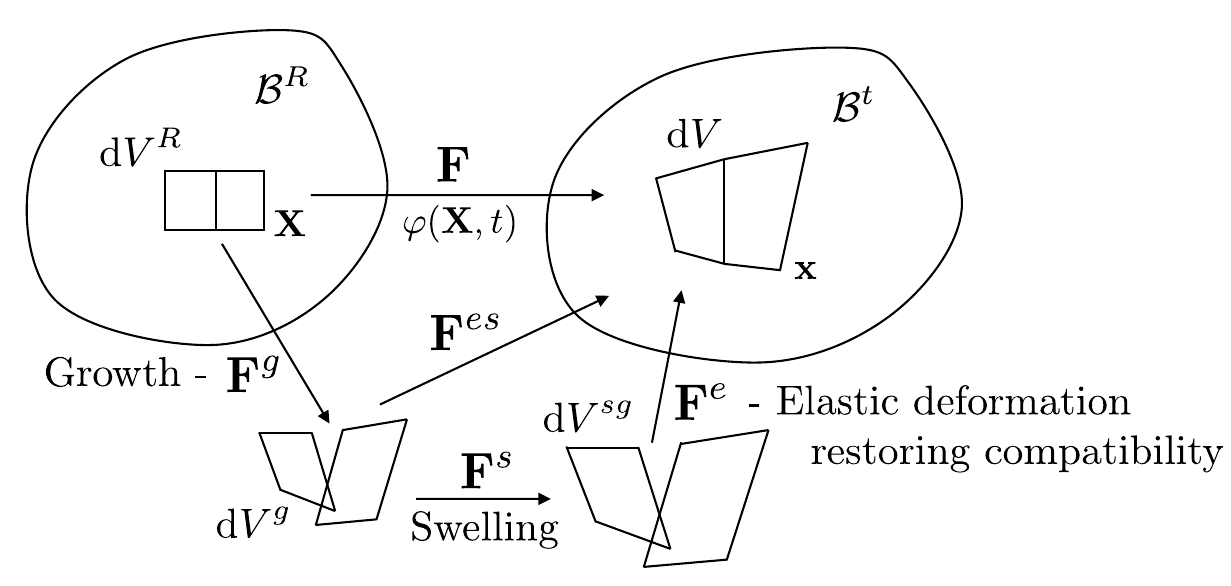}
    \caption{Decomposition of the deformation gradient.}
    \label{fig:Fdecomp}
\end{figure}

Consider an initial compatible dry reference configuration of the body in the absence of diffusing species denoted by $\mathcal{B}^R$.
Let $\mathcal{B}^t$ denote the current/observed configuration of the swelling and growing body at time $t$. Consider a material point at the position $\nten{X}$ in the dry reference configuration whose position in the current configuration is given by $\nten{x} = \varphi(\nten{X},t)$.  We define the deformation gradient $\nten{F}$  and its decomposition 
 as follows\footnote{The Lagrangian spatial gradient operator $\nabla(\cdot)$  is defined as $\nabla(\cdot) = \pdv{(\cdot)}{\nten{X}}$} (see \Cref{fig:Fdecomp})
\begin{equation}
\nten{F}= \nabla \bm{\varphi} = \nten{F}^{es} \nten{F}^g = \nten{F}^{e} \nten{F}^{s} \nten{F}^g,\quad \nten{F}^{s} = \lam^s \ \nten{I},\quad \nten{F}^{es} = \nten{F}^e \nten{F}^s \label{eq:Fdecomp}
\end{equation}
where $\nten{F}^g$ is the growth tensor describing the change in solid mass, $\nten{F}^s$ is the swelling tensor describing the change in mass due to diffusing species, and $\nten{F}^e$ is the elastic deformation tensor. The swelling tensor is taken to be isotropic where $\lambda^s$ is the swelling stretch (similar to \cite{chester2010coupled,chester2011thermo}). The growth tensor $\nten{F}^g$ maps the material point at $\nten{X}$ in the dry reference configuration to its relaxed unswollen state that would be obtained if the particle associated with $\nten{X}$ is cut out from the body and drained of its diffusing species content while retaining the mass of the solid skeleton. The density of the solid matrix is assumed to be constant from its value in dry reference space, $\rho_0^m$, through the mapping by $\nten{F}^g$. The tensor $\nten{F}^{es}(\nten{X})$ maps the relaxed unswollen state to the current state of the particle. Note that the configuration of the body in the intermediate spaces mapped by $\nten{F}^g$ and $\nten{F}^s$ do not have any physical requirement of compatibility unlike $\mathcal{B}^t$. The displacement $\nten{u}$ is defined as $\nten{u}(\nten{X},t) = \nten{x} - \nten{X}.$\\

 We define the following volumetric ratios (all assumed to be positive)
\begin{align}
&J^g = \textrm{det}\ {\nten{F}^g}, \quad J^s = \textrm{det}\ {\nten{F}^s} = (\lam^s)^3, \quad J^e = \textrm{det}\ {\nten{F}^e},   \label{eq:determinants_combined}\\
&J^{es} = \textrm{det}\ {\nten{F}^{es}} = J^eJ^s , \quad J =\textrm{det}\ {\nten{F}} = J^gJ^eJ^s, \label{eq:determinants_combined2}
\end{align}
such that the volume of a particle in the dry reference space, $\dd{V}^R$, relates to its current volume $\dd{V}$ and volume in the intermediate grown and swollen-grown spaces, $\dd{V}^g$ and $\dd{V}^{sg}$ respectively, as
\begin{equation}
\dd{V}^g = J^g \dd{V}^R,\quad \dd{V}^{sg} = J^sJ^g \dd{V}^R, \quad \dd{V} = J \dd{V}^R 
\end{equation}
The volume fraction of solid, $\phi$, is then defined as
\begin{equation}
\phi = \frac{\dd{V}^{g}}{\dd{V}^{gs}} = \frac{J^g \dd{V}_R}{J^g J^s \dd{V}_R} = \frac{1}{J^s}
\label{eq:solidvolfrac}
\end{equation}
so that it is inversely related to the swelling ratio $J^s$. We have the physical requirement $\phi \in (0,1]$ or equivalently $1\leq J^s<\infty$. Let $\dd{A}$ and $\dd{A}_R$ denote corresponding area elements in the current and dry reference space with outward pointing normals $\nten{n}$ and $\nten{n}_R$, respectively. They are related by Nanson's formula - $\nten{n} \dd{A} = J \nten{F}^{-T} \nten{n}_R \dd{A}_R$.\\

The following deformation tensors are defined for later use,
\begin{equation}
\nten{C}^e  = {\nten{F}^{e}}^T\nten{F}^{e},\quad \nten{C}^{es}  = {\nten{F}^{es}}^T\nten{F}^{es} = (J^s)^{\frac{2}{3}} \nten{C}^e,\quad \nten{B}^e  = \nten{F}^{e}{\nten{F}^{e}}^T,\quad \nten{B}^{es}  = \nten{F}^{es}{\nten{F}^{es}}^T = (J^s)^{\frac{2}{3}} \nten{B}^e \label{eq:Cedef3}
\end{equation}
Using \cref{eq:Fdecomp}, we can derive the following rate relation 
\begin{equation}
\nten{L} = \dot{\nten{F}} {\nten{F}}^{-1} = \nten{L}^e + \nten{F}^e \nten{L}^g {\nten{F}^e}^{-1} + \frac{\dot{J}^s}{3 J^s} \nten{I} \label{eq:RateFdecomp}
\end{equation}
where an overdot represents material time derivative and 
\begin{equation}
\nten{L}^e = \dot{\nten{F}}^e {\nten{F}^e}^{-1}, \quad \nten{L}^g = \dot{\nten{F}}^g {\nten{F}^g}^{-1} \label{eq:Ls} 
\end{equation}
 The growth rate tensor $\nten{L}^g$ is then decomposed as follows
\begin{equation}
\nten{L}^g = \frac{\Gamma}{3} \nten{I} + \text{dev}(\nten{L}^g),\quad \Gamma  = \text{tr}(\nten{L}^g) = \frac{\dot{J}^g}{J^g}
\label{eq:Lgdecomp}\end{equation}
where $\Gamma$ is the growth rate and $\text{dev}(\cdot)$ represents the deviatoric operator\footnote{$\textrm{dev}(\nten{Z}) = \nten{Z} - \frac{1}{3}\textrm{tr}(\nten{Z})\nten{I}$}. The deviatoric part of $\nten{L}^g$ is associated with the directionality of growth in the dry reference space. When the volumetric growth rate $\Gamma$ is zero, the growth directionality represents remodelling where the grown reference state is changing from processes like cellular rearrangement or changes in micro-structure. \\

\subsection{Balance of mass}
Mass balance of diffusing species can be written as
\begin{equation}
\dot{c}_R +  \text{Div}(\nten{j}_R^M) = - \dot{\xi}_R \label{eq:massbalance}
\end{equation}
where $\xi_R$ is the mass of diffusing species per unit volume of the dry reference configuration that has been consumed during growth and 
the concentration $c_R$ is the mass of remaining diffusing species per unit dry reference volume. The vector $\nten{j}_R^M$ is the referential diffusion flux written in units of mass of diffusing species per unit area per unit time. The mass balance equation \cref{eq:massbalance} can be rewritten in units of volume by multiplying it by $\Omega^f$, the referential volume per unit mass of diffusing species\footnote{In terms of the referential density (no elastic deformation) of diffusing species, $\rho_0^f$, we have the relation $\Omega^f = 1/\rho_0^f$.},
\begin{equation}
\Omega^f \dot{c}_R +  \text{Div}(\nten{j}_R) = - \Omega^f \dot{\xi}_R, \quad  \nten{j}_R = \Omega^f \nten{j}_R^M \label{eq:massbalance_vol}
\end{equation} 
where $\nten{j}_R$ is the referential diffusion flux written in terms of volume of diffusing species per unit area per unit time.We can relate the remaining concentration of diffusing species per unit grown reference volume, $c_g$, to $c_R$ as
\begin{equation}
c_R = c_g J^g  \label{eq:concentrations}
\end{equation}\\

\textit{Swelling constraint:} The change in $J^s$ is assumed to arise entirely due to the change in the remaining diffusing species content, which yields the swelling constraint 
\begin{equation}
\dot{J}^s = \Omega^f \dot{c}_g \label{eq:swellingconstraint}
\end{equation}
Integration of \cref{eq:swellingconstraint} along with the condition that the swelling ratio is unity when the solid is dry ($c_g=0$) results in the following constraint,
\begin{equation}
J^s = 1 + \Omega^f {c}_g  \label{eq:constraintcombined1} 
\end{equation}\\

\textit{Growth constraint:} The consumption of the diffusing species supplies mass for the growth and leads to the following volumetric growth relation {(also see \ref{app:massbalance})}
\begin{equation}
\dot{J}^g = \Omega^m \dot{\xi}_R \label{eq:constraintgrowth}
\end{equation}
where $\Omega^m$ is the referential volume per unit mass of the solid matrix ($\Omega^m = 1/\rho^m_0$). 
The integrated form of \cref{eq:constraintgrowth} reads 
\begin{equation}
J^g = 1 + \Omega^m \xi_R \label{eq:constraintgrowth_integ}  
\end{equation}
where the growth volume ratio is unity when no diffusing species has been consumed ($J^g =1$ when  $\xi_R = 0$). {The constraint \cref{eq:constraintgrowth_integ} ensures mass balance during species conversion.} \\

The \cref{eq:concentrations,eq:swellingconstraint,eq:constraintgrowth} can be used to rewrite \cref{eq:massbalance_vol} as
\begin{equation}
J^g \dot{J}^s +  \dot{J}^g \left(J^s -1 + \frac{\Omega^f}{\Omega^m}\right) + \text{Div}(\nten{j}_R) = 0 \label{eq:massbalanceJsJg}
\end{equation}
The  \cref{eq:constraintcombined1,eq:constraintgrowth_integ,eq:massbalanceJsJg} together ensure mass balance. Since the density of solid matrix is typically close to the density of the fluid components in our systems of interest, here we assume isochoric conversion of the diffusing species into solid so that $\Omega^f = \Omega^m = \Omega$ . Under this assumption, \cref{eq:massbalanceJsJg} simplifies to
\begin{equation}
J^g \dot{J}^s +  J^s \dot{J}^g + \text{Div}(\nten{j}_R) = 0 \label{eq:massbalanceJsJg_isocho}
\end{equation}
See \ref{app:nonisochoric} for the general version of the theory for non-isochoric species conversion.

\subsection{Mechanical equilibrium}
Let $\nten{T}(\nten{x}, t)$ and $\nten{S}(\nten{X}, t)$ denote the Cauchy and Piola stress tensor fields, respectively. \textcolor{black}{Since swelling and growth are usually much slower than the mechanical response of the body, we ignore the 
momentum arising from swelling-growth.} Mechanical
equilibrium in the absence of body forces and inertial effects requires that \textcolor{black}{(see also \cite{ambrosi2004role,xue2016biochemomechanical})}
\begin{equation}
\text{div}\ \nten{T} = \vec{0}, \quad \nten{T}= \nten{T}^T 
\end{equation}
and equivalently that
\begin{equation}
\text{Div}\ \nten{S} = \vec{0}, \quad \nten{S}\nten{F}^{T}= \nten{F}\nten{S}^T, \quad \text{where}\quad \nten{S} = J \nten{T} \nten{F}^{-T} \label{eq:Piolastresseqb}
\end{equation}

\subsection{Dissipation inequality}
Under isothermal conditions and in absence of body forces, the first two laws of thermodynamics collapse into the following free energy imbalance equation over any material subregion P in the body {(see \ref{app:disspineq})}
\begin{equation}
\int_{\text{P}} \dot{\psi}_R\dd{V_R} \le \int_{\partial\text{P}} (\nten{S} \nten{n}_R)\cdot  \dot{\bm{\varphi}} \dd{A_R} + \int_{\partial \text{P}} \mu (-\nten{j}_R \cdot \nten{n}_R) \dd{A_R}   \label{eq:dissipineq_combined}
\end{equation}
where $\psi_R$ is the {Helmholtz} free energy per unit dry reference volume {of the fluid-solid continuum within the dry material region P} and $\mu$ is the chemical potential of the diffusing species\footnote{Typically the chemical potential $\mu$ is written in terms of energy per species unit (mole or molecule) since the flux is written in terms of species units per unit area per unit time, see for example \cite{hong2008theory,chester2010coupled}. However, since the flux in this manuscript is written in terms of referential species volume per unit area per unit time, the chemical potential is in terms of energy per unit referential species volume.}. Localizing \cref{eq:dissipineq_combined}, while employing \cref{eq:massbalanceJsJg_isocho} 
yields the local dissipation inequality (see \ref{app:disspineq})
\begin{equation}
\nten{S} \cdot \dot{\nten{F}} + \mu (J^g \dot{J}^s + J^s \dot{J}^g) - \nten{j}_R \cdot \nabla \mu - \dot{\psi}_R \ge 0 \label{eq:dissip1} 
\end{equation}
It can be shown using \eqref{eq:RateFdecomp} that (see \ref{app:stresspower}) 
\begin{equation}
\nten{S} \cdot \dot{\nten{F}} =  \frac{1}{2}J^sJ^g\nten{T}^e\cdot\dot{\nten{C}^e} + J^g \nten{M}^{es} \cdot \nten{L}^g + \frac{J \dot{J}^s}{3 J^s} \text{tr}(\nten{T})  \label{eq:strespowers}
\end{equation}
where the Mandel stress $\nten{M}^{es}$ and the elastic second Piola stress $\nten{T}^e$ are given by
\begin{equation}
\nten{M}^{es} = J^{es} {\nten{F}^{es}}^T \nten{T} {\nten{F}^{es}}^{-T},\quad \nten{T}^e = J^e {\nten{F}^e}^{-1} \nten{T} {\nten{F}^e}^{-T} \label{eq:stressdefs}
\end{equation}
Using \cref{eq:strespowers} and \cref{eq:Lgdecomp}$_2$ in \cref{eq:dissip1}, we arrive at the following form of the dissipation inequality,
\begin{equation}
J^g\left(\nten{M}^{es} + {J^s}\mu \nten{I}\right)\cdot\nten{L}^g + \frac{1}{2}J^sJ^g\nten{T}^e\cdot\dot{\nten{C}^e} + \left(\mu +  \frac{J^e}{3}\text{tr}(\nten{T})\right){J^g }\dot{J}^s  - \nten{j}_R \cdot \nabla \mu - \dot{\psi}_R \ge 0 \label{eq:dissip2}
\end{equation}
To be able to specify constitutive equations using the dissipation inequality, the form of the free energy is specified in the following section. 

\subsection{Free energy}
We consider a free energy function of the form $\psi_R = \psi_R(J^g,J^s,\nten{C}^e)$, which we decompose as follows
\begin{align}
\psi_R &= J^g {\psi}_g(\nten{C}^e,J^s) + \mu_0^m \Omega^m \xi_R  +  \mu_0^f \Omega^f c_R\label{eq:Freeenergycombined}\\
& = J^g {\psi}_g(\nten{C}^e,J^s) + {\mu_0^m}(J^g-1)  +  {\mu_0^f}J^g  (J^s-1) \qquad (\text{Using}\ \eqref{eq:concentrations}, \eqref{eq:constraintcombined1},\eqref{eq:constraintgrowth_integ})\\
\psi_g &= {\psi}_g^{\text{mech}}(\nten{C}^e,J^s)  + {\psi}_g^{\text{mix}}(J^s) \label{eq:Freeenergy_g}
\end{align}
where \\
\noindent (i) $\mu_0^m$ is the reference chemical potential for the solid and $\mu_0^m \Omega^m \xi_R$ is the chemical energy of formed solid per unit dry reference volume. \\
(ii) $\mu_0^f$ is the reference chemical potential for the diffusing species and $\mu_0^f \Omega^f c_R$ is the biochemical energy of the diffusing species per unit dry reference volume. It represents the reference energy of the diffusing species with all its dissolved nutrients and growth factors.\\
(iii) ${\psi}_g^{\text{mech}}$ is the change in free energy due to the
deformation of the solid, per unit grown reference volume.\\
(iv) $\psi_g^{\text{mix}}$ is the change in free energy due to mixing of
the remaining diffusing species with the solid, per unit grown reference volume.\\

\noindent A specific form of the free energy is provided later. Differentiating \cref{eq:Freeenergycombined} and employing \cref{eq:concentrations,eq:constraintgrowth,eq:Lgdecomp,eq:swellingconstraint,eq:constraintcombined1} yields the following equation,
\begin{equation}
\dot{\psi}_R = J^g\left( \psi_g + \mu_0^m + \mu_0^f(J^s-1)\right)\nten{I}\cdot\nten{L}^g + J^g \pdv{\psi_g}{\nten{C}^e}\cdot\dot{\nten{C}}^e +    J^g\left({\mu_0^f} + \pdv{\psi_g}{J^s}\right)\dot{J}^s \label{eq:psi_Rdot}
\end{equation}

\subsection{Constitutive response}

Substituting \cref{eq:psi_Rdot} in the dissipation inequality \cref{eq:dissip2} yields
\begin{equation}
J^g\nten{T}^g\cdot\nten{L}^g + \left(\frac{1}{2}J^sJ^g\nten{T}^e - J^g\pdv{\psi_g}{\nten{C}^e}\right)\cdot\dot{\nten{C}^e}   
+ {J^g }\left(\mu +  \frac{J^e}{3}\text{tr}(\nten{T}) -  \pdv{\psi_g}{J^s} -\mu_0^f\right)\dot{J}^s  - \nten{j}_R \cdot \nabla \mu \ge 0
\end{equation} 
where $\nten{T}^g$ is the driving stress conjugate to the growth deformation rate tensor $\nten{L}^g$, given by 
\begin{equation}
\nten{T}^g = \nten{M}^{es} + \left(J^s\mu -\mu_0^m - \mu_0^f\left(J^s-1\right)-  \psi_g  \right) \nten{I} \label{eq:Fg1}
\end{equation}
Employing the Coleman-Noll procedure we arrive at the following two constitutive equations for the elastic second Piola stress and chemical potential,
\begin{equation}
 \nten{T}^e =\frac{2}{J^s} \pdv{{\psi}_{g}}{\nten{C}^e}, \quad \mu = \mu_0^f +  \pdv{{\psi}_{g}}{J^s} -  \frac{1}{3} J^e \text{tr}(\nten{T}),  \label{eq:Constit_energetic}
\end{equation}
 respectively. We can then readily derive the constitutive equations for the Cauchy, Piola, and Mandel stresses using \cref{eq:Fdecomp,eq:Piolastresseqb,eq:stressdefs} as 
 \begin{equation}
 \nten{T} = 2\frac{J^g}{J} \nten{F}^e \pdv{{\psi}_{g}}{\nten{C}^e}\/ {\nten{F}^e}^T,\quad \nten{S} = 2 J^g \nten{F}^{es} \pdv{{\psi}_g}{\nten{C}^{es}}\/{\nten{F}^g}^{-T}, \quad \nten{M}^{es} =2 \nten{C}^e \pdv{{\psi}_{g}}{\nten{C}^e} \label{eq:energetic_constit}
 \end{equation}
Further substituting {\cref{eq:Constit_energetic}}$_2$ for the chemical potential in \cref{eq:Fg1} results in the following equation for the growth driving stress,
\begin{equation}
\nten{T}^g = \nten{M}^{es} + \left({\Delta \mu_0} + J^s \left(\pdv{{\psi}_{g}}{J^s} -  \frac{1}{3} J^e \text{tr}(\nten{T})\right)-  \psi_g  \right)\nten{I} \label{eq:Tg}
\end{equation} 
where $\Delta \mu_0 = \mu_0^f -\mu_0^m$ is the energy associated with converting a unit (referential) volume of diffusing species to solid, henceforth referred to as conversion energy. \\
 
 To ensure non-negative dissipation rate, we require the following residual inequality to be satisfied 
 \begin{equation}
 J^g\nten{T}^g\cdot\nten{L}^g   - \nten{j}_R \cdot \nabla \mu \ge 0 \label{eq:dissip3}
 \end{equation}
We enforce the following inequalities to be independently satisfied, 
\begin{equation}
\nten{T}^g\cdot\nten{L}^g \ge 0,\quad \nten{j}_R \cdot \nabla \mu \le 0
\end{equation}
so that the dissipation inequality \eqref{eq:dissip3} is automatically enforced. We choose the following thermodynamically admissible prescription for the diffusion flux,
\begin{equation}
\nten{j}_R = - \nten{M}_{\text{mob}} \nabla \mu \label{eq:diffusionlaw}
\end{equation}
where the mobility tensor $\nten{M}_{\text{mob}}$ is positive semi-definite. The diffusion law in \cref{eq:diffusionlaw} can be written in the current configuration as
\begin{equation}
\nten{j} = -\frac{\nten{F} \nten{M}_{\text{mob}} \nten{F}^T}{J}\ \text{grad}\ \mu \label{eq:diffusionlaw2}
\end{equation}
where $\nten{j} = \nten{F} \nten{j}^R/J$ is the diffusion flux in the current space such that $\nten{j}\cdot\nten{n}\dd{A} = \nten{j}_R\cdot\nten{n}_R\dd{A}_R$, and grad $(\cdot)= \nten{F}^{-T} \nabla (\cdot)$ is the Eulerian spatial gradient operator. The form of the mobility tensor is specified later in \Cref{subsec:spec_constit_mobility}.

\subsubsection{Growth kinetics}
Following the previous constitutive prescriptions, we are left with the following inequality to be satisfied
\begin{equation}
\nten{T}^g\cdot\nten{L}^g =  {\Gamma}\ \frac{\text{tr}(\nten{T}^g)}{3}   +    \text{dev}(\nten{T}^g)\cdot\text{dev}(\nten{L}^g) \ge 0 \label{eq:gr_ineq}
\end{equation}
where we used\footnote{Note the identity $\nten{Y}\cdot\text{dev}(\nten{Z}) = \text{dev}(\nten{Y})\cdot\text{dev}(\nten{Z})$ } \cref{eq:Lgdecomp} to examine the separate contributions of volumetric growth and growth directionality. Any prescription for $\nten{L}^g$ that satisfies this inequality is a thermodynamically admissible growth law.  We enforce the following inequalities separately, 
\begin{equation}
     {\Gamma}\ \frac{\text{tr}(\nten{T}^g)}{3} \ge 0,\quad \text{dev}(\nten{T}^g)\cdot\text{dev}(\nten{L}^g) \ge 0 \label{eq:split_grlaw2}
\end{equation}
such that the inequality in \cref{eq:gr_ineq} is automatically satisfied.\\

\textbf{Volumetric growth rate:} Using the expression for $\nten{T}^g$ in \cref{eq:Tg} along with \cref{eq:stressdefs}$_1$, we can write \cref{eq:split_grlaw2}$_1$ as
\begin{equation}
         \Gamma f_g\ \ge 0, \quad f_g = \frac{\text{tr}(\nten{T}^g)}{3} = \Delta \mu_0 + J^s\pdv{\psi_g}{J^s} - \psi_g \label{eq:fg_first}
\end{equation}
where $f_g$ is the driving stress for volumetric growth. Any thermodynamically admissible growth law for the volumetric growth rate can be  prescribed as follows
\begin{equation}
   \Gamma = \hat{\Gamma}(f_g) \quad \text{such that} \quad \hat{\Gamma}(f_g) f_g \ge 0 \label{eq:vol_ev}
\end{equation} 
Recalling that $\psi_g = \psi_g^{\text{mech}}+\psi_g^{\text{mix}}$, the driving stress $f_g$ in \cref{eq:fg_first}$_2$ can be decomposed 
as follows
\begin{equation}
f_g = \Delta \mu_0 + f_g^{\text{mix}} + f_g^{\text{mech}} \label{eq:fgtotal}
\end{equation}
where $f_g^{\text{mix}}$ and $f_g^{\text{mech}}$ are the parts of the driving stress that arise from the mixing free energy and mechanical free energy respectively, 
\begin{equation}
f_g^{\text{mix}} = J^s\pdv{\psi_g^{\text{mix}}}{J^s} - \psi_g^{\text{mix}},\quad f_g^{\text{mech}} = J^s\pdv{\psi_g^{\text{mech}} }{J^s} - \psi_g^{\text{mech}} \label{eq:fgbreakup}
\end{equation}

\textbf{Growth directionality evolution:} Similarly using the expression for $\nten{T}^g$ in \cref{eq:Tg}, we can write \cref{eq:split_grlaw2}$_2$ as
\begin{equation}
\text{dev}(\nten{M}^{es}) \cdot \text{dev}(\nten{L}^g)  \ge 0 \label{eq:growthdirec}
\end{equation}
Any thermodynamically admissible evolution law for the growth directionality can thus be prescribed as follows
\begin{equation}
    \text{dev}(\nten{L}^g) = \hat{\nten{f}}(\nten{M}^{es}) \quad \text{such that} \quad \hat{\nten{f}}(\nten{M}^{es})\cdot\text{dev}(\nten{M}^{es}) \ge 0 \label{eq:direc_ev}
\end{equation}
Therefore, the Mandel stress drives growth directionality. Morphogenesis, the process by which the growing body acquires its shape, can arise from both spatially varying growth rates and evolution of growth directionality. When the growth rate is spatially uniform, morphogenesis would be driven solely by the Mandel stress.\\

In \Cref{subsec:spec_constit_growth}, we will specify the forms of the evolution functions $\hat{\Gamma}(f_g)$ and $\hat{\nten{f}}(\nten{M}^{es})$ to obtain specialized forms of the growth evolution laws in \cref{eq:vol_ev,eq:direc_ev}.

\subsection{Summary of formulation and boundary conditions}
\label{subsec:Summary}
The governing equations of the swelling-growth theory for isochoric conversion of diffusing species to solid are summarized below. See \ref{app:nonisochoric} for the general equations for non-isochoric species conversion. \\

\noindent \textbf{Governing equations in $\mathcal{B}^R$}
\begin{align}
&\text{Mechanical equilibrium} : &&\text{Div}~\nten{S} = \nten{0}, \quad \nten{S} = 2 J^g \nten{F}^{es} \pdv{{\psi}_g}{\nten{C}^{es}}\/ {\nten{F}^g}^{-T}  \label{eq:linmombal}\\
&\text{Mass balance} : &&J^g \dot{J}^s + J^s \dot{J}^g = \text{Div}(\nten{M}_{\text{mob}} \nabla \mu)  \label{eq:summ_massbal}\\
&\text{Constitutive equations} : &&\mu = \mu_0^f + \pdv{{\psi}_{g}}{J^s} -  \frac{1}{3} J^e \text{tr}(\nten{T}), \quad \nten{T} = 2\frac{J^g}{J} \nten{F}^e \pdv{{\psi}_{g}}{\nten{C}^e}\/ {\nten{F}^e}^T\\
&\text{Kinetic law for volumetric growth} : &&\frac{\dot{J}^g}{J^g} = \hat{\Gamma}(f_g), \quad \hat{\Gamma}(f_g) f_g \ge 0\\
&\text{Kinetic law for growth directionality} : &&    \text{dev}(\nten{L}^g) = \hat{\nten{f}}(\nten{M}^{es}), \quad \hat{\nten{f}}(\nten{M}^{es})\cdot\text{dev}(\nten{M}^{es}) \ge 0
\end{align}

To complete the theory development, initial and boundary conditions need to be prescribed. We consider the following sets of complementary subsurfaces
of the boundary $\partial{\mathcal{B}^R}$ : $\{\partial{\mathcal{B}^R_{t}},\partial{\mathcal{B}^R_{x}}\}$, $\{\partial{\mathcal{B}^R_{j}},\partial{\mathcal{B}^R_{\mu}}\}$, where any two subsurfaces $\partial{\mathcal{B}^R_{a}}$ and $\partial{\mathcal{B}^R_{b}}$ are complementary if $\partial{\mathcal{B}^R_{a}} \cup \partial{\mathcal{B}^R_{b}} = \partial{\mathcal{B}^R}$ and $\partial{\mathcal{B}^R_{a}} \cap \partial{\mathcal{B}^R_{b}} = \emptyset$. Then for a time interval $t \in [0,t_f]$ we consider the following boundary and initial conditions.\\

\noindent \textbf{Boundary conditions}
\begin{align}
&\text{Traction} : &&\nten{S} \nten{n}_R = \nten{\breve{s}} \quad &&&\text{on } \partial{\mathcal{B}^R_{t}} \times [0,t_f] \label{eq:traction_Bc} \\
&\text{Position} : &&\nten{x}(\nten{X}) = \breve{\nten{x}} \quad &&&\text{on } \partial{\mathcal{B}^R_{x}} \times [0,t_f] \label{eq:pos_Bc}\\
&\text{Diffusion flux} : &&- \left(\nten{M}_{\text{mob}} \nabla \mu \right) \cdot {\nten{n}}_R = \breve{j} \quad &&&\text{on } \partial{\mathcal{B}^{R}_{j}} \times [0,t_f] \label{eq:flux_Bc} \\
&\text{Chemical potential} : &&\mu  = \breve{\mu} \quad &&&\text{on } \partial{\mathcal{B}^{R}_{\mu}} \times [0,t_f] \label{eq:mu_Bc}
\end{align}
with $\nten{\breve{s}}, \breve{\nten{x}}, \breve{j}, \breve{\mu}$ being prescribed functions that depend on $\nten{X}$ and $t$. \\

\noindent \textbf{Initial conditions}
\begin{equation} 
\mu(\nten{X},0) = \mu_0(\nten{X}),\quad \nten{F}^g(\nten{X},0) = \nten{F}^g_0(\nten{X}) \label{eq:initialconds}
\end{equation}
where $\mu_0$ and $\nten{F}^g_0$ are prescribed functions. 
For the analysis in this manuscript, we set $\nten{F}^g_0(\nten{X}) =\nten{I}$.\\

This completes the development of the large deformation swelling-growth theory. It is important to note that setting $\nten{F}^g = \nten{I}$ recovers nonlinear poroelastic theories. In the next section, we will choose specific constitutive equations to specialize our theory and also write the dimensionless form of the governing equations.  

\section{Specific constitutive equations and dimensionless formulation}
\label{sec:spec_constit}
In this section, we specialize the equations of the swelling-growth theory developed in the previous section by choosing specific forms for the free energy, mobility tensor, and growth evolution laws. 

\subsection{Mechanical free energy}
\label{subsec:spec_constit_mech}
We limit our attention to isotropic materials.  The following form is chosen for the mechanical free energy 
\begin{equation}
{\psi}_{g}^{\text{{mech}}}(\nten{C}^{e},J^s) = \frac{G}{2} \left({(J^s})^{\frac{2}{3}}\text{tr}(\nten{C}^{e})  - 3 - 2\ln({J}^{es}) \right)  + \frac{K}{2} \left(\ln{J}^e\right)^2,\quad J^e = \sqrt{\text{det}(\nten{C}^e)} \label{eq:freeenergyspec}
\end{equation}
where $G$ and $K$ are the linear shear modulus and the bulk modulus of the dry solid respectively. The first term in \cref{eq:freeenergyspec} is the entropic free energy change due to mechanical stretching of the polymer network, given by classical statistical mechanics model of rubber elasticity \citep{treloar1975physics}. The second term is the energetic component of free energy change due to compressibility of the hydrogel, as also used in \cite{chester2011thermo}. \textcolor{black}{Note that in the absence of swelling, the mechanical free energy is essentially a compressible neo-Hookean model. While there are better tailored free energy functions with more parameters to describe biological tissues, this is a simple first choice for modelling of soft polymers or tissues (for example see tumor modelling in \cite{xue2016biochemomechanical}).}\\

 Hence the Cauchy and Mandel stresses are derived using \cref{eq:energetic_constit,eq:freeenergyspec} as
\begin{equation}
\nten{T} = \frac{1}{J^{es}}\left(G \left(\nten{B}^{es} -\nten{I}\right)  +  K (\ln{J^e}) \nten{I}\right),\quad \nten{M}^{es} = G \left(\nten{C}^{es} -\nten{I}\right) + K (\ln{J^e}) \nten{I}    \label{eq:cauchy_stress_spec}
\end{equation}
The following derivative is also readily derived,
\begin{equation}
\pdv{{\psi}_g^{\text{mech}}}{J^s} = \frac{G }{3 J^s}\left(\text{tr}(\nten{C}^{es}) -3\right) \label{eq:dpsigmechdJs} 
\end{equation}

\subsection{Mixing free energy}
\label{subsec:spec_constit_mix} 
The free energy of mixing is taken to be \citep{flory1942thermodynamics,huggins1941solutions}
\begin{equation}
{\psi}_{g}^{\text{{mix}}} =  \frac{\mu^*}{\phi}\left(\left(1-\phi\right)\ln(1-\phi) +\chi \phi(1-\phi) \right)\quad \text{where}\quad \mu^* = \frac{k_B T}{\omega^f},\ \omega^f = M^f \Omega^f \label{eq:psig_mix}
\end{equation}
where $\omega^f$ and $M^f$ are the molecular volume and mass respectively of the diffusing species, $k_B$ is the Boltzmann constant, $T$ is the constant temperature, and $\chi$ is the Flory–Huggins interaction parameter. The scalar $\mu^*$ will later be used as a characteristic scaling value for the chemical potential when developing the dimensionless version of the theory. The interaction parameter $\chi$ represents the dis-affinity between the solid and the diffusing species; the larger the $\chi$ the smaller the swelling ratio at diffusion equilibrium (at a fixed growth state). For $\chi\leq 0.5$ the minimum of ${\psi}_{g}^{\text{{mix}}}$ occurs at $\phi \to 0$ which means that in the absence of mechanical free energy (say $G,K \to 0$), the solid would want to keep swelling till $J^s \to \infty$ (i.e. $\phi \to 0$). However for $\chi > 0.5$, the minimum of ${\psi}_{g}^{\text{{mix}}}$ occurs at $0<\phi<1$ so that even in the absence of mechanical energy, the solid would swell to a finite swelling ratio at diffusion equilibrium.  Using the definition of solid volume fraction $\phi$ from \cref{eq:solidvolfrac} in \cref{eq:psig_mix}, the following derivative is readily obtained 
\begin{equation}
\pdv{{\psi}^{\text{mix}}_{g}}{J^s} = \mu^{\text{mix}}(\phi) =  \mu^* \left(\ln\left(1-\phi\right) + \phi + {\chi}{\phi^2}\right)  \label{eq:dpsigmixdJs}
\end{equation}
where we have defined an auxiliary function $\mu^{\text{mix}}(\phi)$ for later convenience.\\

Using \cref{eq:Constit_energetic,eq:Freeenergy_g,eq:dpsigmixdJs,eq:dpsigmechdJs,eq:cauchy_stress_spec,eq:Cedef3}, we obtain the following equation for the chemical potential
\begin{equation}
\mu = \mu_0^f +  \mu^{\text{mix}}(\phi) - {K \ln(J^e)}{\phi} \label{eq:mu_spec}
\end{equation}

\subsection{Growth evolution}
\label{subsec:spec_constit_growth}  
Guided by the inequality in \cref{eq:vol_ev}, we choose a piece-wise linear form of the volumetric growth law that enforces irreversibility  of species conversion\footnote{While technically cells can die and decompose back into the diffusing components, the associated processes and kinetics are usually different from that of growth and we avoid such considerations here.},
\begin{equation}
    \Gamma = 
\begin{cases}
    k_g f_g,& \text{if } f_g\geq 0\\
    0,              & \text{if } f_g<0
\end{cases} \label{eq:volgrowthlaw}
\end{equation}
where $k_g$ is a growth constant. 
Further, using \cref{eq:fgbreakup,eq:freeenergyspec,eq:psig_mix,eq:dpsigmechdJs,eq:dpsigmixdJs}, we arrive at the following expressions for different components of $f_g$ in \cref{eq:fgtotal},
\begin{equation}
f_g^{\text{mech}} = G\left( \ln(J^{es}) - \frac{1}{6} \left(\text{tr}(\nten{C}^{es}) -3\right)\right) - \frac{K}{2}\left(\ln(J^e)\right)^2  \label{eq:fgmechspec}
\end{equation}
\begin{equation}
f_g^{\text{mix}}(\phi) = \mu^* \left(1 + \ln\left(1-\phi\right) + \chi \left(2\phi -1\right) \right) \label{eq:fgmixspec}  
\end{equation}\\

Similarly, guided by the inequality in \cref{eq:direc_ev}, we choose the following linear evolution law for the growth directionality
\begin{equation} 
    \text{dev}(\nten{L}^g) = \frac{k_g}{3} \text{dev}(\nten{M}^{es}) = \frac{k_g G}{3} \text{dev}(\nten{C}^{es}) \label{eq:spec_gr_direc}
\end{equation}
where for simplicity we have chosen the same rate parameter $k_g$ as for the volumetric growth evolution law and have used \cref{eq:cauchy_stress_spec}$_2$ for the Mandel stress.   When $f_g \ge 0$, the evolution laws \cref{eq:volgrowthlaw,eq:spec_gr_direc} can be combined and written as $3\nten{L}^g = k_g\nten{T}^g$.

\subsection{Mobility tensor}
\label{subsec:spec_constit_mobility} 
The diffusion is assumed to be be modelled by the following Eulerian law
\begin{equation}
\nten{j} = -\hat{h}(J^s) m^*\ \text{grad}\ \mu, \quad m^* = \frac{D}{\mu^*}\label{eq:mstardef}
\end{equation}
where $D$ is the diffusion coefficient, $m^*$ is the characteristic scaling value for the mobility, and $\hat{h}(J^s)$ describes the dependence of the mobility in the current space on swelling. Using \cref{eq:diffusionlaw2}, the associated mobility tensor is readily shown to be
\begin{equation}
\nten{M}^{\text{mob}} = \hat{h}(J^s) m^* J \nten{C}^{-1} \quad \text{where } \nten{C} = \nten{F}^T \nten{F} 
\label{eq:mobilityspec}\end{equation} 
Different works in the literature have assumed different monotonically increasing functional forms for $\hat{h}$ \citep{baek2004diffusion,hong2008theory,duda2010theory,chester2011thermo, abi2019kinetics}. For example \cite{hong2008theory} employed $\hat{h} = J^s-1$ while \cite{abi2019kinetics} chose $\hat{h} = (J^s-1)/J^s$. However in the absence of any experimental evidence to choose a particular functional form, here we choose $\hat{h} =1$ which suffices to demonstrate all the relevant physics while avoiding zero mobility in the dry state. Further, we note that different choices for $\hat{h}$ preserve the conclusions in this manuscript.

\subsection{Dimensionless equations and approximate limits}
\label{subsec:spec_constit_dimless} 
In this section, a dimensionless version of the theory is developed which aids greatly with the analysis in later sections and in making suitable approximations under different limiting conditions. We first define the following dimensionless operations and quantities,
\begin{equation}
\overline{\text{Div}} = L^*\ \text{Div},\quad \overline{\nabla} = L^*\ \nabla,\quad \bar{\mu} = \frac{\mu}{\mu^*}, \quad \overline{\nten{M}}_{\text{mob}} = \frac{\nten{M}^\text{mob}}{m^*},\quad \bar{t}_d = \frac{t}{\tau_d} \label{eq:charac1}
\end{equation}
where $L^*$ is a characteristic length scale and $\tau_d = {{L^*}^2}/{D}$ is the associated diffusion timescale. 
Using the definitions from \cref{eq:charac1} along with \cref{eq:mobilityspec} and \cref{eq:mstardef}$_2$ in \cref{eq:summ_massbal} yields the following dimensionless diffusion-consumption equation,
\begin{equation}
J^g \dv{J^s}{\bar{t}_d} + J^s \dv{J^g}{\bar{t}_d} = \overline{\text{Div}}\left(\overline{\nten{M}}_{\text{mob}} \overline{\nabla} \bar{\mu}\right),\quad  \overline{\nten{M}}_{\text{mob}} = J \nten{C}^{-1}\label{eq:reacdiff_dimlss}
\end{equation}
The dimensionless counterpart for the constitutive equation \eqref{eq:mu_spec} for chemical potential is
\begin{equation}
\bar{\mu} = \bar{\mu}_0^f +  \bar{\mu}^{\text{mix}}(\phi) - \left(\frac{K}{G}\right)\left(\frac{G}{\mu^*}\right){ \ln(J^e)}{\phi}\quad \text{where}\quad \bar{\mu}_0^f = \frac{\mu_0^f}{\mu^*},\  \bar{\mu}^{\text{mix}}= \frac{\mu^{\text{mix}}}{\mu^*} \label{eq:mu_dimless}
\end{equation}
The Cauchy stress in \cref{eq:cauchy_stress_spec} is non-dimensionalized using the shear modulus $G$ as follows,
\begin{equation}
\bar{\nten{T}} = \frac{\nten{T}}{G} = \frac{1}{J^{es}} \left(\nten{B}^{es} -\nten{I}\right) +  \left(\frac{K}{G}\right) \frac{\ln{J^e}}{J^{es}}  \nten{I} \label{eq:stress_spec_dimless}
\end{equation}
where $K/G$ is a measure of the compressibility of the solid. A perfectly incompressible version of theory is developed in \ref{app:perfecincomp} where $K/G \to \infty$ such that $J^e\to1$. The dimensionless version of the mechanical equilibrium equation \eqref{eq:Piolastresseqb} is written as
\begin{equation}
\overline{\text{Div}}\ \bar{\nten{S}} = \nten{0} \quad \text{where } \bar{\nten{S}} = \frac{\nten{S}}{G} = J \bar{\nten{T}} \nten{F}^{-T} \label{eq:Piola_dimless}
\end{equation}\\

Defining the following additional quantities,
\begin{equation}
 \bar{f}_g = \frac{f_g}{\mu^*}, \quad \tau_g = \frac{1}{k_g \mu^* \bar{f}^*_g}, \quad \bar{t}_g = \frac{t}{\tau_g}, \quad  \bar{\Gamma} =  \tau_g \Gamma =\frac{1}{J^g}\dv{J^g}{\bar{t}_g} \label{eq:charac2}
\end{equation} 
where $\tau_g$ is a characteristic timescale of growth and  $\bar{f}^*_g$ is a dimensionless constant chosen to normalize the growth law, we can rewrite the volumetric growth law \eqref{eq:volgrowthlaw} in dimensionless form as
\begin{equation}
    \bar{\Gamma}  = 
\begin{cases}
    {\bar{f}_g}/{\bar{f}_g^*},& \text{if } \bar{f}_g\geq 0\\
    0,              & \text{if } \bar{f}_g<0
\end{cases} \label{eq:gammabar_dimlss}
\end{equation}
where $\bar{\Gamma}$ is the dimensionless growth rate.
Non-dimensionalizing \cref{eq:fgtotal,eq:fgmechspec,eq:fgmixspec} using $\mu^*$ yields
\begin{equation}
{\bar{f}_g} =  \Delta \bar{\mu}_0 + \bar{f}_g^{\text{mix}} + \bar{f}_g^{\text{mech}}
\end{equation}
where $\Delta \bar{\mu}_0 = \Delta {\mu}_0/\mu^*$ and
\begin{align}
\bar{f}_g^{\text{mix}}(\phi) = \frac{{f}_g^{\text{mix}}(\phi)}{\mu^*} = 1 + \ln\left(1-\phi\right) + \chi \left(2\phi -1\right)  ,  \label{eq:fgbarmix}\\
\bar{f}_g^{\text{mech}} = \frac{{f}_g^{\text{mech}}}{\mu^*} = \frac{G}{\mu^*}\left( \ln(J^{es}) - \frac{1}{2}\left(\frac{K}{G}\right)\left(\ln(J^e)\right)^2 - \frac{1}{6}  \left(\text{tr}(\nten{C}^{es}) -3\right) \right) \label{eq:fgbarmech}
\end{align}
The evolution equation for growth directionality \cref{eq:spec_gr_direc} can be written in dimensionless form as 
\begin{equation}
\text{dev}(\bar{\nten{L}}^g) = \frac{1}{3} \left(\frac{G}{\mu^*}\right)\frac{\text{dev}(\overline{\nten{M}}^{es})}{\bar{f}^*_g} = \frac{1}{3} \left(\frac{G}{\mu^*}\right)\frac{\text{dev}(\nten{C}^{es})}{\bar{f}^*_g} \label{eq:devLg_dimlss}
\end{equation}
where $\overline{\nten{M}}^{es} = {\nten{M}}^{es}/G$ and $\bar{\nten{L}}^g = \tau_g \nten{L}^g$. \\

Note from \cref{eq:devLg_dimlss,eq:cauchy_stress_spec} that for the growth to completely stop ($\nten{L}^g = \nten{0}$), the stress state $\nten{T}$ necessarily needs to be hydrostatic. This is a consequence of the fact that there is no inherent directionality for swelling or species conversion which are the growth driving mechanisms in our theory (all the non-stress terms in the driving stress $\nten{T}^g$ are hydrostatic in \cref{eq:Tg}). However, in the limit $G/\mu^* \to 0$, the system can stop growing even at non-hydrostatic stress states. Essentially, in this limit, the growth stops once the volumetric growth halts as the remodelling is slow at the timescale of volumetric growth.\\  

For our soft growing systems of interest, the value of the shear modulus $G$ is typically low, $\sim$ 1 kPa \citep{garteiser2012mr,zhang2021morphogenesis}, while the value of $\mu^*$ is high for typical swelling fluids at room temperature, $\sim$ 10-100 MPa \citep{hong2008theory,chester2010coupled}. Thus the value of the dimensionless parameter ${G}/{\mu^*}$ is very small ($\sim$ 1e${}^{-5}$ - 1e${}^{-4}$). Consequently it can be inferred from \cref{eq:devLg_dimlss} that the growth directionality will be nearly isotropic. Thus if $\nten{F}^g_0 = \nten{I}$, the growth tensor $\nten{F}^g$ continues to remain nearly spherical/isotropic during growth unless the dimensionless Mandel stress reaches large and highly non-spherical values. \textit{This explains why the assumption of isotropic growth, which is commonly used in the literature for soft growing systems  \citep{ambrosi2002mechanics,ambrosi2004role,kim2011role,kopf2013continuum,kuhl2014growing}, works well}. \textcolor{black}{Note that this is a consequence of our constitutive choice of same growth constant $k_g$ to describe the evolution of both volumetric growth in \cref{eq:volgrowthlaw} and growth directionality in \cref{eq:spec_gr_direc}, combined with the fact that $\text{tr}(\nten{T}_g) \sim \mu^*$ (when $\mu^* >>G$ and dimensionless stresses are not very large non-spherical values) whereas $\text{dev}(\nten{T}_g) \sim G$. However, more generally growth laws can be specified with different growth constants for the volumetric and deviatoric parts of the growth rate tensor (for example if cellular rearrangement is much faster than cell division in a particular growing system) which can lead to more anisotropic growth evolution.} 
Further, it can be shown from \cref{eq:fgbarmix,eq:fgbarmech} that $\bar{f}_g^{\text{mix}} \gg \bar{f}_g^{\text{mech}}$ for small $G/\mu^*$ as long as the dimensionless stresses are not very large\footnote{Note that while the term $\frac{K}{G}\to \infty$ in the incompressible limit, the term $\frac{K}{G}\left(\ln(J^e)\right)^2$ in $\bar{f}_g^{\text{mech}}$ (\cref{eq:fgbarmech}) still approaches zero. See \ref{app:perfecincomp} and \cref{eq:app_fgmechspec} for expression of $f_g^\text{mech}$ in the perfectly incompressible limit. Numerical simulations in \Cref{sec:Results} for the slightly compressible case also confirm $\bar{f}_g^{\text{mix}} \gg \bar{f}_g^{\text{mech}}$.}. Thus the dimensionless driving stress for volumetric growth can typically be approximated very well as
\begin{equation}
{\bar{f}_g} \approx \Delta \bar{\mu}_0 + \bar{f}_g^{\text{mix}}(\phi)
\end{equation}
which is primarily a function of $\Delta \bar{\mu}_0$ (dimensionless conversion energy), and the solid volume fraction, or equivalently the swelling ratio - recall \cref{eq:solidvolfrac}. Thus the swelling can be seen to be an important driver of growth. The approximation also allows us to examine the driving stress and make several key inferences before even solving boundary value problems, in \Cref{subsec:growthdrivingstress}.\\  

Finally, using eqs. \eqref{eq:charac2}$_4$, \eqref{eq:gammabar_dimlss}, and the definitions of $\bar{t}_d$ and $\bar{t}_g$ from \cref{eq:charac1} and \cref{eq:charac2} respectively, we can rewrite \cref{eq:reacdiff_dimlss} as follows
\begin{equation}
J^g \dv{J^s}{\bar{t}_d} + J^s J^g \bar{\Gamma} \left(\frac{\tau_d}{\tau_g}\right) = \overline{\text{Div}}\left(\overline{\nten{M}}_{\text{mob}} \overline{\nabla} \bar{\mu}\right)
\end{equation}
In many growing systems, due to typically large growth timescales relative to small diffusion timescales (due to small sizes), $\tau_g \gg \tau_d$. Thus at the timescale of growth, the diffusion is often assumed to be equilibrated \citep{greenspan1972models,preziosi2003cancer,ambrosi2002mechanics,ambrosi2004role}, and the following equilibrated version of the diffusion-consumption equation can be used
\begin{equation}
J^s J^g \bar{\Gamma} \left(\frac{\tau_d}{\tau_g}\right) = \overline{\text{Div}}\left(\overline{\nten{M}}_{\text{mob}} \overline{\nabla} \bar{\mu}\right) \label{eq:reac_diff_eqbtd}
\end{equation}
In this limit, the net flux of diffusing species (right hand side) is being matched exactly by the consumption for growth (left hand side). Note that for a given growth rate $\bar{\Gamma}$, the larger the body grows (larger $J^gJ^s$), the larger the consumption term. For the tumor growth experiments discussed in \Cref{sec:Overview_tumor}, this is what causes the equilibrium swelling profile (or equivalently concentration profile using \cref{eq:constraintcombined1}) to drop with time inside the tumor for free growth, resulting in reduced growth rates. When the equilibrated equation \eqref{eq:reac_diff_eqbtd} is used, the initial condition for chemical potential field in \cref{eq:initialconds} need not be supplied. \\

When we are not interested in diffusion-consumption limitations due to large growth sizes and instead want to isolate the effect of applied stresses, we can focus on the limit $\tau_g/\tau_d \to \infty$ where the diffusion is infinitely faster than growth. In this limit, the left hand side of \cref{eq:reac_diff_eqbtd} becomes zero and solutions where all spatial fields are uniform are possible for suitable boundary value problems, allowing for analytical tractability. We analyze this limit while enforcing perfect incompressibility ($K/G \to \infty, J^e \to 1$) in \Cref{subsec:uniformswell}. The results are useful in establishing the effects of mechanical constraints on swelling-growth and subsequently in analyzing and explaining the results when we consider added diffusion-consumption limitations (finite $\tau_g/\tau_d$). In the following section, we will use the developed swelling-growth theory to study relevant boundary value problems and model experiments of growing tumors and bacterial biofilms.

\section{Analysis and results}
\label{sec:Results}

This section is organized as follows: We first examine the  growth driving stress in the limit $G/\mu^* \to 0$ in \Cref{subsec:growthdrivingstress} and establish notions of a critical swelling ratio and critical conversion energy that can halt growth. Following this, in \Cref{subsec:uniformswell} we focus on the effects of applied stresses in the absence of diffusion-consumption limitations by studying uniform swelling-growth that emerges in the limit of infinitely fast diffusion ($\tau_g/\tau_d \to \infty$) and uniform applied pressure. By linking the swelling ratio to the applied pressure, we establish  
a free swelling ratio in the absence of stresses and a homeostatic pressure associated with the critical swelling ratio. 
In \Cref{subsubsec:mechconfine}, we demonstrate the above established concepts by studying the problem of uniform swelling-growth against mechanical confinement in the infinitely fast diffusion limit. 
To account for diffusion-consumption effects (finite $\tau_g/\tau_d$), which entails spatially varying field variables, we formulate a spherically symmetric boundary value problem in \Cref{subsec:spherical symmetry}. The formulation is used to study the case of free growth in \Cref{subsec:freegrowth} and the case of uniform applied pressure in \Cref{subsec:appliedpressure}. Finally, the theory 
is used to model experimental results of growing tumors in \Cref{subsec:tumorresults} and of bacterial biofilms in \Cref{subsec:biofilm_model}.

\subsection{Growth driving stress}
\label{subsec:growthdrivingstress}

We consider the limit $G/\mu^* \to 0$ (which is the case for our soft growing systems of interest as discussed in \Cref{subsec:spec_constit_dimless}). In this limit, the growth directionality evolution equation \eqref{eq:devLg_dimlss} becomes trivial and enforces isotropy of growth so that $\nten{F}^g = J^g \nten{I}$. Using \cref{eq:gammabar_dimlss} for the growth rate evolution, the growth process is now primarily determined  by the dimensionless driving stress $\bar{f}_g$. In the limit $G/\mu^* \to 0$, we have 
\begin{equation}
\bar{f}_g =    \Delta \bar{\mu}_0 + \bar{f}_g^{\text{mix}}(\phi) \label{eq:fgbar_zero}
\end{equation}
where $\bar{f}_g^{\text{mix}}(\phi)$ is defined in \cref{eq:fgbarmix} and $\phi$ is related to the swelling ratio $J^s$ through \cref{eq:solidvolfrac}. Thus, any dependence of the growth on applied stresses and diffusion-consumption effects appears through the swelling ratio $J^s$ (note that  $J^s$ depends on the mechanical equilibrium and diffusion-consumption equations in \Cref{subsec:Summary}). We first consider $\Delta \bar{\mu}_0 = 0$ (no contribution from species conversion) and examine the dependence of the dimensionless driving stress $\bar{f}_g$ on the swelling ratio $J^s$ and the Flory-Huggins interaction parameter $\chi$. Note that non-zero values of $\Delta \bar{\mu}_0$ will simply shift $\bar{f}_g$ by a constant and not affect its qualitative dependence on $J^s$ and $\chi$.\\ 

\begin{figure}
    \centering
    \includegraphics[width=\textwidth]{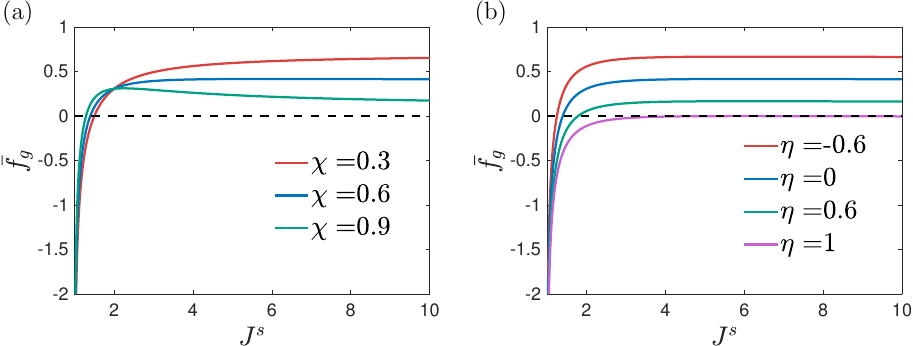}
    \caption{Dimensionless driving stress for volumetric growth, $\bar{f}_g$, as a function of the swelling ratio in the limit ${G}/{\mu^*}\to0$. (a) Plots of $\bar{f}_g$ for different values of Flory-Huggins interaction parameter $\chi$ and zero conversion energy ($\Delta \mu_0 = 0$). (b) Plots of $\bar{f}_g$ for different values of the conversion energy ratio $\eta$ for fixed representative value of $\chi = 0.6$.}
    \label{fig:fgbar}
\end{figure}

\textbf{Dependence on $\bm{J^s}$ :} The dimensionless driving stress $\bar{f}_g$ is plotted as a function of the swelling ratio $J^s$ when $\Delta \bar{\mu}_0 = 0$ in \Cref{fig:fgbar}(a), for different values of the Flory-Huggins interaction parameter $\chi$. This plot shows the functional dependence of $\bar{f}_g$ as a function of the swelling ratio (or equivalently concentration of diffusing species, recall \cref{eq:constraintcombined1}) in the absence of any conversion energy. It can be seen that generally higher $J^s$ leads to a larger driving stress for volumetric growth and that the driving stress saturates at high values of $J^s$.  There exists a critical value of swelling ratio, $J^s_c$, below which $\bar{f_g}<0$. Based on our growth law in \cref{eq:gammabar_dimlss}, this implies that growth halts for $J^s < J^s_c$, while for $J_s>J^s_c$, the growth rate linearly scales with $\bar{f}_g$.  
 \textit{Thus our theory already offers a kinetic basis for the experimentally observed dependence of the growth rate on concentration of diffusing species, wherein the growth rate increases with concentration, saturates at high values and halts below a critical value.} \\

\textbf{Effect of $\bm{\chi}$ : } The influence of $\chi$ on the driving stress is also shown in \Cref{fig:fgbar}(a). 
 The asymptotic value of $\bar{f}_g^{\text{mix}}$ as $J^s \to \infty$ is given by $1-\chi$ and thus decreases for increasing $\chi$ (larger dis-affinity between the solid and diffusing species). For $\chi <0.5$, $\bar{f}_g$ is a monotonically increasing function of $J^s$ and reaches its maximum value at $J^s_{\text{max}} \to \infty$. For $\chi \ge 0.5$, the driving stress $\bar{f}_g$ monotonically increases with $J^s$ till it reaches a maximum value at $J^s = J^s_{\text{max}} = 2\chi/(2\chi-1)$ and subsequently decreases to its asymptotic value as $J^s \to \infty$.  This decrease is typically mild, except for very high values of $\chi$ (such as $\chi=0.9$ in \Cref{fig:fgbar}(a)), which are typically unphysical and represent extreme dis-affinity between solid and diffusing species. The maximum value of the driving stress $\bar{f}_g^{\text{mix}}$  is derived to write
\begin{equation}
   \max_{1 \le J^s < \infty}  \bar{f}_g^{\text{mix}}(J^s) = \begin{cases}
    \chi - \ln\left({2 \chi}\right),& \text{for } \chi > 0.5\\
    1-\chi,              & \text{for } 0<\chi<0.5
\end{cases} \label{eq:maxfg}
\end{equation}
which is always a positive value (as also seen in \Cref{fig:fgbar}(a)).\\

\begin{figure}
    \centering
    \includegraphics[width=\textwidth]{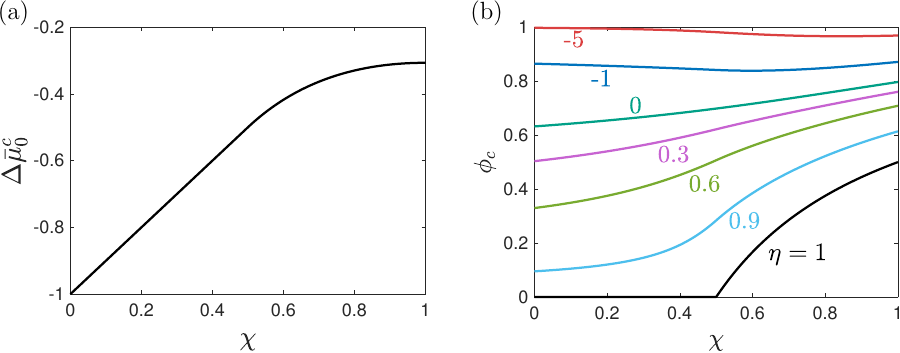}
    \caption{(a) Plot of the dimensionless critical conversion energy that prohibits volumetric growth for any swelling ratio as a function of the Flory-Huggins interaction paremeter $\chi$ in the limit $G/\mu^* \to0$. (b) Plot of the critical solid value fraction $\phi_c$, which is the value of solid volume fraction that stops growth ($\bar{f}_g(\phi_c)=0$), versus $\chi$ for various values of the conversion energy ratio $\eta$, in the limit $G/\mu^* \to0$.\label{fig:fig4}  }
\end{figure}

\textbf{Effect of $\bm{\Delta \bar{\mu}_0}$ :} Now we consider the effect of non-zero dimensionless conversion energy ${\Delta \bar{\mu}_0}$. A positive value of ${\Delta \bar{\mu}_0}$ corresponds to remaining energy after species conversion that accelerates growth whereas a negative value corresponds to an energy penalty that deters growth. We can define a critical dimensionless conversion energy, $\Delta \bar{\mu}_0^c$, as the maximum value of ${\Delta \bar{\mu}_0}$ such that $\bar{f}_g \leq 0$ for all values of $J^s$. Using \cref{eq:fgbar_zero}, we can write
\begin{equation}
  \Delta \bar{\mu}_0^c  =  - \max_{1 \le J^s < \infty}  \bar{f}_g^{\text{mix}}(J^s) \label{eq:maxcost}
\end{equation}
When $\Delta \bar{\mu}_0 < \Delta \bar{\mu}_0^c$, growth is not possible irrespective of the swelling ratio since $\bar{f}_g\le 0$ for any value of the swelling ratio $J^s$. Using \cref{eq:maxfg,eq:maxcost}, the critical dimensionless conversion energy for growth, $\Delta \bar{\mu}_0^c$, is plotted in \Cref{fig:fig4}(a) as a function of $\chi$. The value of $\Delta \bar{\mu}_0^c$ is always negative, thus there is a maximum energy penalty for the species conversion, above which growth will always be unfavourable no matter the swelling. The larger the $\chi$, 
the smaller the conversion penalty required to stop growth. We can then define a conversion energy ratio $\eta = {\Delta \bar{\mu}_0}/{\Delta \bar{\mu}_0^c}$ such that $\eta \le 1$ for growth to be feasible.  A plot of $\bar{f}_g$ as a function of $J^s$ for varying values of $\eta$ is plotted in \Cref{fig:fgbar}(b) for $\chi =0.6$. The larger the value of $\eta$, the smaller the dimensionless driving stress for any given $J^s$, and for $\eta =1$, the growth is seen to be always unfeasible. 
Since $\Delta \bar{\mu}_0^c<0$, a positive value of $\eta$ corresponds to an energy penalty which is seen to deter growth (reduced $\bar{f}_g$) compared to $\eta =0$ while a negative value corresponds to excess energy that can be seen to accelerate growth.\\

\textbf{Critical swelling ratio dependence:} As noted earlier, there exists a critical value of swelling ratio, $J^s_c$, such that growth halts for $J^s < J^s_c$. The swelling ratio in the absence of any constraints has a preferred value (established in the next section) and it can fall due to applied stresses and diffusion-consumption effects. If it falls to or below the critical swelling ratio, growth halts, and thus it is a critical parameter in capturing the effects of mechanical and diffusion-consumption constraints on growth. From \Cref{fig:fgbar}, it can be seen that the critical swelling ratio $J^s_c$ 
depends on both $\chi$ and $\eta$. The critical solid volume fraction $\phi_c = 1/J^s_c$ can be defined as the maximum value of the solid volume fraction above which growth ceases and has been plotted in \Cref{fig:fig4}(b) as a function of $\chi$ for different values of $\eta$. 
For a fixed value of $\chi$, increasing values of $\eta$ lead to lower values of $\phi_c$, or equivalently higher values of $J^s_c$.
At a given value of $\Delta \bar{\mu}_0$, increasing values of $\chi$ lead to smaller values\footnote{The dependence of $\phi_c$ on $\chi$ for a fixed $\eta$ can be non-monotonic as seen in \Cref{fig:fig4}(b) since $\Delta \bar{\mu}_0^c$ is a function of $\chi$ and thus a fixed value of $\eta$ does not necessarily correspond to a fixed value of $\Delta \bar{\mu}_0$ for varying $\chi$.} of $J^s_c$. 
 In the limit $\eta \to -\infty$, that is the conversion energy being extremely favourable for growth, $\phi_c \to 1$, which means that the solid would have to be completely dry to stop growing and even a small amount of swelling would lead to growth.  In the limit $\eta \to 1$, it can be shown that $\phi_c = 1/{J^s_{\text{max}}}$.

\subsection{Uniform growth without diffusion-consumption limitations ($\tau_g/\tau_d \to \infty$)}
\label{subsec:uniformswell} 

 We have shown in the previous section that the growth in systems with small $G/\mu^*$ is primarily dependent on the swelling ratio and thus, to study the effect of applied stresses and diffusion-consumption limitations on growth, we would have to study their effect on the swelling ratio. Accounting for diffusion-consumption effects requires a full spatial solution of the coupled governing equations in \Cref{subsec:Summary}, which follows in later sections. Here we isolate the effects of mechanical constraints on growth by considering uniform swelling-growth that emerges in the infinitely fast diffusion limit ($\tau_g/\tau_d \to \infty$) for the case of uniform applied pressure and chemical potential on the boundary. 
 The results from this section also aid analysis in later sections where we include diffusion-consumption limitations.\\

 Setting $\tau_g/\tau_d \to \infty$, the equilibrated diffusion-consumption equation \eqref{eq:reac_diff_eqbtd} reduces to
\begin{equation}
\overline{\text{Div}}\left(\overline{\nten{M}}_{\text{mob}} \overline{\nabla} \bar{\mu}\right) = 0 \label{eq:app_reac_diff_eqbtd}
\end{equation}
Consider a boundary value problem where the chemical boundary condition is purely in terms of an applied chemical potential of value $\mu_0^f$ (which corresponds to a bath of diffusing species maintained at its reference chemical potential) so that $\partial{\mathcal{B}^{R}_{j}} = \emptyset$, $\partial{\mathcal{B}^{R}_{\mu}} = \partial{\mathcal{B}^R},$ and $\breve{\mu} = \mu_0^f$ in \cref{eq:flux_Bc,eq:mu_Bc}. Additionally, assume that the mechanical boundary condition is purely in terms of an applied Cauchy pressure of value $P_b$ so that $\partial{\mathcal{B}^R_{x}} = \emptyset$, $\partial{\mathcal{B}^R_{t}}=\partial{\mathcal{B}^R}$, and $\nten{\breve{s}} = - J P_b \nten{F}^{-T} \nten{n}_R$ in \cref{eq:traction_Bc,eq:pos_Bc}. It can be shown that the solution fields for the deformation tensors and the stresses are spatially uniform and spherical and that the chemical potential is spatially uniform, such that
\begin{equation}
\nten{F}^s = (J^s)^{\frac{1}{3}} \nten{I} = \phi^{-\frac{1}{3}} \nten{I},\quad \nten{F}^g = (J^g)^{\frac{1}{3}} \nten{I},\quad \nten{F}^e = (J^e)^{\frac{1}{3}} \nten{I}, \quad \nten{T} = -P_b \nten{I}, \quad \mu = \mu_0^f \label{eq:uniformsol}
\end{equation} 
Using \cref{eq:uniformsol,eq:cauchy_stress_spec,eq:mu_spec}, the uniform elastic and swelling deformations can be obtained as the solution to the following coupled set of equations, 
\begin{align}
  \mu^{\text{mix}}(\phi) - {K \ln(J^e)}{\phi} = 0\\
   G\left((J^e)^{\frac{2}{3}}\phi^{\frac{1}{3}} - \phi\right) + \mu^{\text{mix}}(\phi) + J^e P_b = 0
\end{align}
where $\mu^{\text{mix}}(\phi)$ is defined in \cref{eq:dpsigmixdJs}. If we further restrict ourselves to the limit of perfect incompressibility so that $K/G \to \infty$, then $J^e \to 1$, and the two equations reduce to one equation that relates the applied pressure and the swelling ratio (\ref{sec:app_uniformswell_incomp}),
\begin{equation}
\left(\phi^{\frac{1}{3}} - \phi\right) + \left(\frac{\mu^*}{G}\right) \bar{\mu}^{\text{mix}}(\phi) +\bar{P}_b =0 \quad \text{where}\quad \bar{P}_b = \frac{P_b}{G} \label{eq:eqb_sol_incomp}
\end{equation}
 The solution for $\phi$ in \cref{eq:eqb_sol_incomp} is the equilibrium solid volume fraction as a function of the applied pressure in the limit of infinitely fast diffusion and perfect incompressibility, which we denote by $\phi_{eq}(\bar{P}_b)$. The related swelling ratio is given by $J^s_{eq}(\bar{P}_b) = 1/\phi_{eq}$. \\

\begin{figure}[!htb]
    \centering
    \includegraphics[width=\textwidth]{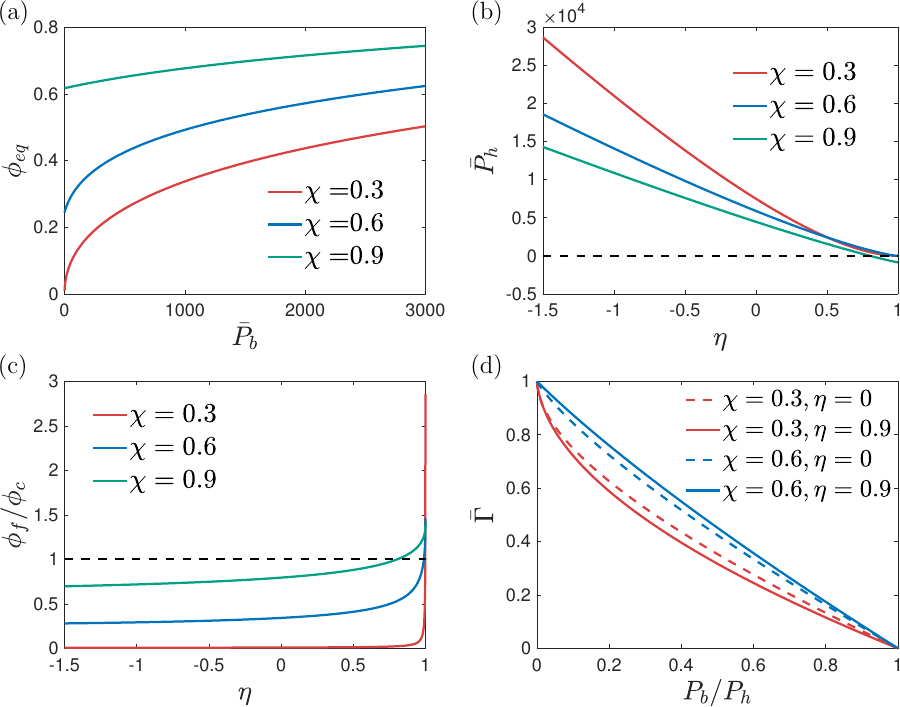}
    \caption{Uniform swelling-growth that emerges for uniform applied pressure $P_b$ and infinitely fast diffusion ($\tau_g/\tau_d \to \infty$), in the limit of perfect incompressibility. Plots, for various $\chi$, of (a) The equilibrium value of solid volume fraction, $\phi_{eq}$, as a function of dimensionless applied pressure $\bar{P}_b$.  (b) The dimensionless homeostatic pressure ($\bar{P}_h$) as a function of the conversion energy ratio ($\eta$). (c) The ratio of the free solid volume fraction ($\phi_{f}$) to the critical solid volume fraction ($\phi_c$), as a function of the conversion energy ratio. (d) Plots, for various $\chi$ and $\eta$, of the dimensionless volumetric growth rate as a function of $P_b/P_h$, the ratio of applied pressure to the homeostatic pressure. }\label{fig:fig5}
\end{figure}

 To solve for the swelling in \cref{eq:eqb_sol_incomp}, we cannot set $G/\mu^* \to 0$ as this leads to $\phi_{eq} \to 0$ (infinite swelling) for $\chi<0.5$. Thus here we set $G/\mu^* = 4 \times 10^{-5}$, which is a representative value for tumors and bacterial biofilm systems that we model later. The equilibrium solid volume fraction $\phi_{eq}$ is plotted as a function of the dimensionless applied pressure $\bar{P}_b$ in \Cref{fig:fig5}(a) for various values of $\chi$. We only plot positive values of pressure, as a solution for $\phi_{eq}$ is not necessarily guaranteed for negative pressures $\bar{P}_b<0$ (diffusion equilibrium might not be possible). It can be seen that the larger the applied pressure, the higher the equilibrium solid volume fraction or consequently the lower the equilibrium swelling ratio as one would intuitively expect. Given that we know from the previous section that the growth rate generally decreases with decreasing swelling ratio, this explains why the growth rate decreases in the presence of mechanical confinement\footnote{It was numerically verified that $J^s_{eq}(\bar{P}_b\geq0) \leq J^s_{\text{max}}$ irrespective of $\chi$ and thus the growth rate necessarily decreases with increasing positive pressure ($\bar{f}_g$ increases monotonically with $J^s$ for $J^s<J^s_{\text{max}}$).}. For a given value of applied pressure it can be seen that higher values of $\chi$, which corresponds to greater dis-affinity between the solid and diffusing species, leads to higher solid volume fraction or lower swelling ratio. Finally, we define the free solid volume fraction $\phi_{f}$ and the free swelling ratio $J^s_{f}$ as follows
 \begin{equation}
\phi_{f} = \phi_{eq}(\bar{P}_b =0), \quad J^s_{f} = 1/\phi_{f}   \label{eq:free_phi}
 \end{equation}
which characterize the swelling state under free-growth conditions in the absence of stresses and diffusion-consumption constraints. \\

The dimensionless driving stress for volumetric growth, $\bar{f}_g$, specializes for uniform isotropic growth in the limit of infinitely fast diffusion and  perfect incompressibility as follows  (\ref{sec:app_uniformswell_incomp})
\begin{equation}
\bar{f}_g = \bar{f}_g^{\infty} = \Delta \bar{\mu}_0  +  \bar{f}_g^{\text{mix}}(\phi) +  \frac{G}{\mu^*} \left(\ln(\phi^{-1})  - \frac{1}{2} \left(\phi^{-\frac{2}{3}} -1\right) \right) \label{eq:exact_fg}
\end{equation}
wherein the driving stress remains a function of the swelling and there is one additional term compared to \cref{eq:fgbar_zero} to account for the finite value of $G/\mu^*$. Note that for the small value of $G/\mu^*$ considered here, $\bar{f}_g$, the critical solid volume fraction $\phi_c$, and the dimensionless critical conversion energy $\Delta \bar{\mu}_0^c$, are practically unchanged from their values at the limit $G/\mu^* \to 0$ from the previous section. Nevertheless, we solve for the exact values of $\Delta \bar{\mu}_0^c$ and $\phi_c$ using the following equations,
\begin{equation}
    \Delta \bar{\mu}_0^c = - \max\limits_{0 \le \phi \le 1} \left(\bar{f}_g^{\infty} - \Delta \bar{\mu}_0\right), \quad \bar{f}_g^\infty(\phi_c) = 0 \label{eq:exact_uniform}
\end{equation}
We can define a homeostatic pressure $P_h$ as the applied pressure that results in an equilibrium solid volume fraction equal to the critical value that stops growth, so that 
\begin{equation}
 \phi_{eq}(\bar{P}_h) = \phi_c \quad \text{where}\quad \bar{P}_h = \frac{P_h}{G} 
\label{eq:Th_sol}\end{equation}
The homeostatic pressure is the preferred pressure state that the system tends to and stops growing at, for the case of uniform applied pressure in absence of diffusion-consumption limitations. Using \cref{eq:uniformsol}, the related homeostatic stress state is given by $\nten{T}_h = -P_h \nten{I}$. The dimensionless homeostatic pressure is plotted as function of the conversion energy ratio $\eta$ in \Cref{fig:fig5}(b). The more favourable the conversion energy (smaller $\eta$) or smaller the value of $\chi$, the larger the homeostatic pressure. Further, the ratio $\phi_{f}/\phi_c = J^s_{c}/J^s_f$ is plotted in \Cref{fig:fig5}(c), the homeostatic pressure is positive when it is smaller than one (since $J^s_f$ is the swelling at zero pressure and the swelling ratio decreases monotonically with increasing applied pressure) while the homeostatic pressure is negative when it is larger than one. Hence, the homeostatic pressure is mostly positive except for small negative values at high values of $\chi$ and $\eta$. 
Thus for the uniform swelling-growth problem considered here,  \textit{our swelling-growth theory offers a kinetic basis for the typically phenomenologically prescribed homeostatic stress, its relation to underlying material parameters, and for why it is typically observed to be compressive for soft growing systems}. However, note that for a general boundary value problem, there is no fixed stress state (dependent on material parameters alone) that stops growth based on our growth laws.  This emphasizes the need to move away from phenomenological prescriptions as experimental observations under specific conditions need not translate to other scenarios. \\

Finally, we show the dependence of the dimensionless growth rate as a function of the applied pressure using \cref{eq:gammabar_dimlss,eq:exact_fg,eq:eqb_sol_incomp}. The growth rate only depends on the solid volume fraction which only depends on the applied pressure here in the absence of diffusion-consumption limitations. 
We choose $\bar{f}_g^* = \bar{f}_g(\phi_{f})$ so that $\bar{\Gamma} = 1$ when $P_b =0$. The dimensionless growth rate $\bar{\Gamma}$ is plotted as a function of $P_b/P_h$ in \Cref{fig:fig5}(d) for different values of $\chi$ and $\eta$ ($P_h>0$ for all combinations of chosen $\chi$ and $\eta$). It can be seen that larger applied pressure (higher values of $P_b/P_h$) always leads to decreasing growth rates. For $P_b/P_h>1$, we have $\bar{\Gamma} = 0$. \textit{Hence, our swelling-growth theory also offers a kinetic basis for the experimentally observed dependence of growth rate on applied stresses.}\\

We note that all the conclusions above also hold for the case of non-isochoric species conversion (see \ref{app:nonisochoric}). Having demonstrated the ability of the theory to qualitatively capture the different experimentally observed dependences of the growth rate on the concentration of the diffusing species and the applied stress (in the absence of diffusion-consumption effects), we will now employ the theory to solve boundary value problems where these dependences become relevant.

\subsubsection{Growth under mechanical confinement without diffusion-consumption limitations}
\label{subsubsec:mechconfine}

First, we consider the problem of growth under mechanical confinement in the limit of perfect incompressibility and infinitely fast diffusion so that we neglect diffusion-consumption limitations. The results will already demonstrate the coupling between swelling and growth and how the notion of homeostatic stress developed in the previous section manifests in a physical boundary value problem.  \\

We consider mechanical confinement in the form of a deformable spherical shell and  assume the growing body is spherical, so that $\partial{\mathcal{B}^R}$ is the surface of a sphere.  The confining shell is taken to have inner and outer radial dimensions, $A$ and $B$ respectively, so that $B/A$ is a thickness measure of the shell.  
It is assumed to be non-growing ($\nten{F}^g=\nten{I}$), non-swelling ($\nten{F}^s=\nten{I}$), perfectly incompressible ($J^e \to 1$) and made up of a strain stiffening material described by a Mooney-Rivlin model
\citep{mooney1940theory,rivlin1948large} so that its reference free energy $\psi_R$ is of the form \citep{boulanger2001finite}
\begin{equation}
    \psi_R = \frac{G_c}{2}\left( (1-n) \left(\text{tr}(\nten{B}) -3\right) + n \left(I_2(\nten{B}) -3\right) \right), \quad {I}_2(\nten{B}) = \frac{1}{2} \left( (\mathrm{tr}(\nten{B}))^2-\mathrm{tr} \left( \nten{B}^2 \right) \right) \label{eq:confinefreener}
\end{equation}
where  $G_c$ is the linear shear modulus of the confinement, $n$ is a stiffening parameter (larger the $n$ the more the stiffening) and $\nten{B}=\nten{B}^e$.  It can be shown \citep{chen2018thin} that the pressure applied by the mechanical confinement upon deformation only depends on the circumferential stretch of its inner boundary, $\lam_a$, so that
\begin{equation}
    \frac{P_b}{G_c} =  \frac{1-n}{2}\left(4\lam_b^{-1} + \lam_b^{-4} - \lam_a^{-4} - 4\lam_a^{-1} \right) + \frac{n}{2}\left(-4\lam_b + 2\lam_b^{-2} + 4\lam_a - 2 \lam_a^{-2}\right) \label{eq:Tb_shell}
\end{equation}
where the circumferential stretch at the outer boundary, $\lam_b$, is expressed in terms of $\lam_a$ due to incompressibility by the relation
\begin{equation}
    \lam_b = \left(1+(\lam_a^3-1)\left(\frac{A}{B}\right)^3\right)^{\frac{1}{3}} \label{eq:lamb_confine}
\end{equation}
We assume that, at $t=0$, the undeformed inner boundary of the shell is just in contact with the initially diffusion equilibrated growing body, so that the circumferential stretch of the confining shell is $\lam_a^3 = J/J_0$, where $J$ is the volume ratio of the uniformly growing body and $J_0 = J(t=0) = J^s_{f}$ is its initial value. The chemical boundary condition is still taken as $\mu = \mu_0^f$ on $\partial{\mathcal{B}^R}$. \\
 
We choose the following representative parameters based on our modelling of growing tumors later in \Cref{subsec:tumorresults},
\begin{equation}
G/\mu^* = 4 \times 10^{-5}, \quad \chi = 0.55,\quad  \eta = 0.95 , \quad n = 0.9 
\end{equation}
The choice of $n=0.9$ corresponds to a strain stiffening confinement where the pressure applied by the confinement increases continuously with deformation. Later, when we model growing bacterial biofilms in \Cref{subsec:biofilm_model}, we will consider neo-Hookean confinement ($n=0$)  wherein the applied pressure attains a maximum value with increasing deformation. \\

\begin{figure}[p]
    \centering
    \includegraphics[width=\textwidth]{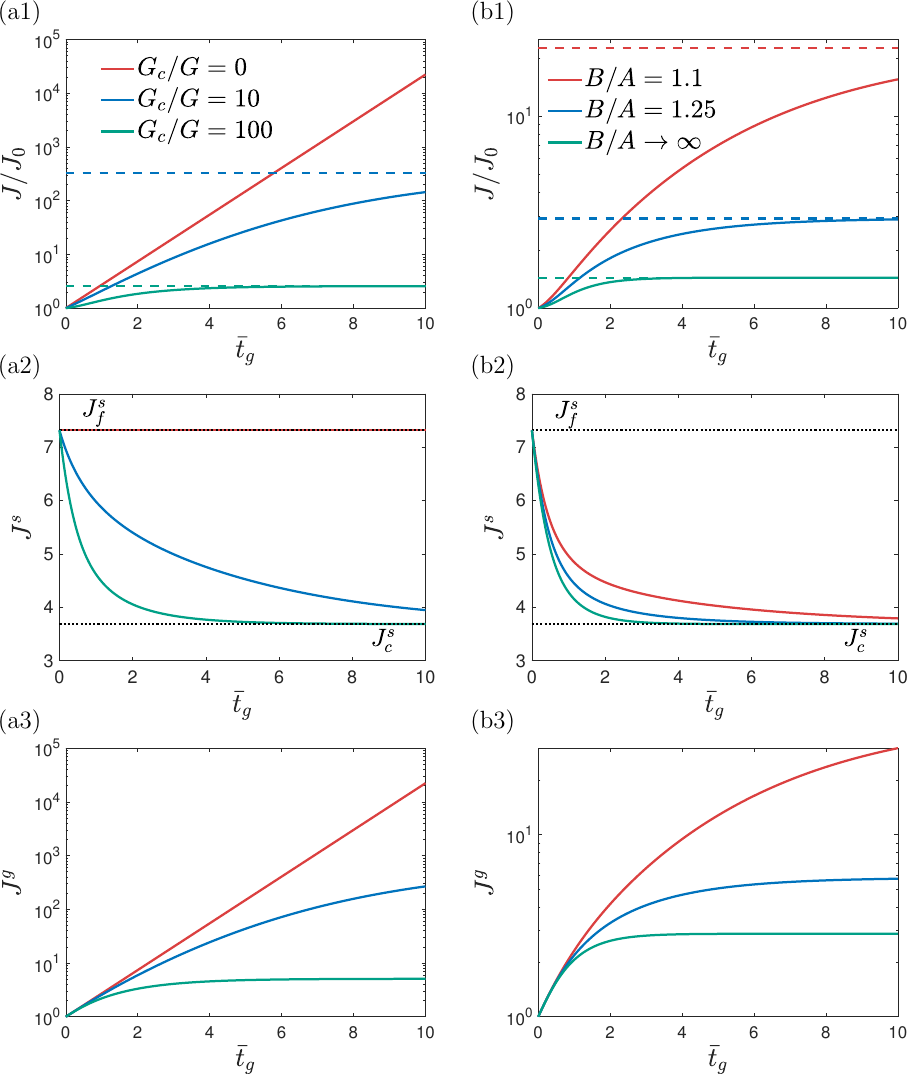}
    \caption{Temporal evolution of variables for uniform swelling-growth against mechanical confinement, in the limit of infinitely fast diffusion ($\tau_g/\tau_d \to \infty$)  and perfect incompressibility, with $n=0.9$. (a1-a3) Plots for a confining medium ($B/A \to \infty$) for varying modulus ratio of confinement to the growing body ($G_c/G$). (b1-b3) Plots for confining shells with varying dimension ratio $B/A$ (outer to inner dimension)   and fixed modulus ratio $G_c/G = 250$. The dashed lines in (a1) and (b1) correspond to the analytic  homeostatic limit. }
    \label{fig:us_shell}
\end{figure}

The form of the solution from the previous section in \cref{eq:uniformsol} continues to apply here. 
To solve the boundary value problem, we need to integrate the growth evolution equation \eqref{eq:gammabar_dimlss} (with $\bar{f}_g$ in \cref{eq:exact_fg}) while satisfying \cref{eq:eqb_sol_incomp,eq:Tb_shell}. This is done numerically and the results are plotted in \Cref{fig:us_shell} as a function of $\bar{t}_g$ for different cases of stiffness ratio $G_c/G$ and confinement dimension ratio $B/A$. Note that the dimensionless results here are independent of dimensional parameters such as the growth constant $k_g$.
In \Cref{fig:us_shell}(a1-a3), the results are plotted for the case of a confining medium ($B/A \to \infty$) with varying stiffness ratio $G_c/G$ to show the effect of confining stiffness on growth (later we model bacterial biofilm growth against a confining medium). It can be seen that in the case of free growth ($G_c/G \to 0$), the swelling ratio remains unchanged during growth at its free value $J^s_{f}$. This leads to constant growth rate of $\bar{\Gamma} = 1$, i.e a straight line in log-log plot of $J^g$ vs $\bar{t}_g$ with slope of unity (remember that $\bar{f}_g^*$ was chosen such that the dimensionless growth rate is unity for free growth). When the confining stiffness is finite, the applied pressure increases with deformation of the medium due to growth of the body which reduces the swelling ratio $J^s$ (\Cref{fig:us_shell}(a2)), thus the swelling is coupled to the growth. This decrease in $J^s$ leads to decrease in the growth rate (\Cref{fig:us_shell}(a3)) from its free growth value (growth is coupled to the swelling). The combined reduction of growth rate and $J^s$ with time from their free growth values leads to a suppressed profile for the evolution of $J/J_0$ compared to free growth (\Cref{fig:us_shell}(a1)) which will be the physically observed evolution of volume of the growing body in experiments.\\

If the confining medium deforms enough that the applied pressure reaches the homeostatic value $P_h$ (or equivalently the swelling ratio reaches the critical value $J^s_c$) discussed in the previous section, the growth will stop and the field variables stop evolving with time. The volume ratio $J/J_0$ at which growth will stop can be analytically evaluated by setting $P_b = P_h$ and $\lam_a^3 = J/J_0$ in \cref{eq:Tb_shell} where $P_h$ is solved using \cref{eq:Th_sol}. These values are plotted using dashed lines in \Cref{fig:us_shell}(a1) and are confirmed to match with simulations. As one would intuitively expect, 
larger confining stiffness lowers the growth rate and leads to smaller steady size of the growing body since the homeostatic stress (or critical swelling ratio) is reached at smaller deformations (\Cref{fig:us_shell}(a1)). Such behaviour has been observed in tumors growing against a confining medium \citep{helmlinger1997solid}. Note that, for a neo-Hookean confinement (i.e non stiffening so that $n=0$), the applied pressure will saturate at a maximum cavitation pressure with increasing deformation. If the homeostatic pressure is larger than this cavitation pressure, the body will never stop growing due to the confinement. We will see such behavior in the bacterial biofilm growth experiments we model later. \\

Next, we demonstrate the effect of varying thickness of the spherical shells for a fixed value of $G_c/G = 250$ in \Cref{fig:us_shell}(b1-b3). The same discussion above for the case of varying stiffness ratio applies again and it can be seen that larger thicknesses of the confinement for a given stiffness ratio lead to smaller steady sizes as one would intuitively expect and as seen in the tumor experiments in \Cref{sec:Overview_tumor}. The assumption of infinitely fast diffusion employed so far prevents accounting for diffusion-consumption effects which are important for large volumetric growth. We relax this assumption in the following sections. 

\subsection{Spherically symmetric equations}
\label{subsec:spherical symmetry}
To account for diffusion-consumption effects as the body is growing, we need to account for spatially varying field variables. We restrict the analysis to a spherically symmetric setting and consider a swelling and growing body subjected to a uniform external pressure and immersed in a bath of diffusing species maintained at a constant reference chemical potential. The dry reference body is described by the radial coordinate $R \in [0, R_0]$ and the current radial coordinates is given by $r = r(R)$ where $r(0)=0$ and $r(R_0)$ is the current radius of the body. Choosing $L^* = R_0$ to non-dimensionalize the problem, we have dimensionless reference coordinates $\bar{R} (= R/R_0) \in [0, 1]$ and dimensionless current coordinates $\bar{r} = r/R_0 = \bar{r}(\bar{R})$. The deformation gradient in spherical basis and associated volume ratio $J$ can then be written as
\begin{equation}
\nten{F} = \text{diag} \left[\lam_r, \lam_\theta, \lam_\theta\right]  = \text{diag} \left[\pdv{\bar{r}}{\bar{R}}, \frac{\bar{r}}{\bar{R}}, \frac{\bar{r}}{\bar{R}}\right],\quad J =  \lam_r \lam_\theta^2 \label{eq:Fsph}
\end{equation}
where $\lam_r$ is the radial stretch and $\lam_\theta$ is the circumferential stretch.
The elastic, swelling, and elasto-swelling deformation gradient tensors $\nten{F}^e,\nten{F}^e,\nten{F}^{es}$, and their associated determinants can be written as
\begin{equation}
\nten{F}^i = \text{diag} \left[\lam_r^i, \lam_\theta^i, \lam_\theta^i\right],\quad J^{i} = \lam_r^i {\lam_\theta^i}^2 \text{ for } i = \{g,e,es\} \label{eq:F_lam_connect}
\end{equation}
where the different stretches are related as follows
\begin{equation}
\lam^e_{i} = \frac{\lam_i}{\lam^s\lam^g_i},\quad \lam^{es}_{i} = \frac{\lam_i}{\lam^g_i}, \quad  \text{ for } i = \{r,\theta\} 
\end{equation}
Thus, given $\bar{r}(\bar{R})$, $\lam_r^g$, $\lam_\theta^g$, and $\lam^s$, all deformation gradient tensors can be determined. 
 The dimensionless Cauchy and Piola stresses in \cref{eq:stress_spec_dimless,eq:Piola_dimless} specialize as
\begin{equation}
\bar{\nten{T}} = \text{diag} \left[\bar{T}_r, \bar{T}_\theta, \bar{T}_\theta\right],\quad \bar{T}_i = \frac{1}{J^{es}} \left(({\lam_i^{es}}^2-1) + \frac{K}{G} \ln(J^e) \right) \text{ for } i =\{r,\theta\},
\end{equation}
\begin{equation}
\text{and} \quad \bar{\nten{S}} = \text{diag} \left[\bar{S}_R, \bar{S}_\Theta, \bar{S}_\Theta\right],\quad \bar{S}_R = \lam_\theta^2 \bar{T}_r,\quad \bar{S}_\Theta = \lam_r \lam_\theta \bar{T}_\theta
\end{equation}
respectively. The dimensionless mechanical equilibrium equation \eqref{eq:Piola_dimless} under spherical symmetry is given by
\begin{equation}
\dv{\bar{S}_R}{\bar{R}} + \frac{2}{\bar{R}}\left(\bar{S}_R - \bar{S}_{\Theta}\right) = 0 \label{eq:spher_eqb}
\end{equation}
The equilibrated diffusion-consumption equation \cref{eq:app_reac_diff_eqbtd} meanwhile specializes as
\begin{equation}
J^s J^g \bar{\Gamma} \left(\frac{\tau_d}{\tau_g}\right) = \frac{1}{\bar{R}^2}\pdv{\left(\bar{R}^2 \bar{M}_\text{mob}^R \pdv{\bar{\mu}}{\bar{R}}\right)}{\bar{R}} \quad \text{where} \quad \bar{M}_\text{mob}^R = \frac{{\lam_\theta}^2}{\lam_r} \label{eq:reacdiff_spherical}
\end{equation}
, the dimensionless chemical potential $\bar{\mu}$ is defined in \cref{eq:mu_dimless}, and the dimensionless mobility tensor is specialized from \cref{eq:reacdiff_dimlss}$_2$.\\

The volumetric growth evolution is given by  \cref{eq:gammabar_dimlss} where $\bar{f}_g$ specializes as
\begin{equation}
\begin{split}
\bar{f}_g =&\ \Delta \bar{\mu}_0 + \bar{f}_g^{\text{mix}}(\phi)  + \frac{G}{\mu^*}\left( \ln(J^{es}) - \frac{K}{2G}\left(\ln(J^e)\right)^2 - \frac{1}{6}  \left({\lam_r^{es}}^2 + 2 {\lam_\theta^{es}}^2 -3\right) \right) 
\end{split}
\end{equation}
We choose $\bar{f}^*_g = \bar{f}_g^\infty(\phi_{f})$ for normalizing the growth law, where $\bar{f}_g^\infty$ is defined in \cref{eq:exact_fg} and $\phi_{f}$ is the free solid volume fraction  for free growth in the limit of infinitely fast diffusion and perfect incompressibility, given by \cref{eq:free_phi}. 
The growth directionality evolution equation \cref{eq:devLg_dimlss} specialize for spherical symmetry as follows
\begin{equation}
\frac{1}{\lam_r^g}\dv{\lam_r^g}{\bar{t}_g} = \frac{\bar{\Gamma}}{3} - \frac{2G}{9 \mu^* \bar{f}_g^*}\left( {\lam_\theta^{es}}^2 - {\lam_r^{es}}^2\right),\quad \frac{1}{\lam_\theta^g}\dv{\lam_\theta^g}{\bar{t}_g} = \frac{\bar{\Gamma}}{3} + \frac{G}{9 \mu^* \bar{f}_g^*}\left( {\lam_\theta^{es}}^2 - {\lam_r^{es}}^2\right) \label{eq:growthlawspher}
\end{equation}
where $\bar{\Gamma}$ is given by \cref{eq:gammabar_dimlss}. It can be shown that only two of the three evolution equations in  \cref{eq:growthlawspher,eq:gammabar_dimlss} are independent and therefore we have two independent evolution equations for the growth (since essentially prescription of two of the three among $\lam_g^r$, $\lam_g^\theta$, and $J^g$ automatically determines the third through \cref{eq:F_lam_connect}$_2$). We thus have four field variables ($\bar{r}$, $\lam_r^g$, $\lam_\theta^g$, and $\lam^s$) and four governing equations (\cref{eq:growthlawspher,eq:spher_eqb,eq:reacdiff_spherical}). The prescription of boundary conditions for \cref{eq:spher_eqb,eq:reacdiff_spherical}, and initial conditions for equations in \eqref{eq:growthlawspher}  complete the definition of a solvable boundary value problem.\\

For the initial conditions for growth evolution, we set $\nten{F}^0_g = \nten{I}$ which reduces to the following choices for spherical symmetry
\begin{equation}
\lam_r^g(\bar{R},t=0) = 1, \quad \lam_\theta^g(\bar{R},t=0) = 1
\end{equation}
The following boundary conditions are prescribed for the radial stress and chemical potential at the outer boundary
\begin{equation}
\bar{T}_r(\bar{R} = 1,t) = -\bar{P}_b(t),\quad \bar{\mu}(\bar{R} = 1,t) = \bar{\mu}_0^f \label{eq:mu_bc} 
\end{equation}
where $\bar{P}_b(t)$ is the applied dimensionless pressure, related to the applied pressure $P_b(t)$ through the relation $\bar{P}_b(t) = P_b(t)/G$.\\

The spherically symmetric equations are solved using a numerical scheme outlined in \ref{app:numerics}. For solving the equations in dimensionless space, the only material parameters of the growing body that need to be specified are $\chi, \Delta \bar{\mu}_0$, $\tau_g/\tau_d$, $G/\mu^*$, and $K/G$. Note from \cref{eq:mu_dimless,eq:reacdiff_spherical,eq:mu_bc} that the value of $\bar{\mu}_0^f$ does not affect any of the solution field variables of the problem except to shift the chemical potential by a constant value. Thus without loss of generality we set $\bar{\mu}_0^f =0$. The parameter $\Delta \bar{\mu}_0$ can be replaced with the conversion energy ratio $\eta = \Delta \bar{\mu}_0/\Delta \bar{\mu}_0^c$ where $\Delta \bar{\mu}_0^c$ is evaluated using \cref{eq:exact_uniform}$_1$.\\ 

Since the fields now spatially vary, we define the following volume averaged parameters, 
\begin{equation}
\bar{J} = 3 \int_{0}^{1} J \bar{R}^2 \dd{\bar{R}}, \quad \bar{J}^s = 3 \int_{0}^{1} J^s \bar{R}^2 \dd{\bar{R}} , \quad \bar{J}^g = 3 \int_{0}^{1} J^g \bar{R}^2 \dd{\bar{R}}, \quad \bar{J}_0 = \bar{J}(t=0) \label{eq:spatialaverage}
\end{equation}
where $\bar{J}$ is the total volume ratio of the body with respect to its dry reference volume, $\bar{J}_0$ is initial value of $\bar{J}$ and $\bar{J}/\bar{J}_0$ is the total volume ratio of the body compared to its initial equilibrium volume (Note that we are solving the diffusion equilibrated version of the governing equations so that $\bar{J}_0 \neq 1$). It can be shown using eqs. \eqref{eq:Fsph} and \eqref{eq:spatialaverage}$_1$ that $\bar{J} = \bar{r}^3(\bar{R}=1)=\lam_{\theta}^3(\bar{R}=1)$. \\

In \Cref{subsec:freegrowth} we consider free growth so that $\bar{P}_b =0$ and in \Cref{subsec:appliedpressure} we consider a constant applied pressure $\bar{P}_b(t)= \bar{P}_b$. For modelling the tumor experiments in \Cref{subsec:tumorresults} and the bacterial biofilm experiments in \Cref{subsec:biofilm_model}, the applied pressure is described as a function of the confinement deformation using \cref{eq:Tb_shell} where $\bar{P}_b = (P_b/G_c)\times(G_c/G)$. The  circumferential stretch at the inner boundary of the confining shell is now given by the following equation
\begin{equation}
    \lam_a^3 = \begin{cases}
    {\bar{J}}/{J_c},& \text{if } {\bar{J}}/{J_c}\geq 1\\
    1,              & \text{if } {\bar{J}}/{J_c}<1 \label{eq:lama_confine}
\end{cases}
\end{equation}
where $J_c$ is the ratio of inner volume of undeformed confinement to the initial dry reference volume of the body. The conditional cases in \cref{eq:lama_confine} reflect the fact that the confinement only starts deforming once the growing body comes in contact with it.

\subsection{Free growth with diffusion-consumption limitations}
\label{subsec:freegrowth}
 \begin{figure}
    \centering
    \includegraphics[width=\textwidth]{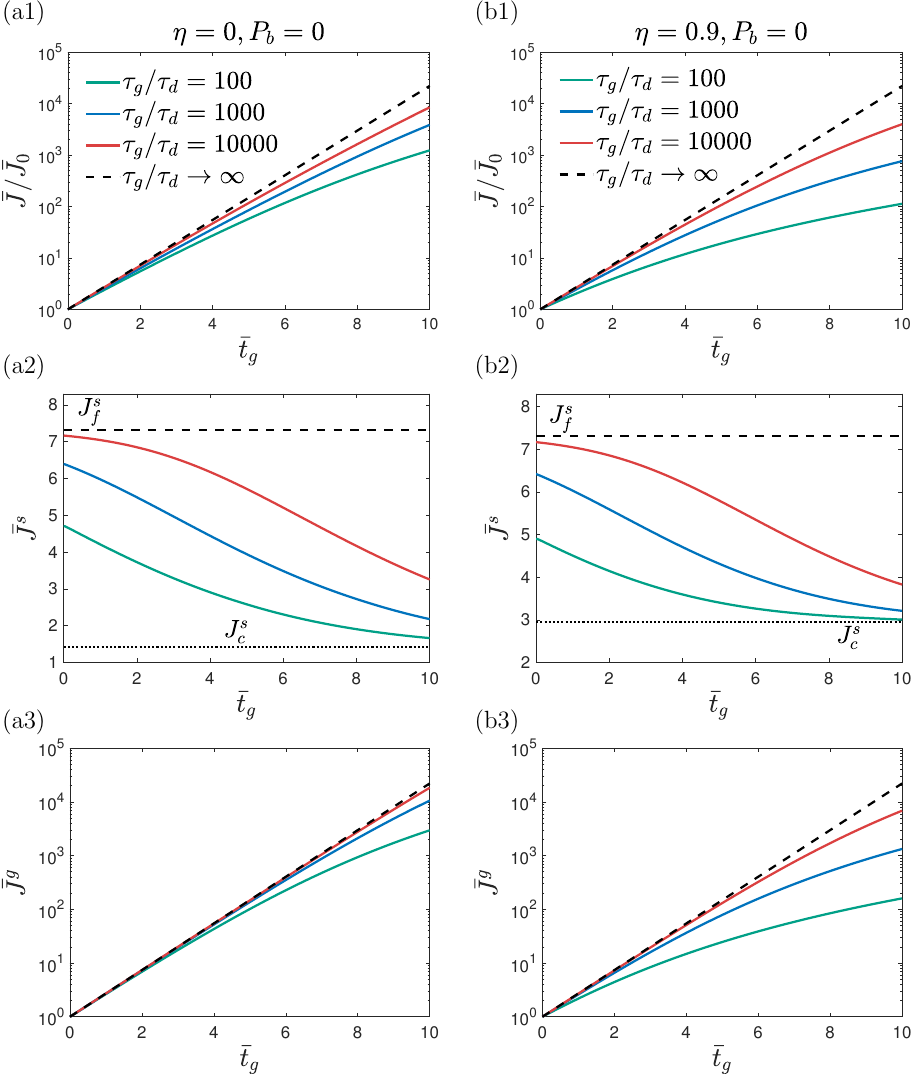}
    \caption{Plots of temporal evolution of spatially averaged quantities for the case of free growth with diffusion-consumption limitations, for various values of $\tau_g/\tau_d$.  (a1-a3) Plots for the case of zero species conversion energy. (b1-b3) Plots for the case of conversion energy ratio $\eta=0.9$. The dashed lines correspond to the limit of infinitely fast diffusion ($\tau_g/\tau_d \to \infty$) and perfect incompressibility ($K/G \to \infty$).}
    \label{fig:freegr}
\end{figure}

In this section, we analyze the problem of free growth in the presence of diffusion-consumption limitations. We
solve the spherically symmetric equations and set $P_b=0$. Once again we consider $\chi =0.55$ and $G/\mu^* = 4\times 10^{-5}$. 
 Near incompressibility is enforced by setting a high value of $K/G=10^6$. We consider different values of the conversion energy ratio $\eta$ and the growth to diffusion timescale ratio $\tau_g/\tau_d$. \\

\begin{figure}
    \centering
    \includegraphics[width=\textwidth]{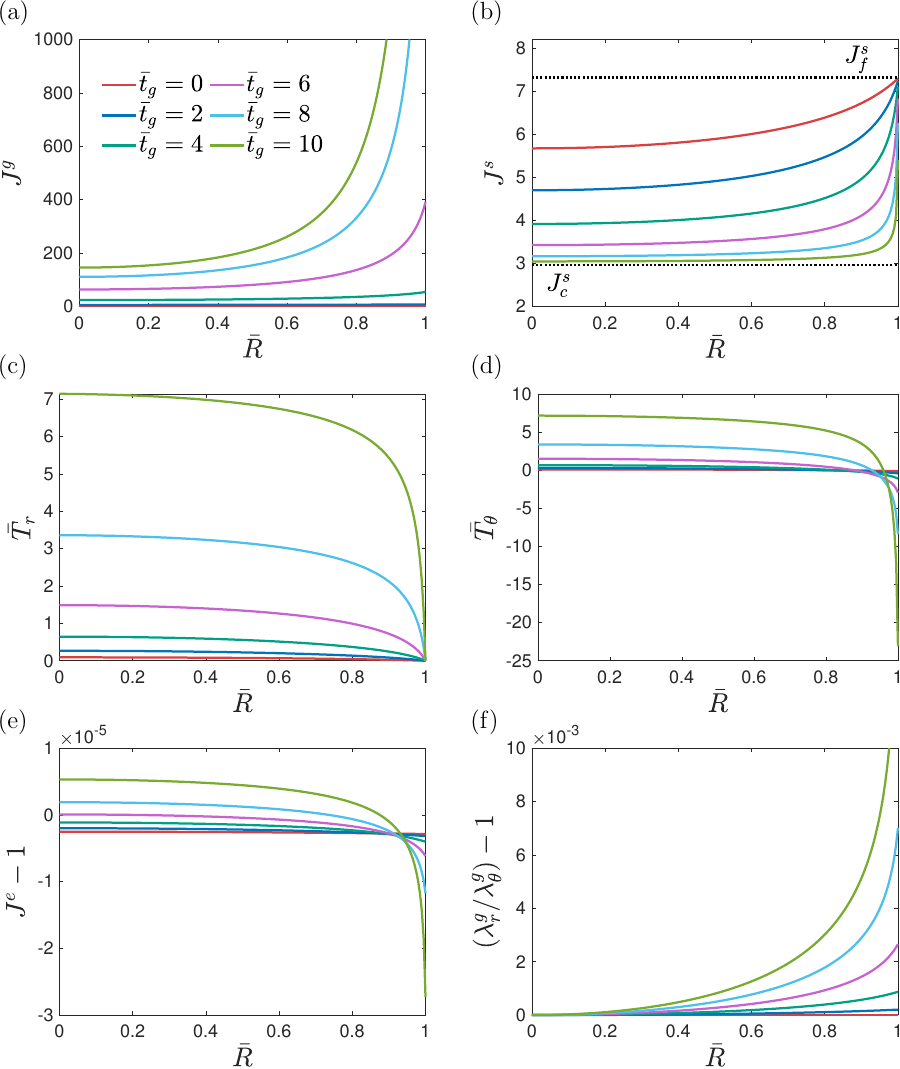}
    \caption{Plots of spatial variation of solution fields (over dimensionless referential radial coordinate $\bar{R}$) for free growth with diffusion-consumption limitations, at various times, for representative case of $\eta=0.9, \tau_g/\tau_d =1000$.}
    \label{fig:spatialfreegr}
\end{figure}

The time evolution of spatially averaged variables defined in \cref{eq:spatialaverage} is plotted in \Cref{fig:freegr} for varying values of $\tau_g/\tau_d$ at fixed conversion energy ratio $\eta$. The plots in \Cref{fig:freegr}(a1-a3) are for the case of $\eta =0$ and those in \Cref{fig:freegr}(b1-b3) are for the case of $\eta =0.9$.  The dashed lines correspond to the case of infinitely fast diffusion ($\tau_g/\tau_d \to \infty$) and perfect incompressibility ($K/G \to \infty$) wherein the swelling ratio is the free value $J^s_f$ at all times and the dimensionless growth rate is unity. The spatial profiles of the field variables are plotted at increasing times in \Cref{fig:spatialfreegr} for a representative case of $\eta =0.9, \tau_g/\tau_d = 1000$ (the timescale ratio used for tumor growth modelling later).

\FloatBarrier

\begin{figure}[!htb]
    \centering
    \includegraphics[width=\textwidth]{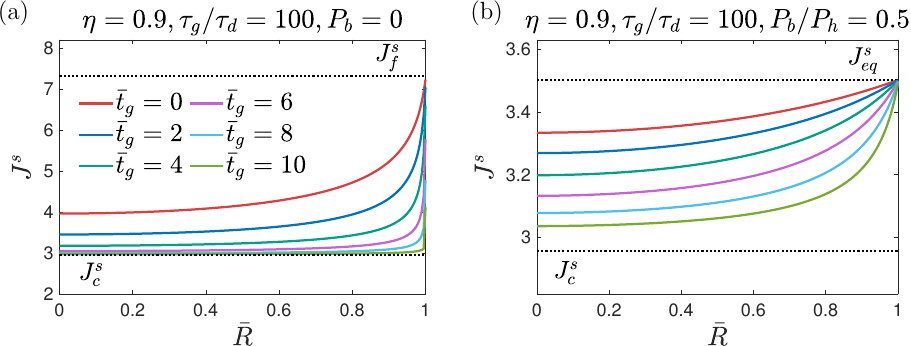}
    \caption{Plots of spatially averaged swelling ratio for growth with diffusion-consumption limitations for the case of conversion energy ratio $\eta =0.9$ and growth to diffusion timescale ratio $\tau_g/\tau_d=100$.  (a) Free growth case (b) Case of applied pressure equal to $50\%$ of the homeostatic pressure. } \label{fig:moreJs}
\end{figure}

 When $\tau_g/\tau_d$ is finite, it is seen from \Cref{fig:spatialfreegr}(b) that the equilibrium swelling ratio profile is spatially decreasing from the outer edge of the growing body to the inside, due to diffusion-consumption balance, and that this profile drops with time as the body swells and grows. This leads to spatially decreasing volumetric growth rates from the outside to inside which all drop with time.  Consequently the spatially averaged swelling ratio ($\bar{J}^s$) drops with time (\Cref{fig:freegr}(a2,b2)), as does the spatially averaged growth rate as seen by the decreasing slope with time of the log-log plot of $\bar{J}^g$ vs $\bar{t}_g$ in \Cref{fig:freegr}(a3,b3). The combined drop in swelling ratio and growth rates lead to decreasing overall volume increase rate, resulting in a decreasing slope with time of the log-log plot of $\bar{J}/\bar{J}_0$ vs $\bar{t}_g$ in \Cref{fig:freegr}(a1,b1). For a given $\eta$, smaller values of $\tau_g/\tau_d$ result in lower swelling ratio profiles at all times as can be seen by comparing \Cref{fig:moreJs}(a) where the spatial profiles of the swelling ratio are plotted for the case of $\eta=0.9, \tau_g/\tau_d = 100$ with the curves in \Cref{fig:spatialfreegr}(b) for $\eta=0.9, \tau_g/\tau_d = 1000$ (due to a smaller consumption term on the left hand side of \cref{eq:reacdiff_spherical}). Consequently for a given $\eta$, smaller values of $\tau_g/\tau_d$ lead to larger decrease with time in spatially averaged swelling ratio, growth rates, and volume increase rate, as seen in \Cref{fig:freegr}. Further, for increasing $\tau_g/\tau_d$, the curves correctly approach the results for infinitely fast diffusion limit ($\tau_g/\tau_d \to \infty$). Thus, the presence of diffusion-consumption effects leads to decreased growth and volume increase rates both with time and in comparison to the constant free growth rate in the infinitely fast diffusion limit. Finally, we note that the higher the conversion energy ratio $\eta$ for a given $\tau_g/\tau_d$, the more pronounced the drop in averaged growth rate and volume increase rate compared to their values in the corresponding infinitely fast diffusion limit\footnote{Note that the value of $\bar{f}^*_g$ chosen to normalize the growth law is different for different $\eta$ (since we enforce dimensionless volumetric growth rate of 1 for free growth with no diffusion-consumption limitations, irrespective of conversion energy). }. This is due to the fact that a larger value of $\eta$ corresponds to a smaller value of $\Delta \mu_0$ (when all other parameters are fixed) which can be shown to lead to a smaller dimensionless growth rate at a given swelling ratio (see \ref{app:gammbar_eta}). \\
 
The spatial variation of the growth volume ratio $J^g$ in \Cref{fig:spatialfreegr}(a) at increasing times is consistent with the earlier discussion wherein the growth rates are spatially decreasing from the outer edge of the body to the inside due to the spatial variation of swelling ratio. The dimensionless radial stress profiles are plotted in \Cref{fig:spatialfreegr}(c). The radial stress is zero on the outer boundary due to the boundary condition for free growth while being increasingly tensile towards the core of the body, with the stress variation becoming more pronounced with increasing time. This is due to the fact that the outer layers are growing at a faster rate and pulling on the slower growing inner layers. The dimensionless circumferential stress plots in  \Cref{fig:spatialfreegr}(d) indicate the circumferential stress is tensile in most of the growing body but is compressive near the outer edge. This is due to mechanical equilibrium arising from the combined effects of swelling and growth and is not intuitive, nevertheless the spatial variation in radial and circumferential stress profiles is consistent with previous studies \citep{ambrosi2002mechanics,xue2016biochemomechanical}. Note that the high tensile stresses in the core at large growth volumes can lead to cavitation in the growing body, see for example \cite{goriely2010elastic,mcmahon2010spontaneous}. \textcolor{black}{In the context of tumors and bacteria, this might manifest in the form of tendency of cells to liquefy.} However, the high surface tension of the diffusing fluids at small length scales such as in tumors and bacteria could also delay such failure. Nevertheless, the ability to account for large deformations in our swelling growth theory allows the possibility to access such nonlinear phenomena which might be critical to understanding growing biological systems.  Next, from  the plots of the elastic volume ratio in \Cref{fig:spatialfreegr}(e), we can verify that the simulations are indeed in the nearly incompressible limit ($J^e -1 \to 0$). The quantity $\lam_r^g/\lam_\theta^g - 1$ is a measure of the anisotropy of growth, the closer it is to zero the more isotropic the growth. This anisotropy measure is plotted in \Cref{fig:spatialfreegr}(f) and the growth is seen to be nearly spherical or isotropic at all times which is consistent with the discussion in \Cref{subsec:spec_constit_dimless} due to low value of $G/\mu^*$ \textcolor{black}{and our choice of growth laws}. Nevertheless, the growth is seen to be more anisotropic towards the outer edge of the body and at increasing times, with the radial growth stretch being higher than the circumferential growth stretch. This is due to the fact that the circumferential stress is always more compressive (or less tensile) than the radial stress which leads to suppressed circumferential growth in comparison to the radial growth, and this stress difference is more pronounced for larger radial coordinate and at increasing times. \\

 Revisiting the spatial swelling profiles in \Cref{fig:spatialfreegr}(b) and \Cref{fig:moreJs}(a), we note that the swelling ratio is seen to not drop below the critical value $J^s_c$. When $J^s$ approaches $J^s_c$ the growth rate $\bar{\Gamma}$ drops to zero which results in zero consumption of the diffusing species locally (see \cref{eq:reac_diff_eqbtd}), thus halting local reduction in swelling. 
At very large times, the vast majority of the body is seen to be at the critical swelling ratio $J^s_c$ which means that the growth is confined to a thin outer rim. We note that the growth can technically never completely stop at the outer edge since, as discussed in \Cref{subsec:spec_constit_dimless}, $\nten{L}^g = \nten{0}$ necessitates a hydrostatic stress state and this requires  $\nten{T}=\nten{0}$ at the outer boundary (since $T_r = 0$). Zero stress ($\nten{T} = \nten{0}$) along with $\mu = \mu_0^f$ at the boundary would necessitate the swelling ratio at the boundary to be the free swelling ratio   
which is not compatible with zero growth rate.  However, since $G/\mu^*$ is very small, the volumetric growth can nearly stop when the swelling ratio approaches the critical value over the entire body from accumulating stresses (the remodelling relieves stresses that can resume volumetric growth but this process is slow since $G/\mu^*$ is very small). We note that as the body grows to very large volumes, the use of the equilibrated version of the diffusion-consumption equation will become less accurate. However the accurate simultaneous integration of growth evolution equations along with non-equilibrated diffusion-consumption equation is numerically prohibitive due to the large $\tau_g/\tau_d$, strong swelling-growth coupling, and extreme growth sizes.

\subsection{Growth under constant applied pressure with diffusion-consumption limitations} 
\label{subsec:appliedpressure}
 \begin{figure}
    \centering
    \includegraphics[width=\textwidth]{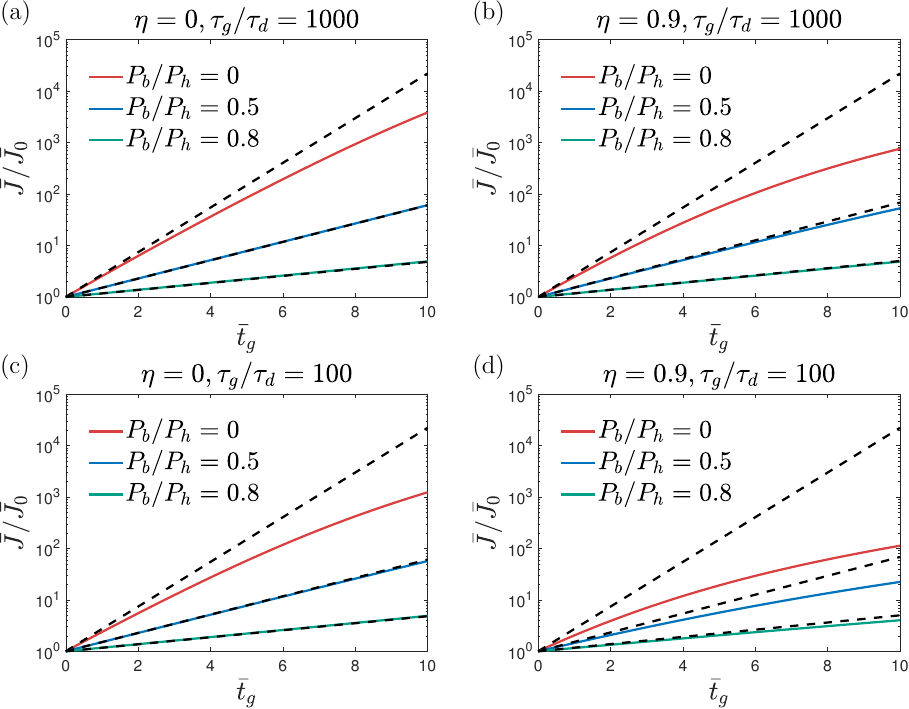}
    \caption{Plots of temporal evolution of total volume ratio with respect to initial equilibrium value ($\bar{J}/\bar{J}_0$) for the case of growth under uniform and constant applied pressure with diffusion-consumption limitations, for different values of conversion energy ratio $\eta$, growth to diffusion timescale ratio ($\tau_g/\tau_d$), and ratio of applied pressure to homeostatic pressure ($P_b/P_h$). The dashed lines represent the limit of infinitely fast diffusion ($\tau_g/\tau_d \to \infty$) and perfect incompressibility, and are shown for comparison. }\label{fig:appliedStress}
\end{figure} 
We now consider the combined effects of applied pressure ($P_b \neq 0$) and diffusion-consumption limitations on the growth. The same parameters from the previous section for free growth have been employed here. 
The applied pressure is taken to be constant so that $\bar{P}_b(t) = \bar{P}_b$. Using the value of homeostatic pressure $P_h$ developed for the limit of infinitely fast diffusion and perfect incompressibility in \Cref{subsec:uniformswell}, we consider different levels of applied pressure $P_b$ by considering different values of the ratio $P_b/P_h$ (which is also equal to $\bar{P}_b/\bar{P}_h$). The results are  plotted in \Cref{fig:appliedStress} for different values of $\eta$ and $\tau_g/\tau_d$. Since much of the physics is similar to the case of free growth,
only the temporal evolution of $\bar{J}/\bar{J}_0$ is shown for sake of brevity, while a representative plot of the spatial profile evolution of $J^s$ is shown in \Cref{fig:moreJs}(b) for the case $\eta =0.9,\tau_g/\tau_d=100, P_b/P_h=0.5$. The dashed lines in \Cref{fig:appliedStress}  correspond to the limit of infinitely fast diffusion and perfect incompressibility wherein the swelling ratio will be spatially and temporally constant with a value $J^s_{eq}(\bar{P}_b)$ (defined in \Cref{subsec:uniformswell}), leading to constant growth rate ($\bar{\Gamma}$ in \Cref{fig:fig5}(d)).\\ 

 Similar to the free growth case, it is seen from \Cref{fig:moreJs}(b) that as the body grows, the swelling ratio decreases everywhere. However the values of $J^s$ are lower compared to the free growth case in \Cref{fig:moreJs}(a) due to applied pressure and the spatial profiles take longer to drop due to the reduced growth rates at smaller swelling ratios. The decreasing swelling ratio profile once again leads to decreasing volume increase rates with time in \Cref{fig:appliedStress} and the decrease is larger for smaller $\tau_g/\tau_d$ when all other parameters are the same. Higher applied pressures are seen to result in reduced growth rates for a given $\eta$ and $\tau_g/\tau_d$ (similar to the infinitely fast diffusion limit as seen by the dashed lines). The reduction in volume increase rates compared to the infinitely fast diffusion limit is less pronounced for higher applied pressure, this is because the growth rate $\bar{\Gamma}$ is lower for higher applied pressure which results in a smaller consumption term on the left hand side of \cref{eq:reacdiff_spherical}. 
 On the other hand, the reduction in volume increase rate 
 compared to the infinitely fast diffusion limit is once again more pronounced for larger values of conversion energy ratio $\eta$ when all other parameters are held fixed. Similar trends also apply for case of mechanical confinement considered in \Cref{subsubsec:mechconfine} when accounting for diffusion-consumption limitations, once again we skip these results for brevity and directly demonstrate application to tumor growth modelling in the next section.

\subsection{Modelling tumor growth}
\label{subsec:tumorresults}

We now model the tumor growth experiments discussed in \Cref{sec:Overview_tumor} using our swelling-growth theory.  We choose $R_0$ such that the initial equilibrated volume (under no applied pressure) is just in contact with the smallest confinement (Case C, inner volume = 0.003mm$^3$). The material parameters for the simulation are listed in \Cref{Table:tumor_matparms} and the associated key dimensionless parameters are 
\begin{equation}
     G/\mu^* = 4 \times 10^{-5}, \quad \tau_g/\tau_d = 1000, \quad \eta = 0.972, \quad K/G = 10^6
\end{equation}
We note that all parameters except the conversion energy $\Delta \mu_0$ are either obtained or directly inferred from experiments/literature or once fitted have been found to be in the range of reported values in the literature. For example, the growth timescale $\tau_g$ is directly inferred from the initial growth rate in the free growth experiments whereas the diffusion timescale $\tau_d$ is fitted for, and the resulting associated diffusion coefficient has a value $D \approx 3.6 \times 10^{-7} \text{cm}{}^2/\text{s}$, which is a typical value for several growth factors in tumor studies \citep{thorne2004diffusion,kim2011role}. The dimensions of the confinement in Cases B-D are listed in \ref{app:numerics_parameters}. For Case D, the breakage of the confinement is modelled by removing the confinement at $t = 10.35$ days. The simulated predictions of the theory are compared with the experiments in \Cref{fig:tumor}(b), and it can be seen that the theory is able to capture all the cases well with a single set of parameters. We emphasize again the fact that unlike most conventional growth models, we have not prescribed any ad hoc functional dependence of growth evolution on concentration including critical concentration for growth. Neither have we introduced an ad hoc homeostatic stress. The kinetic growth law based on the driving stress is automatically able to capture simultaneously the effects of both the mechanical confinement and diffusion-consumption limitations. Further, we draw attention to our relatively small parameter space compared to other tumor growth studies. Our modelling is also fully consistent, and accounts for mass balance.  

\begin{table}[]
\caption{Material parameter values for modelling of tumor growth experiments} 
\begin{tabular}{lll}
\hline
Quantity & Value & Source/Comment \\ \hline
$R_0$       &  56.33 $\mu$m     &   Chosen such that  $\frac{4}{3}\pi {R}_0^3$ $\bar{J}_0 = 0.003$ mm$^3$                \\
$G$        &   1 kPa    &     \cite{stylianopoulos2012causes}                  \\
$K$        &   1 GPa    &     Typical value for nearly incompressible soft polymer                 \\
$\tau_g$         &  1.0317 days     &   Fitted from initial free growth rate       \\
$\Delta \mu_0$       &  -10.15 MPa    &      Fitted, corresponds to $\eta = 0.972$                 \\
 $D$        &   3.6 $\times 10^{-7}$ cm$^2$/s    &   Fitted, close to reported values in literature\\        
  $\chi$       & 0.6      &   Assumed based on free tumor porosity ($\phi^f$) $\sim$ 0.25 \\
  $\omega^f$       &  $1.66 \times 10^{-28}$ m$^3$/molecule     &     \cite{hong2008theory}              \\
  $T$        &   300 K    &   Room temperature        \\
 $G_c$        &   250 kPa    &    \cite{alessandri2013cellular}                     \\
   $n$      &   0.9    &    Fitted, confinement known to be stiffening    \\
   
  \hline 
\end{tabular} 
  \label{Table:tumor_matparms}
\end{table}

\subsection{Modelling bacterial biofilm growth}
\label{subsec:biofilm_model}

 \begin{figure}
    \centering
    \includegraphics[width=\textwidth]{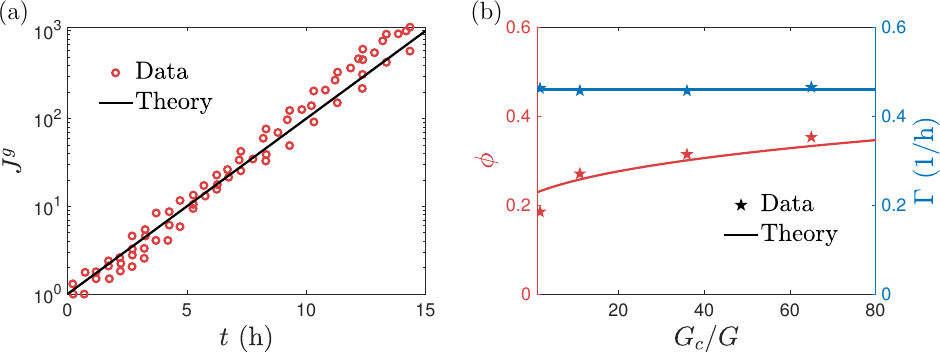}
    \caption{Comparison of experimental data \citep{zhang2021morphogenesis} and theoretical predictions for swelling-growth of bacterial biofilms. (a) Temporal evolution of the growth volume ratio. (b) The solid volume fraction ($\phi$) and growth rate ($\Gamma$) as a function of ratio of shear modulus of confinement to shear modulus of biofilm ($G_c/G$).} \label{fig:biofilms_model}
\end{figure} 

Finally, in this section, we model experiments of swelling and growing 3D bacterial biofilms. Our collaborators at Yale University published experiments of growing 3D \textit{Vibrio cholerae} biofilms embedded in hydrogels, as they grow by orders of magnitude from their initial size \citep{zhang2021morphogenesis}. Using 3D visualization techniques with high spatio-temporal resolution, they were able to capture the growth process at the level of individual cells which allows the separate extraction of growth and swelling data. The biofilms are grown starting from a few cells to tens of thousands, against confining hydrogel medium which is replenished with nutrient filled fluid. The confining stiffness of the hydrogel was varied by about two orders of magnitude. Yet, interestingly it was observed that the growth rate of the biofilms is constant with time (and inferred spatially uniform) and independent of the confining stiffness even though the swelling process was not.  These results are shown in \Cref{fig:biofilms_model}. The time evolution of the growth volume ratio from multiple experiments is overlaid in \Cref{fig:biofilms_model}(a) where it can be seen that the growth rate is constant. The swelling ratio reaches a steady value during growth and the steady value is plotted as function of the stiffness ratio of confinement to the biofilm in \Cref{fig:biofilms_model}(b), along with the growth rate. Note that from our earlier results in \Cref{subsubsec:mechconfine}, we expect the growth rate to be suppressed by increasing stiffness of the confining medium. We will resolve this seeming contradiction in this section and demonstrate the ability of the theory to model the experimental results.  \\

In a previous study, we analysed the morphogenesis in these biofilms as they grow using a energy minimization approach  \citep{li2022nonlinear}. However, we focused our attention to morphogenesis and  used a kinematic prescription for the volumetric growth while neglecting swelling since it was shown experimentally that the growth rate remained constant and spatially uniform. Here instead, we will neglect the morphogenesis and use our swelling growth theory considering spherical geometry (so that we can use \cref{eq:Tb_shell} for the pressure applied by confining medium) to capture the swelling dependence and explain the growth rate behaviour. We first note that since the initial size ($R_0=L_*$) is small ($\sim$ 1 $\mu$m), the diffusion timescale is small compared to the growth timescale ($\tau_g/\tau_d \sim 1 \times 10^5$ for typical $D \sim 10^{-7}$ cm$^2$/s) and we can comfortably employ the infinitely fast diffusion approximation. Since the confinement is a thick medium, we set $B/A\to \infty$ (and thus $\lam_b \to 1$) in \cref{eq:lamb_confine}. Further, since the swelling reaches a steady value during growth, the pressure applied by the confinement must saturate for large deformations. Thus, we assume a neo-Hookean free energy function for the confining medium ($n=0$ in \cref{eq:confinefreener})   such that the applied pressure $P_b$ in \cref{eq:Tb_shell} saturates at a cavitation pressure\footnote{While in reality the pressure might saturate from damage in the confinement, for our analysis here only the peak pressure is relevant for which the cavitation limit of neo-Hookean model is a good approximation} value of $2.5G_c$. Further, the growth rate being independent of the confining stiffness, is possible if the homeostatic pressure of the system is very high compared to the maximum confining stiffness such that $P_h\gg $ max($G_c$). If this is the case, since $P_b \sim G_c$, the pressure ratio $P_b/P_h \to 0$ irrespective of the confining stiffness varying by two orders of magnitude. From the dependence of volumetric growth rate on $P_b/P_h$ in \Cref{fig:fig5}(d), it can be inferred that the dimensionless growth rate will be approximately unity irrespective of the confinement if $P_b/P_h \to 0$. Since the maximum value of stiffness ratio $G_c/G$ is $\sim 100$, all we need to capture the growth rate independence is to require $\bar{P}_h\gg100$.\\

With the above discussion in mind, we choose the material parameters in 
\Cref{table:biofilm} to model the experiments using the swelling growth theory in the infinitely fast diffusion limit (\Cref{subsec:uniformswell}). The steady value of swelling is obtained by setting $\bar{P}_b = 2.5 G_c/G$ in \cref{eq:eqb_sol_incomp}. Predictions of the swelling growth theory agree with the experimental data shown in \Cref{fig:biofilms_model}. We note that the difference from our results in \Cref{subsubsec:mechconfine} arises from two factors - (i) the confining medium is non-stiffening ($n=0$) so that the applied pressure saturates and (ii) we have chosen $\eta = -10$ here, which corresponds to a highly favourable conversion energy in comparison to the choice of $\eta=0.95$ previously. In principle, any value of $\Delta \mu_0$ greater than the value chosen in \Cref{table:biofilm} will also produce the same results. Essentially, if the conversion energy is favourable enough, it maintains the growth rate irrespective of changes in swelling ratio due to mechanical confinement. Or equivalently, if the conversion energy is favourable enough, the homeostatic pressure will be high (see \Cref{fig:fig5}(b)) compared to even the largest confining stiffness. We hypothesize that the underlying cellular mechanism that leads to this high growth favourability is the use of a bacterial strain that is locked in a high cyclic diguanylate level \citep{zhang2021morphogenesis,beyhan2007smooth} and therefore always producing biofilm matrix.   The swelling ratio however is independent of the conversion energy $\Delta \mu_0$ and only depends on the applied pressure (\cref{eq:eqb_sol_incomp}), its dependence on $G_c/G$ has been captured in \Cref{fig:biofilms_model}(b) by fitting for $\chi$. Thus our swelling-growth theory is able to successfully model both tumor and bacterial biofilm swelling-growth. We conclude the manuscript with a summary in the next section.

\begin{table}[]
\caption{Material parameter values for modelling of bacterial biofilm growth experiments}
\centering
\begin{tabular}{lll}
\hline
Quantity & Value & Source/Comment \\ \hline
$G$        &   1kPa    &     \cite{zhang2021morphogenesis}                  \\
$T$        &   300 K    &   Room temperature        \\
$\tau_g$         &  2.17 h      &   Fitted from constant growth rate       \\
  $\chi$       & 0.59      &   Fitted from $\phi$ vs $G_c/G$  \\
  $\Delta \mu_0$       &  105.88 MPa    &      Corresponds to $\eta = -10$                 \\ $\omega^f$       &  $1.66 \times 10^{-28}$ m$^3$/molecule     &     \cite{hong2008theory}              \\ \hline 
\end{tabular}
\label{table:biofilm}
\end{table}

\section{Summary and conclusions} 
\label{sec:Conclusions}

Biological systems exhibit a vast array of growth phenomena that inherently rely on the coupling between large deformations and diffusion of constituents that are needed to feed the growth. Existing models of growth typically employ experimentally motivated phenomenological prescriptions to capture the consequences of this coupling by prescribing the directionality of growth (i.e. anisotropy), its rate, and the conditions required for it to stop (i.e. a homeostatic stress and a critical concentration of diffusing species).   In this work,  we investigate the coupled nature of growth by development of a  thermodynamically consistent and mass conserving  large deformation swelling-growth theory. The theory considers a solid body permeated by a representative diffusing species whose conversion into additional solid material leads to growth while the remaining diffusing species swells the growing solid. 
The mechanics is modelled by treating the mixture as a single homogenized continuum body, which circumvents difficulties of mixture theory.
The driving stress for growth is identified using the dissipation inequality, and a kinetic growth law is prescribed. It is shown that this framework  successfully captures experiments of growing tumors and bacterial biofilms under diverse diffusion-consumption and mechanical constraints without the conventionally employed phenomenological prescriptions.\\ 

Several insights are drawn from the theory. Non-dimensionalization of the equations reveals the key role of the ratio between the shear modulus of the growing material and the characteristic chemical potential (i.e. $G/\mu^*$); when this ratio is small, which is typically the case for soft swelling and growing systems, \textcolor{black}{our choice of growth laws here} predict nearly isotropic growth unless large and highly non-spherical dimensionless stresses are reached. This explains why conventional growth studies that often employ the assumption of isotropic growth for these systems  fare well. 
In the same limit of small $G/\mu^*$, it is shown that the growth driving stress is primarily a function of the species conversion energy per unit volume and the swelling ratio (i.e. $\Delta \mu_0$ and $J_s$) and the existence of a critical swelling ratio that stops growth is established. It is shown that the critical swelling ratio  simultaneously captures the effect of two conventionally imposed  phenomenological prescriptions -  (i) a  critical concentration of diffusing species that stops growth and (ii) a homeostatic pressure that the system tends to and stops growing at for uniform pressure loading in absence of diffusion-consumption constraints. Through the latter, the theory offers a kinetic basis for the homeostatic stress in relation to underlying material parameters, and explains  why it is typically observed to be compressive for soft growing systems driven by swelling. Nonetheless, it is shown that for a general boundary value problem, the kinetic growth law does not specify a fixed stress-state that stops growth,  thus emphasizing the potential pitfalls of phenomenological prescriptions, as experimental observations under specific conditions need not translate to other scenarios. Further, the kinetic growth law based on the driving stress is also able to qualitatively explain experimentally observed dependence of growth on concentration of diffusing species, namely the increase in growth rate with concentration, and saturation of growth rate at high concentration values.  A critical conversion energy below which growth is never possible irrespective of the swelling is also established.\\

The ability of the theory to account for  large deformations in both swelling and growth, in contrast to several studies in the literature, opens up a new avenue for investigation  of the  spontaneous emergence of  nonlinear morphological growth phenomena such as  growth induced wrinkling, buckling, fracture, and cavitation, which have been  conventionally studied via inverse methods  using kinematic assumptions.
However, the theory is not without limitations, for example, directional growth mechanisms are not accounted for and need to be introduced to obtain anisotropic homeostatic stress-states. {Also, the model cannot account for specific cellular biological processes involved in growth which are homogenized as an effective species conversion.}
The conversion energy $\Delta \mu_0$, though physically motivated must be determined experimentally; its estimation from homogenization of experimental data for cellular processes is non-trivial, \textcolor{black}{future work could potentially embed biological models in service of such estimation}.  \textcolor{black}{To describe the growth process in greater detail like in several mixture growth theories, such as by accounting for micro-environment, multiplicity of cell species, and inter diffusion of multiple components including nutrients, growth factors, and drugs, the theory needs to be generalized to account for multiple reaction-diffusion equations.}   Finally, the primary physics predicted by the theory has been established in this manuscript by solving boundary value problems with simple geometry.   Future work should consider more complex geometries   to study morphogenesis under mechanical constraints and breakage of symmetry.


\begin{thebibliography}{100}
\expandafter\ifx\csname natexlab\endcsname\relax\def\natexlab#1{#1}\fi
\expandafter\ifx\csname url\endcsname\relax
  \def\url#1{\texttt{#1}}\fi
\expandafter\ifx\csname urlprefix\endcsname\relax\def\urlprefix{URL }\fi

\bibitem[{Abi-Akl et~al.(2019)Abi-Akl, Abeyaratne, and Cohen}]{abi2019kinetics}
Abi-Akl, R., Abeyaratne, R., Cohen, T., 2019. Kinetics of surface growth with
  coupled diffusion and the emergence of a universal growth path. Proceedings
  of the Royal Society A 475~(2221), 20180465.

\bibitem[{Afshar and Di~Leo(2021)}]{afshar2021thermodynamically}
Afshar, A., Di~Leo, C.~V., 2021. A thermodynamically consistent gradient theory
  for diffusion--reaction--deformation in solids: Application to
  conversion-type electrodes. Journal of the Mechanics and Physics of Solids
  151, 104368.

\bibitem[{Alessandri et~al.(2013)Alessandri, Sarangi, Gurchenkov, Sinha,
  Kie{\ss}ling, Fetler, Rico, Scheuring, Lamaze, Simon,
  et~al.}]{alessandri2013cellular}
Alessandri, K., Sarangi, B.~R., Gurchenkov, V.~V., Sinha, B., Kie{\ss}ling,
  T.~R., Fetler, L., Rico, F., Scheuring, S., Lamaze, C., Simon, A., et~al.,
  2013. Cellular capsules as a tool for multicellular spheroid production and
  for investigating the mechanics of tumor progression in vitro. Proceedings of
  the National Academy of Sciences 110~(37), 14843--14848.

\bibitem[{Amar et~al.(2011)Amar, Chatelain, and Ciarletta}]{amar2011contour}
Amar, M.~B., Chatelain, C., Ciarletta, P., 2011. Contour instabilities in early
  tumor growth models. Physical review letters 106~(14), 148101.

\bibitem[{Ambrosi et~al.(2011)Ambrosi, Ateshian, Arruda, Cowin, Dumais,
  Goriely, Holzapfel, Humphrey, Kemkemer, Kuhl,
  et~al.}]{ambrosi2011perspectives}
Ambrosi, D., Ateshian, G.~A., Arruda, E.~M., Cowin, S., Dumais, J., Goriely,
  A., Holzapfel, G.~A., Humphrey, J.~D., Kemkemer, R., Kuhl, E., et~al., 2011.
  Perspectives on biological growth and remodeling. Journal of the Mechanics
  and Physics of Solids 59~(4), 863--883.

\bibitem[{Ambrosi and Guana(2007)}]{ambrosi2007stress}
Ambrosi, D., Guana, F., 2007. Stress-modulated growth. Mathematics and
  mechanics of solids 12~(3), 319--342.

\bibitem[{Ambrosi and Guillou(2007)}]{ambrosi2007growth}
Ambrosi, D., Guillou, A., 2007. Growth and dissipation in biological tissues.
  Continuum Mechanics and Thermodynamics 19~(5), 245--251.

\bibitem[{Ambrosi and Mollica(2002)}]{ambrosi2002mechanics}
Ambrosi, D., Mollica, F., 2002. On the mechanics of a growing tumor.
  International journal of engineering science 40~(12), 1297--1316.

\bibitem[{Ambrosi and Mollica(2004)}]{ambrosi2004role}
Ambrosi, D., Mollica, F., 2004. The role of stress in the growth of a multicell
  spheroid. Journal of mathematical biology 48~(5), 477--499.

\bibitem[{Ambrosi and Preziosi(2002)}]{ambrosi2002closure}
Ambrosi, D., Preziosi, L., 2002. On the closure of mass balance models for
  tumor growth. Mathematical Models and Methods in Applied Sciences 12~(05),
  737--754.

\bibitem[{Ambrosi et~al.(2010)Ambrosi, Preziosi, and
  Vitale}]{ambrosi2010insight}
Ambrosi, D., Preziosi, L., Vitale, G., 2010. The insight of mixtures theory for
  growth and remodeling. Zeitschrift f{\"u}r angewandte Mathematik und Physik
  61~(1), 177--191.

\bibitem[{Anand(2023)}]{anand2023}
Anand, L., 2023. Introduction to Coupled Theories in Solid Mechanics.
  Unpublished MIT 2.077 course notes.

\bibitem[{Araujo and McElwain(2004)}]{araujo2004history}
Araujo, R.~P., McElwain, D.~S., 2004. A history of the study of solid tumour
  growth: the contribution of mathematical modelling. Bulletin of mathematical
  biology 66~(5), 1039--1091.

\bibitem[{Ateshian(2007)}]{ateshian2007theory}
Ateshian, G.~A., 2007. On the theory of reactive mixtures for modeling
  biological growth. Biomechanics and modeling in mechanobiology 6~(6),
  423--445.

\bibitem[{Ateshian et~al.(2012)Ateshian, Morrison~III, Holmes, and
  Hung}]{ateshian2012mechanics}
Ateshian, G.~A., Morrison~III, B., Holmes, J.~W., Hung, C.~T., 2012. Mechanics
  of cell growth. Mechanics research communications 42, 118--125.

\bibitem[{Ateshian et~al.(2014)Ateshian, Nims, Maas, and
  Weiss}]{ateshian2014computational}
Ateshian, G.~A., Nims, R.~J., Maas, S., Weiss, J.~A., 2014. Computational
  modeling of chemical reactions and interstitial growth and remodeling
  involving charged solutes and solid-bound molecules. Biomechanics and
  modeling in mechanobiology 13, 1105--1120.

\bibitem[{Ateshian and Ricken(2010)}]{ateshian2010multigenerational}
Ateshian, G.~A., Ricken, T., 2010. Multigenerational interstitial growth of
  biological tissues. Biomechanics and modeling in mechanobiology 9, 689--702.

\bibitem[{Azeloglu et~al.(2008)Azeloglu, Albro, Thimmappa, Ateshian, and
  Costa}]{azeloglu2008heterogeneous}
Azeloglu, E.~U., Albro, M.~B., Thimmappa, V.~A., Ateshian, G.~A., Costa, K.~D.,
  2008. Heterogeneous transmural proteoglycan distribution provides a mechanism
  for regulating residual stresses in the aorta. American Journal of
  Physiology-Heart and Circulatory Physiology 294~(3), H1197--H1205.

\bibitem[{Baek and Srinivasa(2004)}]{baek2004diffusion}
Baek, S., Srinivasa, A., 2004. Diffusion of a fluid through an elastic solid
  undergoing large deformation. International Journal of non-linear Mechanics
  39~(2), 201--218.

\bibitem[{Bertuzzi et~al.(2010)Bertuzzi, Fasano, Gandolfi, and
  Sinisgalli}]{bertuzzi2010necrotic}
Bertuzzi, A., Fasano, A., Gandolfi, A., Sinisgalli, C., 2010. Necrotic core in
  emt6/ro tumour spheroids: Is it caused by an atp deficit? Journal of
  theoretical biology 262~(1), 142--150.

\bibitem[{Beyhan and Yildiz(2007)}]{beyhan2007smooth}
Beyhan, S., Yildiz, F.~H., 2007. Smooth to rugose phase variation in vibrio
  cholerae can be mediated by a single nucleotide change that targets c-di-gmp
  signalling pathway. Molecular microbiology 63~(4), 995--1007.

\bibitem[{Biot(1941)}]{biot1941general}
Biot, M.~A., 1941. General theory of three-dimensional consolidation. Journal
  of applied physics 12~(2), 155--164.

\bibitem[{Biot and Temple(1972)}]{biot1972theory}
Biot, M.~A., Temple, G., 1972. Theory of finite deformations of porous solids.
  Indiana University Mathematics Journal 21~(7), 597--620.

\bibitem[{Bistri and Di~Leo(2023)}]{bistri2023continuum}
Bistri, D., Di~Leo, C.~V., 2023. A continuum electro-chemo-mechanical gradient
  theory coupled with damage: Application to li-metal filament growth in
  all-solid-state batteries. Journal of the Mechanics and Physics of Solids
  174, 105252.

\bibitem[{Bouklas and Huang(2012)}]{bouklas2012swelling}
Bouklas, N., Huang, R., 2012. Swelling kinetics of polymer gels: comparison of
  linear and nonlinear theories. Soft Matter 8~(31), 8194--8203.

\bibitem[{Boulanger and Hayes(2001)}]{boulanger2001finite}
Boulanger, P., Hayes, M., 2001. Finite-amplitude waves in Mooney-Rivlin and
  Hadamard materials. Springer.

\bibitem[{Bryers(2008)}]{bryers2008medical}
Bryers, J.~D., 2008. Medical biofilms. Biotechnology and bioengineering
  100~(1), 1--18.

\bibitem[{Byrne et~al.(2003)Byrne, King, McElwain, and Preziosi}]{byrne2003two}
Byrne, H.~M., King, J.~R., McElwain, D.~S., Preziosi, L., 2003. A two-phase
  model of solid tumour growth. Applied Mathematics Letters 16~(4), 567--573.

\bibitem[{Carpio et~al.(2019)Carpio, Cebri{\'a}n, and
  Vidal}]{carpio2019biofilms}
Carpio, A., Cebri{\'a}n, E., Vidal, P., 2019. Biofilms as poroelastic
  materials. International Journal of Non-Linear Mechanics 109, 1--8.

\bibitem[{Casciari et~al.(1992)Casciari, Sotirchos, and
  Sutherland}]{casciari1992variations}
Casciari, J.~J., Sotirchos, S.~V., Sutherland, R.~M., 1992. Variations in tumor
  cell growth rates and metabolism with oxygen concentration, glucose
  concentration, and extracellular ph. Journal of cellular physiology 151~(2),
  386--394.

\bibitem[{Chalut and Janmey(2014)}]{chalut2014clamping}
Chalut, K.~J., Janmey, P.~A., 2014. Clamping down on tumor proliferation.
  Biophysical journal 107~(8), 1775.

\bibitem[{Chatelain et~al.(2011)Chatelain, Balois, Ciarletta, and
  Amar}]{chatelain2011emergence}
Chatelain, C., Balois, T., Ciarletta, P., Amar, M.~B., 2011. Emergence of
  microstructural patterns in skin cancer: a phase separation analysis in a
  binary mixture. New Journal of Physics 13~(11), 115013.

\bibitem[{Chen(2018)}]{chen2018thin}
Chen, Z., 2018. From a thin membrane to an unbounded solid: dynamics and
  instabilities in radial motion of nonlinearly viscoelastic spheres. Ph.D.
  thesis, Massachusetts Institute of Technology.

\bibitem[{Chester and Anand(2010)}]{chester2010coupled}
Chester, S.~A., Anand, L., 2010. A coupled theory of fluid permeation and large
  deformations for elastomeric materials. Journal of the Mechanics and Physics
  of Solids 58~(11), 1879--1906.

\bibitem[{Chester and Anand(2011)}]{chester2011thermo}
Chester, S.~A., Anand, L., 2011. A thermo-mechanically coupled theory for fluid
  permeation in elastomeric materials: application to thermally responsive
  gels. Journal of the Mechanics and Physics of Solids 59~(10), 1978--2006.

\bibitem[{Ciarletta et~al.(2013)Ciarletta, Ambrosi, Maugin, and
  Preziosi}]{ciarletta2013mechano}
Ciarletta, P., Ambrosi, D., Maugin, G., Preziosi, L., 2013.
  Mechano-transduction in tumour growth modelling. The European Physical
  Journal E 36~(3), 1--9.

\bibitem[{Ciarletta et~al.(2011)Ciarletta, Foret, and
  Ben~Amar}]{ciarletta2011radial}
Ciarletta, P., Foret, L., Ben~Amar, M., 2011. The radial growth phase of
  malignant melanoma: multi-phase modelling, numerical simulations and linear
  stability analysis. Journal of the Royal Society Interface 8~(56), 345--368.

\bibitem[{Croix et~al.(1996)Croix, Rak, Kapitain, Sheehan, Graham, and
  Kerbel}]{croix1996reversal}
Croix, B.~S., Rak, J.~W., Kapitain, S., Sheehan, C., Graham, C.~H., Kerbel,
  R.~S., 1996. Reversal by hyaluronidase of adhesion-dependent multicellular
  drug resistance in mammary carcinoma cells. JNCI: Journal of the National
  Cancer Institute 88~(18), 1285--1296.

\bibitem[{Curatolo et~al.(2017)Curatolo, Gabriele, and
  Teresi}]{curatolo2017swelling}
Curatolo, M., Gabriele, S., Teresi, L., 2017. Swelling and growth: a
  constitutive framework for active solids. Meccanica 52~(14), 3443--3456.

\bibitem[{Dervaux and Amar(2011)}]{dervaux2011buckling}
Dervaux, J., Amar, M.~B., 2011. Buckling condensation in constrained growth.
  Journal of the Mechanics and Physics of Solids 59~(3), 538--560.

\bibitem[{DiCarlo and Quiligotti(2002)}]{dicarlo2002growth}
DiCarlo, A., Quiligotti, S., 2002. Growth and balance. Mechanics Research
  Communications 29~(6), 449--456.

\bibitem[{Doi(2009)}]{doi2009gel}
Doi, M., 2009. Gel dynamics. Journal of the Physical Society of Japan 78~(5),
  052001.

\bibitem[{Duda et~al.(2010)Duda, Souza, and Fried}]{duda2010theory}
Duda, F.~P., Souza, A.~C., Fried, E., 2010. A theory for species migration in a
  finitely strained solid with application to polymer network swelling. Journal
  of the Mechanics and Physics of Solids 58~(4), 515--529.

\bibitem[{Faghihi et~al.(2020)Faghihi, Feng, Lima, Oden, and
  Yankeelov}]{faghihi2020coupled}
Faghihi, D., Feng, X., Lima, E.~A., Oden, J.~T., Yankeelov, T.~E., 2020. A
  coupled mass transport and deformation theory of multi-constituent tumor
  growth. Journal of the Mechanics and Physics of Solids 139, 103936.

\bibitem[{Flory(1942)}]{flory1942thermodynamics}
Flory, P.~J., 1942. Thermodynamics of high polymer solutions. The Journal of
  chemical physics 10~(1), 51--61.

\bibitem[{Fraldi and Carotenuto(2018)}]{fraldi2018cells}
Fraldi, M., Carotenuto, A.~R., 2018. Cells competition in tumor growth
  poroelasticity. Journal of the Mechanics and Physics of Solids 112, 345--367.

\bibitem[{Fung(2013)}]{fung2013biomechanics}
Fung, Y.-c., 2013. Biomechanics: motion, flow, stress, and growth. Springer
  Science \& Business Media.

\bibitem[{Garikipati et~al.(2004)Garikipati, Arruda, Grosh, Narayanan, and
  Calve}]{garikipati2004continuum}
Garikipati, K., Arruda, E.~M., Grosh, K., Narayanan, H., Calve, S., 2004. A
  continuum treatment of growth in biological tissue: the coupling of mass
  transport and mechanics. Journal of the Mechanics and Physics of Solids
  52~(7), 1595--1625.

\bibitem[{Garteiser et~al.(2012)Garteiser, Doblas, Daire, Wagner, Leitao,
  Vilgrain, Sinkus, and Van~Beers}]{garteiser2012mr}
Garteiser, P., Doblas, S., Daire, J.-L., Wagner, M., Leitao, H., Vilgrain, V.,
  Sinkus, R., Van~Beers, B.~E., 2012. Mr elastography of liver tumours: value
  of viscoelastic properties for tumour characterisation. European radiology
  22, 2169--2177.

\bibitem[{Goriely et~al.(2010)Goriely, Moulton, and
  Vandiver}]{goriely2010elastic}
Goriely, A., Moulton, D.~E., Vandiver, R., 2010. Elastic cavitation, tube
  hollowing, and differential growth in plants and biological tissues.
  Europhysics Letters 91~(1), 18001.

\bibitem[{Greenspan(1972)}]{greenspan1972models}
Greenspan, H., 1972. Models for the growth of a solid tumor by diffusion.
  Studies in Applied Mathematics 51~(4), 317--340.

\bibitem[{Helmlinger et~al.(1997)Helmlinger, Netti, Lichtenbeld, Melder, and
  Jain}]{helmlinger1997solid}
Helmlinger, G., Netti, P.~A., Lichtenbeld, H.~C., Melder, R.~J., Jain, R.~K.,
  1997. Solid stress inhibits the growth of multicellular tumor spheroids.
  Nature biotechnology 15~(8), 778--783.

\bibitem[{Hlatky et~al.(1988)Hlatky, Sachs, and Alpen}]{hlatky1988joint}
Hlatky, L., Sachs, R.~K., Alpen, E.~L., 1988. Joint oxygen-glucose deprivation
  as the cause of necrosis in a tumor analog. Journal of cellular physiology
  134~(2), 167--178.

\bibitem[{Hong et~al.(2008)Hong, Zhao, Zhou, and Suo}]{hong2008theory}
Hong, W., Zhao, X., Zhou, J., Suo, Z., 2008. A theory of coupled diffusion and
  large deformation in polymeric gels. Journal of the Mechanics and Physics of
  Solids 56~(5), 1779--1793.

\bibitem[{Huggins(1941)}]{huggins1941solutions}
Huggins, M.~L., 1941. Solutions of long chain compounds. The Journal of
  chemical physics 9~(5), 440--440.

\bibitem[{Humphrey and Rajagopal(2002)}]{humphrey2002constrained}
Humphrey, J., Rajagopal, K., 2002. A constrained mixture model for growth and
  remodeling of soft tissues. Mathematical models and methods in applied
  sciences 12~(03), 407--430.

\bibitem[{Jain(2005)}]{jain2005normalization}
Jain, R.~K., 2005. Normalization of tumor vasculature: an emerging concept in
  antiangiogenic therapy. Science 307~(5706), 58--62.

\bibitem[{Jain et~al.(2014)Jain, Martin, and Stylianopoulos}]{jain2014role}
Jain, R.~K., Martin, J.~D., Stylianopoulos, T., 2014. The role of mechanical
  forces in tumor growth and therapy. Annual review of biomedical engineering
  16, 321--346.

\bibitem[{Kerbel et~al.(1994)Kerbel, Rak, Kobayashi, Man, Croix, and
  Graham}]{kerbel1994multicellular}
Kerbel, R., Rak, J., Kobayashi, H., Man, M., Croix, B.~S., Graham, C., 1994.
  Multicellular resistance: a new paradigm to explain aspects of acquired drug
  resistance of solid tumors. In: Cold Spring Harbor symposia on quantitative
  biology. Vol.~59. Cold Spring Harbor Laboratory Press, pp. 661--672.

\bibitem[{Kim et~al.(2011)Kim, Stolarska, and Othmer}]{kim2011role}
Kim, Y., Stolarska, M.~A., Othmer, H.~G., 2011. The role of the
  microenvironment in tumor growth and invasion. Progress in biophysics and
  molecular biology 106~(2), 353--379.

\bibitem[{Kobayashi et~al.(1993)Kobayashi, Man, Graham, Kapitain, Teicher, and
  Kerbel}]{kobayashi1993acquired}
Kobayashi, H., Man, S., Graham, C.~H., Kapitain, S.~J., Teicher, B.~A., Kerbel,
  R.~S., 1993. Acquired multicellular-mediated resistance to alkylating agents
  in cancer. Proceedings of the National Academy of Sciences 90~(8),
  3294--3298.

\bibitem[{Koike et~al.(2002)Koike, McKee, Pluen, Ramanujan, Burton, Munn,
  Boucher, and Jain}]{koike2002solid}
Koike, C., McKee, T., Pluen, A., Ramanujan, S., Burton, K., Munn, L., Boucher,
  Y., Jain, R., 2002. Solid stress facilitates spheroid formation: potential
  involvement of hyaluronan. British journal of cancer 86~(6), 947--953.

\bibitem[{Konica and Sain(2020)}]{konica2020thermodynamically}
Konica, S., Sain, T., 2020. A thermodynamically consistent chemo-mechanically
  coupled large deformation model for polymer oxidation. Journal of the
  Mechanics and Physics of Solids 137, 103858.

\bibitem[{K{\"o}pf and Pismen(2013)}]{kopf2013continuum}
K{\"o}pf, M.~H., Pismen, L.~M., 2013. A continuum model of epithelial
  spreading. Soft matter 9~(14), 3727--3734.

\bibitem[{Kuhl(2014)}]{kuhl2014growing}
Kuhl, E., 2014. Growing matter: a review of growth in living systems. Journal
  of the Mechanical Behavior of Biomedical Materials 29, 529--543.

\bibitem[{Levitas and Attariani(2014)}]{levitas2014anisotropic}
Levitas, V.~I., Attariani, H., 2014. Anisotropic compositional expansion in
  elastoplastic materials and corresponding chemical potential: Large-strain
  formulation and application to amorphous lithiated silicon. Journal of the
  Mechanics and Physics of Solids 69, 84--111.

\bibitem[{Li et~al.(2022)Li, Kothari, Chockalingam, Henzel, Zhang, Li, Yan, and
  Cohen}]{li2022nonlinear}
Li, J., Kothari, M., Chockalingam, S., Henzel, T., Zhang, Q., Li, X., Yan, J.,
  Cohen, T., 2022. Nonlinear inclusion theory with application to the growth
  and morphogenesis of a confined body. Journal of the Mechanics and Physics of
  Solids 159, 104709.

\bibitem[{Liu et~al.(2015)Liu, Toh, and Ng}]{liu2015advances}
Liu, Z., Toh, W., Ng, T.~Y., 2015. Advances in mechanics of soft materials: A
  review of large deformation behavior of hydrogels. International Journal of
  Applied Mechanics 7~(05), 1530001.

\bibitem[{Loeffel and Anand(2011)}]{loeffel2011chemo}
Loeffel, K., Anand, L., 2011. A chemo-thermo-mechanically coupled theory for
  elastic--viscoplastic deformation, diffusion, and volumetric swelling due to
  a chemical reaction. International Journal of Plasticity 27~(9), 1409--1431.

\bibitem[{Lubarda and Hoger(2002)}]{lubarda2002mechanics}
Lubarda, V.~A., Hoger, A., 2002. On the mechanics of solids with a growing
  mass. International journal of solids and structures 39~(18), 4627--4664.

\bibitem[{Lucantonio et~al.(2013)Lucantonio, Nardinocchi, and
  Teresi}]{lucantonio2013transient}
Lucantonio, A., Nardinocchi, P., Teresi, L., 2013. Transient analysis of
  swelling-induced large deformations in polymer gels. Journal of the Mechanics
  and Physics of Solids 61~(1), 205--218.

\bibitem[{Mattei et~al.(2018)Mattei, Frunzo, D’acunto, Pechaud, Pirozzi, and
  Esposito}]{mattei2018continuum}
Mattei, M., Frunzo, L., D’acunto, B., Pechaud, Y., Pirozzi, F., Esposito, G.,
  2018. Continuum and discrete approach in modeling biofilm development and
  structure: a review. Journal of mathematical biology 76~(4), 945--1003.

\bibitem[{McMahon et~al.(2010)McMahon, Goriely, and
  Tabor}]{mcmahon2010spontaneous}
McMahon, J., Goriely, A., Tabor, M., 2010. Spontaneous cavitation in growing
  elastic membranes. Mathematics and mechanics of solids 15~(1), 57--77.

\bibitem[{Menzel and Kuhl(2012)}]{menzel2012frontiers}
Menzel, A., Kuhl, E., 2012. Frontiers in growth and remodeling. Mechanics
  research communications 42, 1--14.

\bibitem[{Monod(1949)}]{monod1949growth}
Monod, J., 1949. The growth of bacterial cultures. Annual review of
  microbiology 3~(1), 371--394.

\bibitem[{Mooney(1940)}]{mooney1940theory}
Mooney, M., 1940. A theory of large elastic deformation. Journal of applied
  physics 11~(9), 582--592.

\bibitem[{Mpekris et~al.(2015)Mpekris, Angeli, Pirentis, and
  Stylianopoulos}]{mpekris2015stress}
Mpekris, F., Angeli, S., Pirentis, A.~P., Stylianopoulos, T., 2015.
  Stress-mediated progression of solid tumors: effect of mechanical stress on
  tissue oxygenation, cancer cell proliferation, and drug delivery.
  Biomechanics and modeling in mechanobiology 14~(6), 1391--1402.

\bibitem[{Myers and Ateshian(2014)}]{myers2014interstitial}
Myers, K., Ateshian, G.~A., 2014. Interstitial growth and remodeling of
  biological tissues: tissue composition as state variables. Journal of the
  mechanical behavior of biomedical materials 29, 544--556.

\bibitem[{Nadell et~al.(2015)Nadell, Drescher, Wingreen, and
  Bassler}]{nadell2015extracellular}
Nadell, C.~D., Drescher, K., Wingreen, N.~S., Bassler, B.~L., 2015.
  Extracellular matrix structure governs invasion resistance in bacterial
  biofilms. The ISME journal 9~(8), 1700--1709.

\bibitem[{Narayanan et~al.(2009)Narayanan, Arruda, Grosh, and
  Garikipati}]{narayanan2009micromechanics}
Narayanan, H., Arruda, E., Grosh, K., Garikipati, K., 2009. The micromechanics
  of fluid--solid interactions during growth in porous soft biological tissue.
  Biomechanics and modeling in mechanobiology 8, 167--181.

\bibitem[{Narayanan et~al.(2010)Narayanan, Verner, Mills, Kemkemer, and
  Garikipati}]{narayanan2010silico}
Narayanan, H., Verner, S., Mills, K., Kemkemer, R., Garikipati, K., 2010. In
  silico estimates of the free energy rates in growing tumor spheroids. Journal
  of Physics: Condensed Matter 22~(19), 194122.

\bibitem[{Oden et~al.(2016)Oden, Lima, Almeida, Feng, Rylander, Fuentes,
  Faghihi, Rahman, DeWitt, Gadde, et~al.}]{oden2016toward}
Oden, J.~T., Lima, E.~A., Almeida, R.~C., Feng, Y., Rylander, M.~N., Fuentes,
  D., Faghihi, D., Rahman, M.~M., DeWitt, M., Gadde, M., et~al., 2016. Toward
  predictive multiscale modeling of vascular tumor growth. Archives of
  Computational Methods in Engineering 23~(4), 735--779.

\bibitem[{Olive and Durand(1994)}]{olive1994drug}
Olive, P.~L., Durand, R.~E., 1994. Drug and radiation resistance in spheroids:
  cell contact and kinetics. Cancer and Metastasis Reviews 13~(2), 121--138.

\bibitem[{Preziosi(2003)}]{preziosi2003cancer}
Preziosi, L., 2003. Cancer modelling and simulation. CRC Press.

\bibitem[{Rivlin(1948)}]{rivlin1948large}
Rivlin, R.~S., 1948. Large elastic deformations of isotropic materials iv.
  further developments of the general theory. Philosophical transactions of the
  royal society of London. Series A, Mathematical and physical sciences
  241~(835), 379--397.

\bibitem[{Rodriguez et~al.(1994)Rodriguez, Hoger, and
  McCulloch}]{rodriguez1994stress}
Rodriguez, E.~K., Hoger, A., McCulloch, A.~D., 1994. Stress-dependent finite
  growth in soft elastic tissues. Journal of biomechanics 27~(4), 455--467.

\bibitem[{Roose et~al.(2003)Roose, Netti, Munn, Boucher, and
  Jain}]{roose2003solid}
Roose, T., Netti, P.~A., Munn, L.~L., Boucher, Y., Jain, R.~K., 2003. Solid
  stress generated by spheroid growth estimated using a linear poroelasticity
  model. Microvascular research 66~(3), 204--212.

\bibitem[{Sacco et~al.(2017)Sacco, Causin, Lelli, and
  Raimondi}]{sacco2017poroelastic}
Sacco, R., Causin, P., Lelli, C., Raimondi, M.~T., 2017. A poroelastic mixture
  model of mechanobiological processes in biomass growth: theory and
  application to tissue engineering. Meccanica 52~(14), 3273--3297.

\bibitem[{Salvadori et~al.(2018)Salvadori, McMeeking, Grazioli, and
  Magri}]{salvadori2018coupled}
Salvadori, A., McMeeking, R., Grazioli, D., Magri, M., 2018. A coupled model of
  transport-reaction-mechanics with trapping. part i--small strain analysis.
  Journal of the Mechanics and Physics of Solids 114, 1--30.

\bibitem[{Sarntinoranont et~al.(2003)Sarntinoranont, Rooney, and
  Ferrari}]{sarntinoranont2003interstitial}
Sarntinoranont, M., Rooney, F., Ferrari, M., 2003. Interstitial stress and
  fluid pressure within a growing tumor. Annals of biomedical engineering 31,
  327--335.

\bibitem[{Seminara et~al.(2012)Seminara, Angelini, Wilking, Vlamakis, Ebrahim,
  Kolter, Weitz, and Brenner}]{seminara2012osmotic}
Seminara, A., Angelini, T.~E., Wilking, J.~N., Vlamakis, H., Ebrahim, S.,
  Kolter, R., Weitz, D.~A., Brenner, M.~P., 2012. Osmotic spreading of bacillus
  subtilis biofilms driven by an extracellular matrix. Proceedings of the
  National Academy of Sciences 109~(4), 1116--1121.

\bibitem[{Stylianopoulos et~al.(2012)Stylianopoulos, Martin, Chauhan, Jain,
  Diop-Frimpong, Bardeesy, Smith, Ferrone, Hornicek, Boucher,
  et~al.}]{stylianopoulos2012causes}
Stylianopoulos, T., Martin, J.~D., Chauhan, V.~P., Jain, S.~R., Diop-Frimpong,
  B., Bardeesy, N., Smith, B.~L., Ferrone, C.~R., Hornicek, F.~J., Boucher, Y.,
  et~al., 2012. Causes, consequences, and remedies for growth-induced solid
  stress in murine and human tumors. Proceedings of the National Academy of
  Sciences 109~(38), 15101--15108.

\bibitem[{Taber(1998)}]{taber1998model}
Taber, L.~A., 1998. {A Model for Aortic Growth Based on Fluid Shear and Fiber
  Stresses}. Journal of Biomechanical Engineering 120~(3), 348--354.

\bibitem[{Thorne et~al.(2004)Thorne, Hrabetov{\'a}, and
  Nicholson}]{thorne2004diffusion}
Thorne, R.~G., Hrabetov{\'a}, S., Nicholson, C., 2004. Diffusion of epidermal
  growth factor in rat brain extracellular space measured by integrative
  optical imaging. Journal of neurophysiology 92~(6), 3471--3481.

\bibitem[{Treloar(1975)}]{treloar1975physics}
Treloar, L.~G., 1975. The physics of rubber elasticity. OUP Oxford.

\bibitem[{Ward and King(1997)}]{ward1997mathematical}
Ward, J.~P., King, J.~R., 1997. Mathematical modelling of avascular-tumour
  growth. Mathematical Medicine and Biology: A Journal of the IMA 14~(1),
  39--69.

\bibitem[{Xue et~al.(2016)Xue, Li, Feng, and Gao}]{xue2016biochemomechanical}
Xue, S.-L., Li, B., Feng, X.-Q., Gao, H., 2016. Biochemomechanical poroelastic
  theory of avascular tumor growth. Journal of the Mechanics and Physics of
  Solids 94, 409--432.

\bibitem[{Xue et~al.(2018)Xue, Yin, Li, and Feng}]{xue2018biochemomechanical}
Xue, S.-L., Yin, S.-F., Li, B., Feng, X.-Q., 2018. Biochemomechanical modeling
  of vascular collapse in growing tumors. Journal of the Mechanics and Physics
  of Solids 121, 463--479.

\bibitem[{Yan et~al.(2017)Yan, Nadell, Stone, Wingreen, and
  Bassler}]{yan2017extracellular}
Yan, J., Nadell, C.~D., Stone, H.~A., Wingreen, N.~S., Bassler, B.~L., 2017.
  Extracellular-matrix-mediated osmotic pressure drives vibrio cholerae biofilm
  expansion and cheater exclusion. Nature communications 8~(1), 1--11.

\bibitem[{Zhang et~al.(2021)Zhang, Li, Nijjer, Lu, Kothari, Alert, Cohen, and
  Yan}]{zhang2021morphogenesis}
Zhang, Q., Li, J., Nijjer, J., Lu, H., Kothari, M., Alert, R., Cohen, T., Yan,
  J., 2021. Morphogenesis and cell ordering in confined bacterial biofilms.
  Proceedings of the National Academy of Sciences 118~(31), e2107107118.

\end{thebibliography}

\section*{Acknowledgements}         
The authors acknowledge our collaborators at Yale University, and in particular Professor Jing Yan, whose experiments motivated this study.
 We would also like to acknowledge Professor Lallit Anand for drawing attention to some useful references. Additionally, we thank Joseph Bonavia and Hudson Borja da Rocha for feedback on the manuscript. {Finally we thank the reviewers who were helpful in bringing greater clarity to the mass balance and entropy inequality.}
\appendix 

\section{Theory }
\label{app:theory}

\subsection{Mass balance during growth}
\label{app:massbalance}
{Consider the current mass density of the solid matrix, $\rho$, whose evolution is given by the equation 
\begin{equation}
\pdv{\rho}{t} + \text{div}\left(\rho \nten{v}\right) = \dot{\rho} + \rho\ \text{div}\ {\nten{v}} = \tilde{\rho}
\end{equation}
where $\tilde{\rho}$ is the mass density supply rate to the solid matrix from the diffusing species and $v = \dot{\nten{x}}$ is the velocity of the material point. Based on standard 
kinematics of a continuum, it is known that $\dot{J} = J\ \text{div}\ \nten{v}$, whose substitution into the mass balance equation above produces
\begin{equation}
\dot{\rho}_R =\tilde{\rho}_R, \label{eq:app_rhodot}
\end{equation}
where $\rho_R = \rho J$ is the current solid mass per unit dry reference volume and $\tilde{\rho}_R = \tilde{\rho} J$ is the referential mass density supply rate to the solid matrix (mass supplied per unit dry reference volume per unit time). Thus $\rho_R$ only changes during growth ($\tilde{\rho}_R\neq0$). 
Since we assumed that the density of the solid matrix remains constant (at value $\rho_0^m$) during the growth process through the mapping by $\nten{F}^g$, we can write $\rho_R = \rho_0^m J^g$, the substitution of which in \cref{eq:app_rhodot} yields
\begin{equation}
\rho_0^m \dot{J}^g = \tilde{\rho}_R \label{eq:app_growthconstraint}
\end{equation}
Recognizing $\rho_0^m \equiv 1/\Omega^m$ and $\tilde{\rho}_R \equiv \dot{\xi}_R$, we see that \cref{eq:app_growthconstraint} is equivalent to \cref{eq:constraintgrowth}.
}

\subsection{Dissipation inequality}
\label{app:disspineq}
Under isothermal conditions, the entropy inequality can be written as
\begin{equation}
{{\int_{\text{P}} \dot{\eta}_R\dd{V_R}}} \ge \int_{\partial \text{P}} \mu_{\eta} (-\nten{j}_R \cdot \nten{n}_R) \dd{A_R} \label{eq:dummyyy2}
\end{equation}
where $\eta_R$ is the entropy per unit dry reference volume {of the fluid-solid continuum within the dry material region P} and $\mu_{\eta}$ is the entropy supply per unit reference volume of the diffusing species, {into P through the boundary}. 
Neglecting kinetic energy (or inertial effects) and assuming no body forces, the balance law for energy under isothermal conditions can be written as 
\begin{equation}
 {{\int_{\text{P}} \dot{\epsilon}_R\dd{V_R}}} = \int_{\partial\text{P}} (\nten{S} \nten{n}_R)\cdot  \dot{\bm{\varphi}} \dd{A_R}  + \int_{\partial \text{P}} \mu_{\epsilon} (-\nten{j}_R \cdot \nten{n}_R) \dd{A_R}  \label{eq:dummyyy4}
\end{equation}
where $\epsilon_R$ is the internal energy per unit dry reference volume {of the fluid-solid continuum within the dry material region P} and $\mu_{\epsilon}$ is the internal energy supply per unit dry reference volume of the diffusing species, {into P through the boundary}. 
Introducing the Helmholtz free energy per unit dry reference volume {of the fluid-solid continuum}, $\psi_R = \epsilon_R - \theta \eta_R$, where $\theta$ is the constant absolute temperature, we can write the following rate relation under isothermal conditions
\begin{equation}
\dot{\psi}_R = \dot{\epsilon}_R - \theta \dot{\eta}_R \label{eq:dummyyy5}
\end{equation} 
Defining the chemical potential $\mu$ as follows (see also \cite{salvadori2018coupled,anand2023}),
\begin{equation}
\mu = \mu_{\epsilon} - \theta \mu_{\eta}
\end{equation}
 eqs. \eqref{eq:dummyyy2}, \eqref{eq:dummyyy4}, and \eqref{eq:dummyyy5} can be combined to write the dissipation inequality in \cref{eq:dissipineq_combined}.
Next, we show the localization of \cref{eq:dissipineq_combined} into \cref{eq:dissip1}. We rewrite the first term on the right hand side of \cref{eq:dissipineq_combined} as follows
\begin{equation}
\int_{\partial\text{P}} (\nten{S} \nten{n}_R)\cdot  \dot{\bm{\varphi}} \dd{A_R} = \int_{\partial\text{P}}\left(\nten{S}\cdot \pdv{\dot{\bm{\varphi}}}{\nten{X}} + \left(\text{Div}\ \nten{S}\right)\cdot\dot{\bm{\varphi}}\right)\dd{V_R} = \int_{\partial\text{P}}\nten{S}\cdot \dot{\nten{F}} \dd{V_R} \label{eq:app_dissip1}
\end{equation}
where  we have employed the divergence theorem, the mechanical equilibrium equation \eqref{eq:Piolastresseqb}, and the definition of $\nten{F}$ in \cref{eq:Fdecomp}. Next, consider the second term on the right hand side of \cref{eq:dissipineq_combined},
\begin{equation}
\int_{\partial \text{P}} \mu (-\nten{j}_R \cdot \nten{n}_R) \dd{A_R}  = \int_{\partial \text{P}} \left(-\nten{j}_R \cdot \nabla \mu- \mu\ \text{Div}\ \nten{j}_R \right) \dd{V_R} = \int_{\partial \text{P}} \left(-\nten{j}_R \cdot \nabla \mu + \mu \left(J^g \dot{J}^s +  J^s \dot{J}^g\right)\right) \dd{V_R} \label{eq:app_dissip2}
\end{equation}  
where we have used the divergence theorem and the isochoric diffusion-consumption equation \eqref{eq:massbalanceJsJg_isocho}. Plugging \cref{eq:app_dissip1,eq:app_dissip2} into \cref{eq:dissipineq_combined} and localizing the integral gives the required inequality in \cref{eq:dissip1} {(see also \cite{loeffel2011chemo,levitas2014anisotropic,konica2020thermodynamically,afshar2021thermodynamically,bistri2023continuum})}.

\subsection{Stress power}
\label{app:stresspower}
We prove the stress power equivalence stated in \cref{eq:strespowers} below. First, using \cref{eq:Ls,eq:RateFdecomp,eq:Piolastresseqb}, we write
\begin{equation}
\nten{S} \cdot \dot{\nten{F}} = \left(\nten{S}\nten{F}^T\right)\cdot \nten{L} = J\nten{T}\cdot{\nten{L}} =  J\nten{T}\cdot\left(\nten{L}^e + \nten{F}^e \nten{L}^g {\nten{F}^e}^{-1} + \frac{\dot{J}^s}{3 J^s} \nten{I}\right) \label{eq:app_sp}
\end{equation}
We now consider the first term in \cref{eq:app_sp},
\begin{align}
J\nten{T}\cdot\nten{L}^e &= J\nten{T}\cdot\left(\dot{\nten{F}}^e {\nten{F}^e}^{-1}\right) = \left(J\nten{T}{\nten{F}^e}^{-T}\right)\cdot\dot{\nten{F}}^e  = J^sJ^g\left(J^e\nten{T}{\nten{F}^e}^{-T}\right)\cdot\left({\nten{F}^e}^{-T} {\nten{F}^e}^T\dot{\nten{F}}^e\right)\label{eq:app_sp01}\\
&= J^sJ^g\left(J^e{\nten{F}^e}^{-1}\nten{T}{\nten{F}^e}^{-T}\right)\cdot\left( {\nten{F}^e}^T\dot{\nten{F}}^e\right) = J^sJ^g \nten{T}^e\cdot\text{sym}\left( {\nten{F}^e}^T\dot{\nten{F}}^e\right) =\frac{1}{2} J^sJ^g \nten{T}^e \cdot \dot{\nten{C}}^e \label{eq:app_sp1}
\end{align}
where we have used \cref{eq:Ls,eq:determinants_combined2}, the definition of $\nten{T}^e$ in \cref{eq:stressdefs}$_2$ and its symmetry\footnote{$\nten{Z}$ is symmetric if $\nten{Z} = \text{sym}(\nten{Z})$ where $\text{sym}(\nten{Z}) = \frac{1}{2}(\nten{Z} + \nten{Z}^T).$ If $\nten{Z}$ is symmetric, $\nten{Z}\cdot\nten{Y} = \nten{Z}\cdot\text{sym}(\nten{Y})$.}.   Next, we consider the second term in \cref{eq:app_sp},
\begin{equation}
J\nten{T}\cdot\left(\nten{F}^e \nten{L}^g{\nten{F}^e}^{-1} \right) = \left(J{\nten{F}^e}^T \nten{T} {\nten{F}^e}^{-T}\right)\cdot\nten{L}^g = J^g\left(J^{es}{\nten{F}^{es}}^T \nten{T} {\nten{F}^{es}}^{-T}\right)\cdot\nten{L}^g = J^g\nten{M}^{es}\cdot\nten{L}^g
\label{eq:app_sp2}\end{equation}
where we have used \cref{eq:Fdecomp,eq:determinants_combined2}, and the definition of $\nten{M}^{es}$ in \cref{eq:stressdefs}$_1$.  Finally, we consider the third term in \cref{eq:app_sp},
\begin{equation}
J\nten{T}\cdot\left(\frac{\dot{J}^s}{3 J^s} \nten{I}\right) = \frac{J\dot{J}^s}{3J^s}\text{tr}(\nten{T}) \label{eq:app_sp3}
\end{equation}
Using \cref{eq:app_sp01,eq:app_sp1,eq:app_sp2,eq:app_sp3} in \cref{eq:app_sp} gives us the desired relation in \cref{eq:strespowers}.

\subsection{Non-isochoric conversion reaction}
\label{app:nonisochoric}
In this section, we generalize the theory to non-isochoric species conversion ($\Omega^f \neq \Omega^m$). Following the same derivation procedure in the manuscript while using the non-isochoric diffusion-consumption equation \eqref{eq:massbalanceJsJg} instead of the isochoric version in \cref{eq:massbalanceJsJg_isocho} gives us the desired theory. Skipping the straightforward math, we present here directly the key equations that will differ for the non-isochoric version of the theory from the equations for the isochoric case in the manuscript. The only extra parameter that shows up in the theory is the ratio $\alpha = \Omega^f/\Omega^m = \rho_0^m/\rho_0^f$ which is the referential density ratio of the solid to diffusing species. The growth driving stress $\nten{T}^g$ in \cref{eq:Tg} is now instead
\begin{equation}
\nten{T}^g = \nten{M}^{es} + \left(  {\Delta \mu_0^\text{nis}} + \left(J^s+\alpha-1\right) \left(\pdv{{\psi}_{g}}{J^s} -  \frac{1}{3} J^e \text{tr}(\nten{T})\right)-  \psi_g  \right)\nten{I}   \label{eq:app_Tg2}
\end{equation}
where $\Delta \mu_0^\text{nis} = {\Delta \mu_0} + \mu_0^f\left(\alpha-1\right)$ is the chemical conversion energy per unit referential solid volume for conversion of unit mass of diffusing species to solid. 
The driving stress for volumetric growth in \cref{eq:fg_first}$_2$ is now instead
\begin{align}
f_g &= \left(\alpha-1\right)\left(\pdv{{\psi}_{g}}{J^s} -  \frac{1}{3} J^e \text{tr}(\nten{T})\right) + {\Delta \mu_0^\text{nis}} + J^s\pdv{\psi_g}{J^s} - \psi_g \\
&=\left(\alpha-1\right)(\mu-\mu_0^f) + {\Delta \mu_0^\text{nis}} + J^s\pdv{\psi_g}{J^s} - \psi_g \label{eq:app_Gamma_evln2} 
\end{align} 
where the chemical potential $\mu$ from \cref{eq:Constit_energetic}$_2$ remains unchanged along with the growth evolution equations  \eqref{eq:volgrowthlaw},\eqref{eq:spec_gr_direc}. The specialized form of $f_g$ for the chosen constitutive functions in \Cref{sec:theory} is given by
\begin{equation}
f_g = \Delta \mu_0^{\text{nis}} + \left(\alpha-1\right)\left(\mu^{\text{mix}}(\phi) - {K \ln(J^e)}{\phi}\right) + f_g^{\text{mix}}(\phi) + f_g^{\text{mech}}
\end{equation}
where expressions for $\mu^{\text{mix}}(\phi)$, $f_g^{\text{mix}}$, and $f_g^{\text{mech}}$ can be found in \cref{eq:dpsigmixdJs,eq:fgmechspec,eq:fgmixspec}.  \\

For the boundary value problem in \Cref{subsec:uniformswell} where we consider the limit of infinitely fast diffusion, we have $\mu = \mu_0^f$ and thus $f_g$ in \cref{eq:app_Gamma_evln2} reduces to
\begin{equation}
\quad f_g = \Delta \mu_0^\text{mis} + J^s\pdv{\psi_g}{J^s} - \psi_g
\end{equation}
Thus it can be seen that the driving stress $f_g$ in this case has essentially the same functional form as for the isochoric case in \cref{eq:fg_first}$_2$ except that $\Delta \mu_0$ has been replaced by $\Delta \mu_0^\text{mis}$. Thus all the conclusions devised for the isochoric case in \Cref{subsec:uniformswell} also hold for the non-isochoric case. 

\subsection{$\bar{\Gamma}$ dependence on $\eta$}
\label{app:gammbar_eta}

Now we demonstrate that in the limit of small $G/\mu^*$, a larger value of the conversion energy ratio $\eta$, when all other parameters are fixed, leads to a smaller dimensionless growth rate for a given $\phi>\phi^f$. In the limit of small $G/\mu^*$, the dimensionless growth rate can be written using \cref{eq:fgbar_zero,eq:gammabar_dimlss}, for the choice of $\bar{f}_g^* = \bar{f}_g(\phi^f)$, as
\begin{equation}
    \bar{\Gamma} = \frac{\Delta \bar{\mu}_0 + \bar{f}_g^{\text{mix}}(\phi)}{\Delta \bar{\mu}_0 + \bar{f}_g^{\text{mix}}(\phi^f)} = 1 + \frac{\bar{f}_g^{\text{mix}}(\phi) - \bar{f}_g^{\text{mix}}(\phi^f)}{\Delta \bar{\mu}_0 + \bar{f}_g^{\text{mix}}(\phi^f)} \label{eq:app_ratedep}
\end{equation}
We first note from \Cref{subsec:growthdrivingstress,subsec:uniformswell} that $\phi^f$ and $\Delta \bar{\mu}_0^c$ are independent of $\eta$ when $\chi$ is fixed. Thus when all other parameters are fixed, a larger value of $\eta$ corresponds to a smaller value of $\Delta \mu_0$, and for any given $\phi$ for which $\bar{f}_g^{\text{mix}}(\phi)<\bar{f}_g^{\text{mix}}(\phi^f)$, the dimensionless growth rate will be smaller for larger $\eta$ using \cref{eq:app_ratedep}. Hence, to demonstrate the claim at the start of the section, it only remains to show that $\bar{f}_g^{\text{mix}}$ is a monotonically decreasing function of $\phi$ for $\phi>\phi^f$ such that $\phi>\phi^f$ necessitates  $\bar{f}_g^{\text{mix}}(\phi)<\bar{f}_g^{\text{mix}}(\phi^f)$. In \Cref{subsec:growthdrivingstress}, it was established that $\bar{f}_g^{\text{mix}}$ is always monotonically decreasing with $\phi$ for $\chi\leq0.5$. Defining $\phi_{\text{max}} = 1/J^s_{\text{max}}$, where $J^s_{\text{max}}$ was defined in \Cref{subsec:growthdrivingstress} as the swelling ratio at which $\bar{f}_g$ attains its maximum value, we note that $\bar{f}_g^{\text{mix}}$ is a monotonically decreasing function of $\phi$ for $\phi>\phi_{\text{max}}$.  For $\chi>0.5$, we numerically verified that $\phi_f>\phi_{\text{max}}$  in the limit $G/\mu^* \to 0$. Thus $\bar{f}_g^{\text{mix}}$ is always a monotonically decreasing function of $\phi$ for $\phi>\phi^f$, irrespective of the value of $\chi$, and hence the claim at the start of the section holds true.

\section{Perfectly incompressible limit}
\label{app:perfecincomp} 

Here we consider the perfectly incompressible limit of the theory so that $K/G \to \infty$ and $J^e \to 1$. In rate form this means that  $\dot{J}^e = J^e \text{tr}(\dot{\nten{F}}^e{\nten{F}^e}^{-1}) = {\nten{F}^e}^{-T} \cdot  \dot{\nten{F}}^e = 0$. We do the following manipulation to rewrite this constraint
\begin{equation}
\begin{split}
{\nten{F}^e}^{-T} \cdot  \dot{\nten{F}}^e  &= {\nten{F}^e}^{-T}\cdot\left({\nten{F}^e}^{-T} {\nten{F}^e}^T\dot{\nten{F}}^e\right)  = \left({\nten{F}^e}^{-1}{\nten{F}^e}^{-T}\right)\cdot\left( {\nten{F}^e}^T\dot{\nten{F}}^e\right) \\ &= {\nten{C}^e}^{-1}\cdot\left( {\nten{F}^e}^T\dot{\nten{F}}^e\right)  = {\nten{C}^e}^{-1}\cdot\text{sym}\left( {\nten{F}^e}^T\dot{\nten{F}}^e\right) = {\nten{C}^e}^{-1} \cdot \dot{\nten{C}}^e = 0
\end{split}
\end{equation}
where we have used the definition of $\nten{C}^e$ in \cref{eq:Cedef3} and its symmetry.
We can thus add an arbitrary scalar times ${\nten{C}^e}^{-1} \cdot \dot{\nten{C}}^e$ to the stress power in \cref{eq:strespowers},
\begin{equation}
\nten{S} \cdot \dot{\nten{F}} =  \frac{1}{2}J^sJ^g\nten{T}^e\cdot\dot{\nten{C}^e} + J^g \nten{M}^{es} \cdot \nten{L}^g + \frac{J \dot{J}^s}{3 J^s} \text{tr}(\nten{T}) + \frac{1}{2}J^g \left(P {\nten{C}^e}^{-1}\right)\cdot\dot{\nten{C}^e} \label{eq:app_strespowers}
\end{equation}
where $P$ is an arbitrary scalar field (the Lagrange multiplier associated with the incompressibility constraint). Following the same process in the manuscript, we can now show that
\begin{equation}
\nten{T} = \frac{1}{J^s}\left(2\nten{F}^e \pdv{{\psi}_{g}}{\nten{C}^e}\/ {\nten{F}^e}^T - P \nten{I}\right),\quad \nten{M}^{es} = J^{s} {\nten{F}^{es}}^T \nten{T} {\nten{F}^{es}}^{-T},\quad \nten{T}^e = {\nten{F}^e}^{-1} \nten{T} {\nten{F}^e}^{-T}
\end{equation} 
while the Piola stress is given by \eqref{eq:Piolastresseqb}$_3$. All other quantities and prescriptions in \Cref{sec:theory} apart from the above stresses remain unchanged. \\

The specific form the mechanical free energy in \cref{eq:freeenergyspec} now reads
\begin{equation}
\hat{\psi}_{g}^{\text{{mech}}}(\nten{C}^{e},J^s) = \frac{G}{2} \left({(J^s})^{\frac{2}{3}}\text{tr}(\nten{C}^{e})  - 3 - 2\ln({J}^{s}) \right)   \label{eq:app_freeenergyspec}
\end{equation}
The modified expressions in \Cref{sec:spec_constit} for the incompressible limit are written below\footnote{A quick way to obtain the incompressible equations from the compressible theory is to set $K \ln(J^e) \to (G-P)$.} (unchanged expressions are not listed again for brevity),
\begin{align}
\nten{T} =  \frac{1}{J^{s}}\left(G \nten{B}^{es}   - P \nten{I}\right),\quad \nten{M}^{es} = G \nten{C}^{es} -P \nten{I}    \label{eq:app_cauchy_stress_spec}\\
\mu = \mu_0^f +  \mu^* \left(\ln\left(1-\phi\right) + \phi + {\chi}{\phi^2}\right) - G{\phi} + P\phi \label{eq:app_mu_spec}\\
f_g^{\text{mech}} = G \left(\ln(J^{s})  - \frac{1}{6} \left(\text{tr}(\nten{C}^{es}) -3\right) \right) \label{eq:app_fgmechspec}
\end{align}
For the case of non-isochoric species conversion, the total driving stress $f_g$ for volumetric growth is given by
\begin{equation}
f_g = \Delta \mu_0^{\text{mis}} + f_g^{\text{mix}} + f_g^{\text{mech}} + \left(\alpha-1\right)\left(\mu^\text{mix}(\phi) - G{\phi} + P\phi\right)  
\end{equation}
where $f_g^{\text{mech}}$ in the incompressible limit is expressed in \cref{eq:app_fgmechspec}.

\subsection{Uniform spherical fields in limit of infinitely fast diffusion}
\label{sec:app_uniformswell_incomp}

Consider the uniform swelling-growth problem in \Cref{subsec:uniformswell} where we assume infinitely fast diffusion, now with the added constraint of perfect incompressibility. The uniform elastic deformation field is now $\nten{F}^e = \nten{I}$ since $J^e =1$. Using this along with \cref{eq:uniformsol}$_{4,5}$ in \cref{eq:app_mu_spec,eq:app_cauchy_stress_spec}, we arrive at the following equations
\begin{equation}
-P_b =  \phi\left(G \phi^{\frac{-2}{3}}   - P \right), \quad \mu^{\text{mix}}(\phi) - G{\phi} + P\phi = 0
\end{equation} 
Eliminating $P$ from the two equations and non dimensionalising the resulting equation yields \cref{eq:eqb_sol_incomp}. Further, substituting $\nten{F}^e =\nten{I}$ in \cref{eq:app_fgmechspec} results in the following equation
\begin{equation}
f_g^{\text{mech}} = G \left(\ln(J^{s})  - \frac{1}{2} \left((J^s)^{\frac{2}{3}} -1\right) \right)
\end{equation}
so that the driving force is purely a function of the swelling ratio or solid volume fraction for the uniform swelling-growth problem in the infinitely fast diffusion limit with perfect incompressibility, 
\begin{equation}
f_g = f_g^\infty = \Delta \mu_0^{\text{mis}} +  {f}_g^{\text{mix}}(\phi) +  G \left(\ln(\phi^{-1})  - \frac{1}{2} \left(\phi^{-\frac{2}{3}} -1\right) \right)
\end{equation}
In the case of isochoric species conversion, $\Delta \mu_0^{\text{mis}} =  \Delta \mu_0$. Non-dimensionalizing the result yields \cref{eq:exact_fg}.

\section{Numerical implementation}
\label{app:numerics}

Here we outline the numerical procedure used to solve the spherically symmetric governing equations in \Cref{subsec:spherical symmetry}, which is written formally in \Cref{algo:FDalgo}. The spatial domain $\bar{R}$ was discretized using $N=200$ equally spaced points. The mechanical and swelling equilibrium equations were solved using finite difference schemes that solve a nonlinear system of equations using Matlab's `fsolve' solver while supplying the Jacobian matrix (derivative of residual with respect to the variables) using the `SpecifyObjectiveGradient' option. An exact analytical expression for the Jacobian matrix  was coded for the mechanical equilibrium solver while an approximate Jacobian matrix was supplied for the swelling equilibrium solver (which was sufficient for fast enough convergence). Simultaneous mechanical and swelling equilibrium is ensured using a staggered scheme that is run till convergence. 
 The evolution equations for growth were integrated using a fourth order Runge-Kutta (RK4) scheme. After a convergence analysis in time, a conservative value of $\dd{\bar{t}_g} \sim 10^{-3}$ was chosen for the stepping size of the time integration.

\algdef{SE}[DOWHILE]{Do}{doWhile}{\algorithmicdo}[1]{\algorithmicwhile\ #1}%

\begin{algorithm}
\caption{Spherically Symmetric Swelling Growth}\label{algo:FDalgo}
\begin{algorithmic}
\Procedure{SwellingGrowth}{}
\State $\bar{t}_g,\vv{\lam_r^g},\vv{J^g}  \gets 0, \vv{1}, \vv{1}$ 
\While{$\bar{t}_g<\bar{t}_g^f$}
\State $\vv{\lam_r^g},\vv{J^g} \gets $$\Call{GrowthStep}{\vv{\lam_r^g},\vv{J^g},\dd{\bar{t}_g}}$
\State $\vv{r},\vv{J}^s \gets\Call{MechSwellEqb}{\vv{\lam_r^g},\vv{J^g}}$
\State $\bar{t}_g \gets \bar{t}_g + \dd{\bar{t}_g}$
\EndWhile
\EndProcedure\\
\Procedure{MechSwellEqb}{$\vv{\lam_r^g},\vv{J^g}$} \Comment{Staggered solver for equilibrium}
\State $\vv{r} \gets \Call{MechEqb}{\vv{J}^s,\vv{\lam_r^g},\vv{J^g}} $ \Comment{Finite difference scheme that solves \eqref{eq:spher_eqb},\eqref{eq:mu_bc}$_1$}
  \Do
\State $\vv{J}^s \gets \Call{SwellEqb}{\vv{r},\vv{\lam_r^g},\vv{J^g}} $ \Comment{Finite difference scheme that solves \eqref{eq:reacdiff_spherical},\eqref{eq:mu_bc}$_2$}
\State $\vv{\mu}_1 \gets \Call{ChemPot}{\vv{\lam_r^g},\vv{J^g},\vv{r},\vv{J}^s}$ \Comment{Calculate $\bar{\mu}$ using \eqref{eq:mu_dimless}}
\State $\vv{r} \gets \Call{MechEqb}{\vv{J}^s,\vv{\lam_r^g},\vv{J^g}} $
\State $\vv{\mu}_2 \gets \Call{ChemPot}{\vv{\lam_r^g},\vv{J^g},\vv{r},\vv{J}^s}$ 
\doWhile{$\norm{\vv{\mu}_{2} - \vv{\mu}_1} < \epsilon $} \Comment{$\epsilon$ is a small parameter to check convergence} 
\State \textbf{return} $\vv{r},\vv{J}^s$
\EndProcedure\\

\Procedure{GrowthStep}{$\vv{\lam_r^g},\vv{J^g},\dd{\bar{t}_g}$} \Comment{RK4 integration scheme}
\State $\vv{y} \gets \left[\vv{\lam_r^g},\vv{J^g}\right]$
\State $\vv{k_1} \gets \Call{dydt}{\vv{y}}$
\State $\vv{k_2} \gets \Call{dydt}{\vv{y} + \frac{1}{2} \vv{k_1} \ {\dd{\bar{t}_g}}}$
\State $\vv{k_3} \gets \Call{dydt}{\vv{y} + \frac{1}{2} \vv{k_2} \ {\dd{\bar{t}_g}}}$
\State $\vv{k_4} \gets \Call{dydt}{\vv{y} + \vv{k_3}\ {\dd{\bar{t}_g}}}$
\State $\vv{y} \gets \vv{y} + {\dd{\bar{t}_g}}/{6} \left(\vv{k_1} + 2\vv{k_2} + 2\vv{k_3} + \vv{k_4}\right)$
\State $\vv{\lam_r^g},\vv{J^g} \gets \vv{y}(1:N), \vv{y}(N+1,2:N)$ \Comment{Spatial domain $\bar{R}$ is discretized using $N$ points}
\State \textbf{return} $\vv{\lam_r^g},\vv{J^g}$
\EndProcedure\\

\Procedure{dydt}{$\vv{y}$} 
\State $\vv{\lam_r^g},\vv{J^g} \gets \vv{y}(1:N),\vv{y}(N+1:2N)$
\State $\vv{r},\vv{J}^s \gets \Call{MechSwellEqb}{\vv{\lam_r^g},\vv{J^g}}$
\State $\vv{z}_1 \gets \Call{DLamrgDt}{\vv{\lam_r^g},\vv{J^g},\vv{r},\vv{J}^s}$ \Comment{Calculates $\dv{\lam_r^g}{\bar{t}_g}$ using \eqref{eq:growthlawspher}$_1$}
\State $\vv{z}_2 \gets \Call{DJgDt}{\vv{\lam_r^g},\vv{J^g},\vv{r},\vv{J}^s}$ \Comment{Calculates $\dv{J^g}{\bar{t}_g}$ using \eqref{eq:charac2}$_4$ and \eqref{eq:gammabar_dimlss} }
\State \textbf{return} $\left[\vv{z}_1, \vv{z}_2\right]$
\EndProcedure
\end{algorithmic}
\end{algorithm}

\subsection{Confinement dimensions for tumor growth simulations}
\label{app:numerics_parameters}

The dimensions of the confining shells used for simulating Cases B-D of the tumor growth experiments are listed in \Cref{table:dimensions} where the inner volume of the confinement is given by $\frac{4}{3}\pi A^3$. The capsule volumes have been obtained from the dashed lines of Fig. 2E in \cite{alessandri2013cellular}, a slight correction has been made for Case D such that the confinement volume is chosen based on the volume at which the tumor growth rate starts deviating from the free growth curve. The estimation of dimensions of the confining shells in \cite{alessandri2013cellular} is approximate and is inconsistent with the capsule volumes. Thus we have chosen representative dimension ratios ($B/A$) based on their reported values for thick and thin capsules.  
\newpage

\begin{table}[H]
\caption{Dimensions of confinement in tumor growth experiments}
\centering
\begin{tabular}{llll}
\hline
Case & Inner Volume (mm$^3$) & $B/A$  & $J_c/\bar{J}_0$ \\ \hline
B    & 0.12                  & 1.25 & 40                               \\
C    & 0.003                 & 1.25 & 1                                \\
D    & 0.0034                & 1.12 & 1.135                            \\ \hline
\end{tabular}
\label{table:dimensions}
\end{table}

\end{document}